\documentclass[useAMS,usenatbib]{mn2e}

\usepackage{aas_macros}
\usepackage{dcolumn}
\usepackage{amsmath}
\usepackage{amssymb}
\usepackage{graphicx,times}

\begin{document}

\title[The Stellar Populations of Bright Coma Cluster Galaxies]{The Stellar Populations of Bright Coma Cluster Galaxies}
\author[J. Price et al.]
{J. Price$^{1}$\thanks{james.price@bristol.ac.uk},
S. Phillipps$^{1}$, 
A. Huxor$^{1}$, R.J. Smith$^2$, J.R. Lucey$^2$ \\
$^{1}$Astrophysics Group, H.H. Wills Physics Laboratory, University of Bristol, Tyndall Avenue,  Bristol BS8 1TL, UK \\
$^{2}$Department of Physics, University of Durham, Durham, DH1 3LE, UK }

\date{MNRAS accepted}

\pagerange{\pageref{firstpage}--\pageref{lastpage}} \pubyear{2010}

\maketitle

\label{firstpage}

\begin{abstract}
In this paper we study the stellar populations of 356 bright, $M_{r}$ $\leq$ -19, Coma cluster members located in a 2 degree field centred on the cluster core using SDSS DR7 spectroscopy. We find $\sim$ 31\% of the sample have significant emission in H$\beta$, [OIII]5007, H$\alpha$ or [NII]6585, due to star-formation or AGN/LINER activity. The remaining portion of the sample we describe as passive or quiescent. Using line-ratio diagnostics, we find the fraction of galaxies displaying AGN/LINER type emission increases with increasing galaxy luminosity while the star-forming fraction decreases. For the quiescent galaxies we find strong correlations between absorption line index strength and velocity dispersion ($\sigma$) for CN2, C4668, Mgb and H$\beta$. Employing a planar analysis technique that factors out index correlations with $\sigma$, we find significant cluster-centric radial gradients in H$\beta$, Mgb and C4668 for the passive galaxies. We use state-of-the-art stellar population models \citep{schiavon07} and the measured absorption line indices to infer the single-stellar-population-equivalent (SSP-equivalent) age and [Fe/H] for each galaxy, as well as their abundance patterns in terms of [Mg/Fe], [C/Fe], [N/Fe] and [Ca/Fe]. For the passive galaxy subsample we find strong evidence for ``archaeological downsizing", with age $\propto \sigma^{0.90 \pm 0.06}$. This trend is shown to be robust against variations in sample selection criteria (morphologically early-type vs spectroscopically quiescent), emission-line detection thresholds, index velocity broadening corrections and the specific SSP model employed. Weaker positive correlations are obtained between $\sigma$ and all other measured stellar population parameters. We recover significant cluster-centric radial stellar population gradients for the passive sample in SSP-equivalent age, [Mg/Fe], [C/Fe] and [N/Fe]. These trends are in the sense that, at fixed velocity dispersion, passive galaxies on the outskirts of the cluster are 24\% $\pm$ 9\% younger with lower [Mg/Fe] and [N/Fe] but higher [C/Fe] than those in the cluster core. We find no significant increase in cluster-centric radial stellar population gradients when fitting to a passive galaxy subset selected to cover the cluster core and South-West region, which contains the NGC 4839 subgroup. Thus we conclude that the NGC 4839 in-fall region is not unique, at least in terms of the stellar populations of bright galaxies. We speculate that the more pronounced cluster-centric radial gradients seen by other recent studies may be attributed to the luminosity range spanned by their samples, rather than to limited azimuthal coverage of the cluster. Finally, for our passive sample we have found an age-metallicity anti-correlation which cannot be accounted for by correlated errors.
 
\end{abstract}


\section{Introduction}

Rich galaxy clusters have long been the target of studies attempting to address the fundamental questions of galaxy formation and evolution. They are ideal laboratories for this task as they harbour large numbers of galaxies distributed across a wide range in local density and at a common distance. The most direct route to answer these questions is to observe clusters at a range of redshifts and trace out the assembly history of their galaxy populations. Unfortunately this method is observationally inefficient, requiring significant amounts of telescope time, and of course often relies on data of lower signal-to-noise (S/N). An alternative approach is to conduct detailed analysis of the stellar populations of large samples of galaxies in nearby clusters and then use these characteristics as a function of their mass, environment or other key properties, to try and reconstruct their formation and evolutionary channels.

The latter of the methods described above, which may be termed a type of galaxy archaeology, was pioneered by works which identified the environmental dependence of galaxy morphology in clusters \citep{dressler80}, confirmed that red galaxies, predominately of early-type, follow a well-defined sequence in colour-magnitude space while later types occupy a blue cloud \citep{aaronson81,bower92,strateva01}, and have analysed the scaling relations seen for early-type galaxies \citep{kormendy77,dressler87,djorgovski87,bender92}. 

In a similar vein, the comparison of integrated photometric and spectroscopic observations of unresolved galaxies with stellar population models has become an increasingly valuable tool for galaxy archaeology. While it is possible, at least partially, to break the well known age-metallicity degeneracy by combining appropriate sets of broadband optical and near-infrared colours \citep{james06,carter09}, the standard technique typically relies on using spectral line indices which target a small number of information-rich absorption features \citep{burstein84,rose85}, with \cite{faber85} introducing the now de facto standard Lick indices. In particular \cite{worthey94} demonstrated how effective pairs of such indices can be at disentangling the effects of age and metallicity in unresolved optical spectra. 

Even prior to this seminal work, evidence had been gathering that nearby early-type galaxies had non-solar $\alpha$-element abundances \citep{peletier89,worthey92,trager97}, an effect that manifests itself as different metallic line indices implying different metallicities and ages for a given galaxy. Typically this has been found to be more prevalent in giant early-type galaxies and is considered evidence for shorter star-formation episodes in more massive galaxies \citep{thomas05}. This conclusion is driven by the enrichment history of the interstellar medium out of which the current generation of stars formed. On timescales $\lesssim$ 1 Gyr this is dominated by Type II supernova, yielding more $\alpha$-elements (normally traced via Mg) than Fe relative to the solar neighbourhood, but on longer timescales by Type Ia supernova which redress the balance. With this in mind significant effort has recently been devoted to including non-solar alpha element abundance patterns in stellar population models \citep{trager00a,proctor02,thomas03,thomas04,schiavon07,dotter07}.

At $\sim$ 100 Mpc Coma is the nearest rich and dense galaxy cluster and so the stellar populations of its galaxies have been the subject of extensive study for some time. Below we review the key findings of a selection of relevant studies on or involving Coma members.

Using multifibre spectroscopic observations of 125 early-type galaxies with -20.5 $\lesssim$ M$_{B}$ $\lesssim$ -16 located in two fields, one centred on the cluster core and another $\sim$ 40$^\prime$ to the southwest, \cite{caldwell93} identified a number of galaxies with what they termed ``abnormal" spectra in the latter field. In this case they defined abnormal to refer to spectra that show signs of recent star formation and/or nuclear activity. They also note that these galaxies are closely associated with an area of enhanced x-ray emission, now generally accepted to be indicative of the NGC 4839 group merging with the main bulk of the cluster \citep{briel92}.

\cite{jorgensen99} analysed the stellar population parameters of 115 early-type cluster members with Gunn r $\lesssim$ 15, although only 71 galaxies in her sample had velocity dispersions and all relevant indices measured. Using the stellar population models of \cite{vazdekis96} she derived a relatively low median age of $\sim$ 5 Gyr with significant intrinsic scatter ($\pm$ 0.18 dex) and found [Fe/H] $\sim$ 0.08, again with sizable scatter ($\pm$ 0.19). She also concluded that age and [Fe/H] were not correlated with luminosity or velocity dispersion. However, strong correlations with these descriptors were found for [Mg/Fe].

The spectroscopic survey of \cite{poggianti01} observed 278 Coma members (257 with no detectable emission by their criteria), across the exceptionally wide magnitude range of -20.5 $\lesssim$ M$_{B}$ $\lesssim$ -14, located in two 32.5 x 50.8 arcmin$^{2}$ fields, one targeted on the cluster core and one on the NGC 4839 subgroup. They found that metallicity positively correlates with luminosity but with a substantial scatter and that, interestingly, the scatter increases for fainter galaxies. In addition they identify an age-metallicity anticorrelation in any given luminosity bin, although as they do not take into account correlated errors \citep[see e.g.][]{kuntschner01} the strength of this conclusion is somewhat limited. Further still, using the same dataset, \citet[][hereafer C02]{carter02} find evidence for a cluster-centric radial gradient in the stellar populations of their sample galaxies. They interpret this trend as a variation in metallicity in the sense that galaxies in the cluster core are more metal rich than those in the outer regions.

Contemporaneously with the Poggianti et al. study, \cite{moore01} obtained spectroscopy for 87 bright early-type galaxies (M$_{B}$ $\lesssim$ -17) in the core of the cluster. Using the \cite{worthey94} models and a multiple hypothesis testing technique, he found this sample to have a sizable metallicity spread, -0.55 $\leq$ [Fe/H] $\leq$ +0.92, but a uniform 8 Gyr age of formation with an intrinsic scatter of $\pm$ 0.3 dex (4 to 16 Gyr). He identified no correlation between age and velocity dispersion but did find [Fe/H]-$\sigma$ and [Mg/Fe]-$\sigma$ relations.

More recently, studies such as those of \cite{thomas05} and \cite{sanchezblaz06b} have included Coma members in their high density samples when attempting to constrain the environmental dependence of galaxy stellar populations. The former report that all of the stellar population parameters they measure, age, total metallicity and [$\alpha$/Fe], are correlated with velocity dispersion regardless of local density and that the ages of galaxies in low density environments appear systematically lower than those in high density regions. By contrast \cite{sanchezblaz06b} find a less well defined picture. Only their low density sample has a significant metallicity-$\sigma$ correlation when using Mg$b$ to estimate metallicity, as opposed to a positive correlation regardless of metallicity indicator for their high density subset. Even more interestingly they observe age-$\sigma$, or so-called ``downsizing", relations for low density regions and the Virgo cluster but not in their Coma galaxies. Finally they find no age-metallicity anticorrelation for their high density sample.

\citet[][hereafter T08]{trager08} conducted a detailed analysis of 12 early-type Coma cluster members in the magnitude range -21.5 $\lesssim$ M$_{B}$ $\lesssim$ -16.5 based on Keck/LRIS spectra. They concluded that their sample is consistent with having a remarkably young uniform SSP-equivalent age of 5.2 $\pm$ 0.2 Gyr and find no indication of an age-$\sigma$ relation, a result which they suggest is supported by the majority of previous work on the stellar populations of Coma early-type galaxies.

In the most recent study, at the time of writing, focusing on the stellar populations of bright Coma members, \cite{matkovic09} observed 74 early-type galaxies with -22 $\leq$ M$_{R}$ $\leq$ -17.5 ($\sim$ -20.5 $\leq$ M$_{B}$ $\leq$ -16) in the cluster core. Their sample spans the velocity dispersion range 30 $\leq$ $\sigma$ $\leq$ 260 km s$^{-1}$, 32 galaxies having $\sigma$ $\geq$ 100 km s$^{-1}$. Due to large uncertainties in their derived ages they are unable to confirm an age-$\sigma$ relation but find that on average lower $\sigma$ galaxies are indeed younger with lower metallicity and [$\alpha$/Fe] than cluster members with higher velocity dispersions.

Finally, while somewhat fainter in luminosity coverage than this work, the most detailed study to date focusing primarily on Coma dwarf galaxies is that of \cite{smith08} (hereafter S08) and \cite{smith09a}. They observed 89 red cluster members with -18.5 $\lesssim$ M$_{B}$ $\lesssim$ -15.75 split across two fields, one targeting the cluster core and one a degree to the south-west of the cluster centre. They found evidence for a strong cluster-centric radial gradient in galaxy age with galaxies in the core typically having older ages than systems in the outer region. A similar trend is also reported for [Mg/Fe], with stronger Mg enhancement seen in the cluster core than the outer region, while no such correlation is observed for [Fe/H].

As such, it is clear that even in this well studied cluster the stellar population parameters of its host galaxies and how they scale with other galaxy properties are still somewhat in debate. In this current work we aim to resolve many of the issues discussed above and to clarify the effects of velocity dispersion and environment on bright Coma cluster members in as homogeneous a fashion as possible. To this end we employ spectroscopy from the Sloan Digital Sky Survey\footnote{http://www.sdss.org} Data Release 7 \citep{abazajian09} and the up-to-date stellar population models of \citet[][hereafter referred to as the Schiavon models]{schiavon07}, which allow the determination of elemental abundance patterns in addition to SSP-equivalent age and metallicity.

The paper is outlined as follows. In section 2 we briefly overview the SDSS observations and detail our sample selection. In section 3 we perform emission line detection and measure our galaxies' velocity dispersions and absorption line indices. In section 4 we derive and analyse their stellar population parameters as a function of environment and galaxy velocity dispersion. In section 5 we discuss our findings and in section 6 we review our conclusions.

\section{The Sample}

\subsection{SDSS Data}

The SDSS uses a dedicated 2.5 m telescope at Apache Point Observatory with a large format mosaic CCD camera for imaging and a pair of fibre-fed double spectrographs each with 320 fibres. Both spectrographs have a blue and red channel that when combined cover $\sim$ 3800-9100 \AA\ at a resolution of R $=$ 1850-2200 \citep[for further details see][]{york00}. The spectra are flux calibrated by the SDSS spectral reduction pipeline using 16 spectroscopic standards on each plate, colour selected to be F8 subdwarfs. The SDSS fibre diameter of 3$^{\prime\prime}$ equates to 1.4 kpc at Coma assuming H$_{0}$ = 71 km s$^{-1}$ Mpc$^{-1}$, $\Omega_{M}$ = 0.27 and $\Omega_{\Lambda}$ = 0.73 \citep{hinshaw09}. We adopt a distance modulus $m-M = 35.0$ for Coma. 

All spectroscopy for this work comes from the SDSS Main Galaxy Sample, the target selection of which is detailed in \cite{strauss02}. Briefly, SDSS imaging is used initially for star-galaxy separation and then to measure an $r$-band Petrosian magnitude and define a circular aperture containing half a galaxy's Petrosian flux. The Main Galaxy Sample then consists of galaxies with $r_{\mathrm{petro}}$ $\leq$ 17.77 and mean surface-brightness $<\!\mu_{r}\!>_{50}$ $\leq$ 24.5 mag arcsec$^{-2}$ with r$_{\mathrm{fibre}}$ $<$ 19, after correcting for Galactic extinction. Objects with $r_{\mathrm{fibre}}$ $<$ 15 are rejected to limit the effects of cross talk when extracting the spectra of neighbouring faint objects, as are targets that are flagged as saturated, bright or blended and not deblended by the photometric pipeline. For reference, applying the above constraints to the SDSS photometric database we estimate $\sim$ 96 \% of galaxies in our Coma field (see below) have SDSS spectra. It is more than likely the remaining fraction result from having a neighbour within 55$^{\prime\prime}$, the minimum separation permitted between fibres.

\subsection{Sample Selection}

\begin{figure}
\flushleft
\scalebox{0.44}[0.44]{\includegraphics{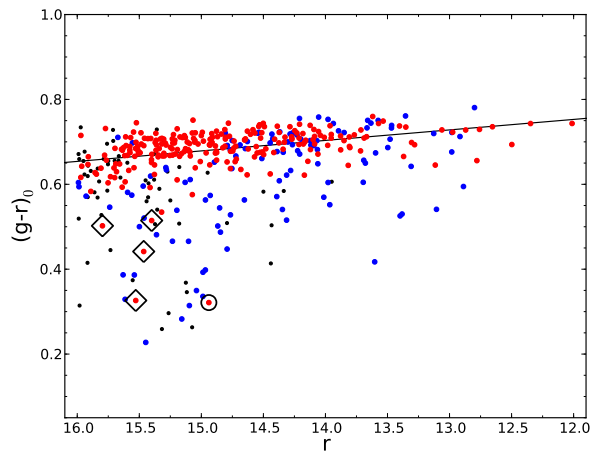}}
\caption{Colour-magnitude diagram for the 417 Coma members in our sample (see text for selection criteria). Red and blue filled circles are non-emission line and emission line galaxies (see Section \ref{emidetect}) respectively with S/N $\geq$ 20 per \AA. Black points are galaxies which did not meet our S/N cut or, in the case of detected emission, were not cleaned sufficiently (see Section \ref{emidetect}). The black diamonds highlight galaxies which are significantly bluer than the majority of our non-emission sample, determined here by them having a (g-r) colour more than 3 times bluer than the rms scatter of an unweighted fit to the full non-emission sample at their luminosity. The black ringed galaxy, GMP 2946, has a number of flags that indicate problems with its Petrosian magnitude (see text for details).}
\label{cmr}
\end{figure}

The rationale behind our sample selection is simple, we aim to include as many Coma members as possible and extend the recent work of \cite{smith08} and \cite{smith09a} to brighter magnitudes. To this end we include all galaxies with $r_{\mathrm{petro}}$ $\leq$ 16 within 2$^{\circ}$ ($\sim$ 3.3 Mpc) of our designated cluster centre at RA = 12:59:48, Dec = +27:58:50, which is approximately half way between NGC 4889 and NGC 4874, that have spectroscopy in SDSS DR7. We note that all SDSS photometric values quoted throughout the rest of this paper are Petrosian magnitudes unless otherwise specified and have been corrected for Galactic extinction and k-correction using the provided parameters from the SDSS database. In addition we opt for an inclusive redshift cut of $0.01 \leq z \leq 0.04$ based on $\sim$ 4$\sigma$ limits of the cluster velocity dispersion from \cite{colless96} and taking Coma to be at $z$ = 0.0231\footnote{http://nedwww.ipac.caltech.edu/}. These criteria result in our sample containing 417 confirmed Coma members. We note that a more conservative 3$\sigma$ cut only removes six galaxies and does not significantly affect our results.

Our magnitude cut is set by a trade off between maximising sample size, and completeness, and obtaining galaxy spectra that have sufficient S/N; in this case we opt for a median S/N per \AA\ $\geq$ 20 between 4000-6000 \AA, to suitably constrain stellar population parameters (see Section~\ref{paramerrors}). Effectively this amounts to a cut of $r_{\mathrm{fibre}}$ $\lesssim$ 18, which results in $\sim$ 87\% (363) of our sample having the desired S/N. Of course the majority of our incompleteness, due to our required S/N, occurs in the final 15.75 $\leq r \leq$ 16.0 bin where we are $\sim$ 44\% complete.

In Fig. \ref{cmr} we plot the colour-magnitude diagram for our sample. As detailed in Section~\ref{emidetect}, galaxies are, at least initially, only subdivided further based on whether they show detectable emission and as such it is interesting to see from Fig. \ref{cmr} that almost all spectroscopically quiescent galaxies (98\%) that meet our selection requirements are compatible with being on the cluster's red sequence. The remaining five galaxies are worth further comment. 

\begin{table}
\centering
\caption{Classification data for the four blue passive galaxies highlighted by black diamonds in Fig. \ref{cmr}. Morphology is taken from NED and the quoted H$\delta$ equivalent widths, in units of \AA, are those measured by the SDSS pipeline.}
\label{k+a}
\begin{tabular}{cccc}
\hline
ID & (g-r)$_{0}$ & Morphology & EW(H$\delta$)$_{sdss}$ \\
\hline
GMP 2640 & 0.33 & S0p & 7.9 \\
GMP 3892 & 0.44 & SB0/a & 5.6 \\
GMP 3439 & 0.50 & SB0 & 2.3 \\
GMP 4974 & 0.51 & SA0 & 3.75 \\
\hline
\end{tabular}
\end{table}

The galaxy highlighted by a black circle in Fig. \ref{cmr} is GMP 2946 \citep{gmp83}. This object is projected onto the halo of the giant galaxy NGC 4889, and its SDSS photometry is erroneous. More careful treatment shows that it lies on the cluster's red sequence. We tabulate the relevant details of the four other blue passive galaxies, denoted by black diamonds in Fig. \ref{cmr}, in Table \ref{k+a} where the morphological information is taken from NED and the H$\delta$ absorption line equivalent widths are those measured by the SDSS pipeline {\sc spectro1d}. We note that the \cite{poggianti04} classification requirements for post-starburst k+a galaxies are no emission and EW(H$\delta$) $>$ 3 \AA. Thus three out of four of these galaxies are compatible with being k+a systems. Furthermore, all three are located toward the faint end of our sample's magnitude range, a fact which is perhaps not surprising given the lack of bright blue k+a galaxies in Coma as reported by \cite{poggianti04}. Hereafter we will refer to these four galaxies as the blue subset of the passive sample and where comparisons to literature studies involving colour selected samples are necessary, we intend to mask this subset.

\section{Spectroscopic Measurements}
\subsection{Emission line detection}
\label{emidetect}

\begin{figure*}
\begin{minipage}{170mm}
\flushleft
\scalebox{0.65}[0.55]{\includegraphics{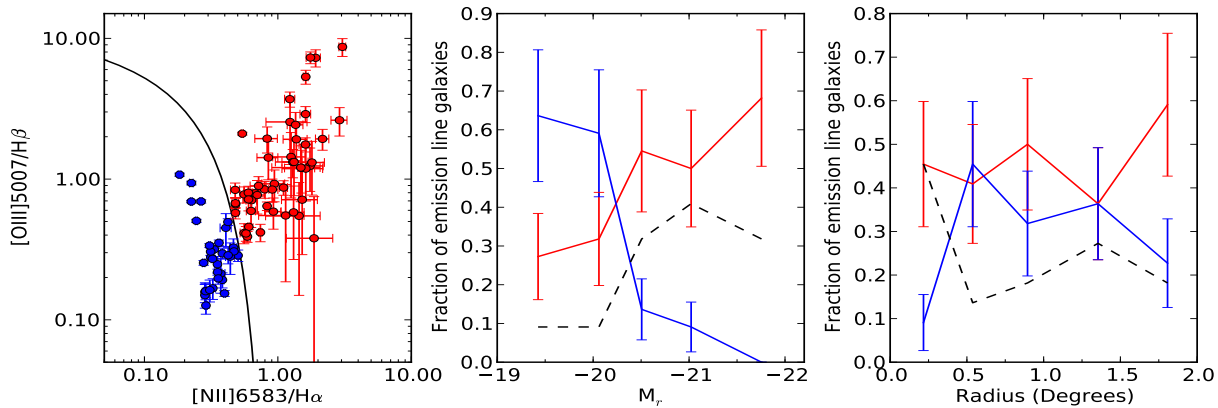}} 
\caption{Emission-line analysis diagrams: The left panel displays the line ratio diagnostic space of \citet{bpt81} with the demarcation line of \citet{kauffmann03} used to classify the 110 galaxies in our emission-line sample. Red circles are AGN/LINER while blue circles are star-forming galaxies. The middle and right panels show the fraction of emission line galaxies which are AGN/LINER (red lines) or star-forming (blue lines) as a function of luminosity and cluster-centric radius. The black dashed lines indicate the fraction of galaxies for which a definite classification could not be made (referred to as ambiguous emission driver, or AED, galaxies in the text).}
\label{bpt}
\end{minipage}
\end{figure*}

To conduct our stellar population analysis we must first identify and reject, or correct \citep[e.g.][]{trager00a}, galaxies with nebular emission which acts to infill the stellar H$\beta$ absorption, our key Balmer line age indicator, and so results in overestimated stellar ages. We could, of course, opt to use the higher order Balmer lines H$\gamma$ or H$\delta$ which in turn suffer less from emission infill. However, here we prefer H$\beta$ for its relative stability against abundance pattern variations and, hence, its essential role in the fitting code we shall employ (see Section \ref{paramerrors}). We also wish to avoid including galaxies with emission in [NI]5198 which acts to perturb the Mgb index \citep{g&e96}. Fortunately the broad wavelength coverage of the SDSS spectra allows access to H$\alpha$ and [NII]6583 which are ideal for flagging galaxies in our sample that host active galactic nuclei (AGN or LINER) or are currently undergoing star formation \citep[see e.g.][]{smith07}.

As a first pass we make use of the emission line measurements taken by {\sc spectro1d}. The code uses a median/mean filter to fit each spectrum's continuum with emission and absorption features masked. Emission, and pseudo-Lick absorption, lines are then measured on the continuum subtracted spectrum as constrained single gaussians, with lines in close proximity to one another being fit as a blend. Here we deem galaxies with equivalent widths (EW) in any of H$\beta$, [OIII]5007, H$\alpha$ or [NII]6585 $>$ 0, detected at the 3$\sigma$ level, to be emission-line candidates. This cut selects 186 galaxies with S/N $\geq$ 20. 

The method of detecting and measuring emission lines has been significantly improved recently with the introduction of {\sc Gandalf} by \cite{sarzi06}, itself an extension of the pixel fitting code of \cite{cap&em04} which will be detailed later. Instead of requiring the observed spectra to be continuum subtracted prior to analysis, they show that it is possible to simultaneously fit the stellar component and nebular emission lines. This approach acts to remove the inherent bias caused by the masking of emission line regions, which often also contain important absorption features, when fitting the stellar continuum in the former method. Indeed we note that as the technique used by the SDSS does not make any attempt to disentangle absorption and emission components, we seek to check the robustness of the survey's emission line detections.  

As input, {\sc Gandalf} requires a set of template spectra with higher spectral resolution than the observations. Here we use 75 randomly selected spectra of types F to M from the Indo-U.S. Stellar Library \citep{valdes04}, which have a wavelength coverage of 3460-9464 \AA\ observed at FWHM $\sim$ 1 \AA. We also include another 10 templates from the same library which are specifically chosen to have strong Mgb indices and so further help minimise template mismatch in systems with strong magnesium absorption. The code then finds the best linear combination of the template spectra and emission lines, again parameterised as gaussians, in a $\chi^{2}$ sense with respect to the observed data. 

We opt to fit our sample in the wavelength range 4000-6800 \AA\ as this includes H$\alpha$ and [NII]6585, our primary emission indicators, and, if detected, allows for the emission features which affect the Lick indices relevant to this work to be corrected. Following \cite{sarzi06} we require that bona-fide emission lines be detected at an amplitude-to-noise ratio (A/N) $\geq$ 4, where noise is defined as the residual-noise level between the best fit and observed spectrum and therefore this cut actively screens against template mismatch generating fake emission lines. As before, a galaxy with this level of emission in any of H$\beta$, [OIII]5007, H$\alpha$ or [NII]6585 is flagged. By comparing EW as a function of A/N for each line, we empirically estimate that A/N = 4 corresponds to EW(H$\alpha$) $\sim$ 0.5 \AA, EW([NII]6585) $\sim$ 0.6 \AA\ and EW([OIII]5007) $\sim$ 0.5 \AA. In the case of H$\beta$ we use H$\alpha$ to identify emission as we do not expect to detect the former without the latter, such that A/N$_{H\alpha}$ = 4 corresponds to EW(H$\beta$) $\sim$ 0.25 \AA. Thus these represent the emission line detection limits for our data. This method returns 117 galaxies with significant emission, 63\% of the initial 186 flagged using the SDSS data. 

We then use {\sc Gandalf} to correct, by subtracting off the best gaussian fit, any line with A/N $\geq$ 4. For galaxies with detectable H$\alpha$ emission but no significant H$\beta$ emission we choose to force a H$\beta$ correction. We have checked and found that all H$\alpha$/H$\beta$ flux ratios are compatible, within errors, with Case B recombination \citep{osterbrock05}. During the testing of this technique we found an issue whereby the emission lines of certain galaxies were poorly fit, often, but not always, because the line in question was exceptionally strong, and as such left large residuals in the cleaned spectra. To quantify the quality of each fit we compute a $\chi^{2}_{\nu}$ between the best fit and observed spectrum in the immediate region around each line and visually inspect the corrected spectra to find the maximum acceptable $\chi^{2}_{\nu}$. In practise the most important line to adequately clean is H$\beta$ as it is our primary age indicator and, together with [OIII]5007, one of the strongest emission lines in the wavelength range covered by the Lick indices. With this in mind we determined $\chi^{2}_{H\beta}$ $\leq$ 1.1 to provide suitable cleaning.

To summarise, our final sample consists of 246 Coma members with S/N $\geq$ 20 \AA$^{-1}$, median S/N = 35 \AA$^{-1}$, and no emission down to the limits defined above, and 110 with S/N $\geq$ 20 \AA$^{-1}$, median S/N = 34 \AA$^{-1}$  and $\chi^{2}_{H\beta}$ $\leq$ 1.1 with detectable emission in any of H$\beta$, [OIII]5007, H$\alpha$ or [NII]6585. We note that both spectral classifications, passive or emission line, are based on the spectrum obtained from the region of each galaxy sampled by the 3$^{\prime\prime}$ (1.4 kpc) SDSS fibre and may not apply more globally to the galaxy as a whole.

\subsection{Emission line Analysis}
\label{emiclass}

\begin{figure}
\flushleft
\scalebox{0.36}[0.36]{\includegraphics{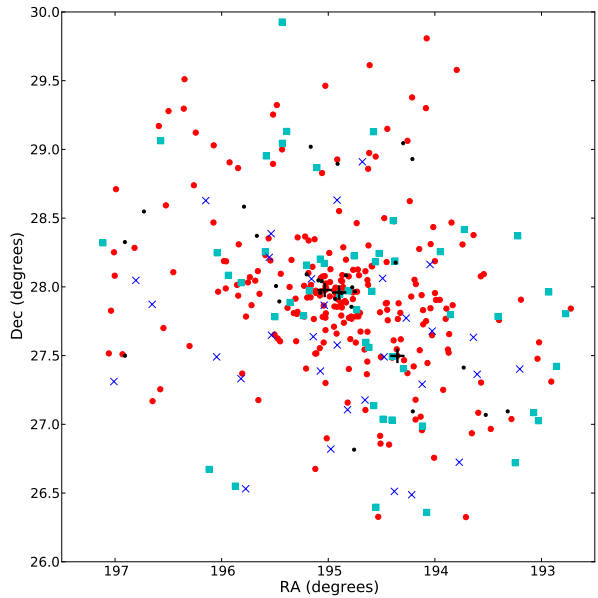}}
\caption{Distribution on the sky of the 356 galaxies in our final sample (see text for details). Galaxies are split into passive (red filled circles), AGN/LINER (cyan), star forming (blue crosses) and those with an ambiguous emission driver (black points) based on the approach presented in Section \ref{emiclass}. The large black symbols highlight the positions of NGC 4874, NGC 4889 and NGC 4839.}
\label{skycover}
\end{figure}

Having determined which galaxies in our sample show emission within the spatial coverage of the SDSS fibre, we next seek to discern what is driving said emission. To this end, in Fig. \ref{bpt} we plot our {\sc Gandalf} derived line ratios on a version of the \cite{bpt81} or BPT diagnostic diagram. Next we use the empirical relation of \cite{kauffmann03}, derived using SDSS spectroscopy, to delineate between star-formation and AGN/LINERs as the primary emission driver. As not all of our galaxies have significant detected emission in all of the four indicators, we only attempt to classify those with A/N $\geq$ 3 in H$\beta$, [OIII]5007, H$\alpha$ and [NII]6585 and galaxies with A/N $\geq$ 3 in H$\alpha$ and [NII]6585 for which it is possible to comment using [NII]/H$\alpha$ alone. Applying these constraints we assign 51 galaxies (46\% of the emission line sample defined previously) to be AGN/LINERs, 32 (29\%) to have ongoing star-formation and 27 (25\%) to be ambiguous based on this data (hereafter referred to as galaxies with an ambiguous emission driver or AED galaxies). In Fig. \ref{skycover} we show the distribution of our sample on the sky delineated by spectral classification.

The left panel in Fig. \ref{bpt} is qualitatively very similar to the equivalent diagram presented in figure 7, top left panel, of \cite{smith07} who studied a large sample of galaxies in the Shapley supercluster, and, as with this work, imposed no selection criteria based on colour or morphology. Aside from a small number of star-forming galaxies toward the central left region of the BPT diagram in Fig. \ref{bpt}, the majority of cluster members with emission define a continuous sequence from low ratios to high ratios in both [NII]/H$\alpha$ and [OIII]/H$\beta$. As commented by \cite{smith07}, this makes the assignment of a star-forming or AGN/LINER classification very sensitive to the demarcation relation employed.

Also in Fig. \ref{bpt} we plot the fraction of emission line galaxies which are determined to be star-forming or have AGN/LINERs relative to the entire emission sample, as a function of r-band absolute magnitude and cluster-centric radius with $\sim$ 20 galaxies per bin. For reference these panels also display the fraction of AED galaxies. A well defined trend is seen with luminosity in the sense that the fraction of AGN/LINERs increases with increasing brightness whereas the fraction of star-formers decreases, a result in agreement with \cite{kauffmann03} and \cite{smith07}. Interestingly no such correlation is observed as a function of radius with the fractional distributions remaining roughly constant, except for the inner-most bin where the star-forming fraction appears to diminish. However, this drop coincides with a rapid increase in those galaxies whose emission driver is undetermined making firm conclusions difficult.

\subsection{Velocity dispersions} 

\begin{figure*}
\flushleft
\scalebox{0.7}[0.7]{\includegraphics{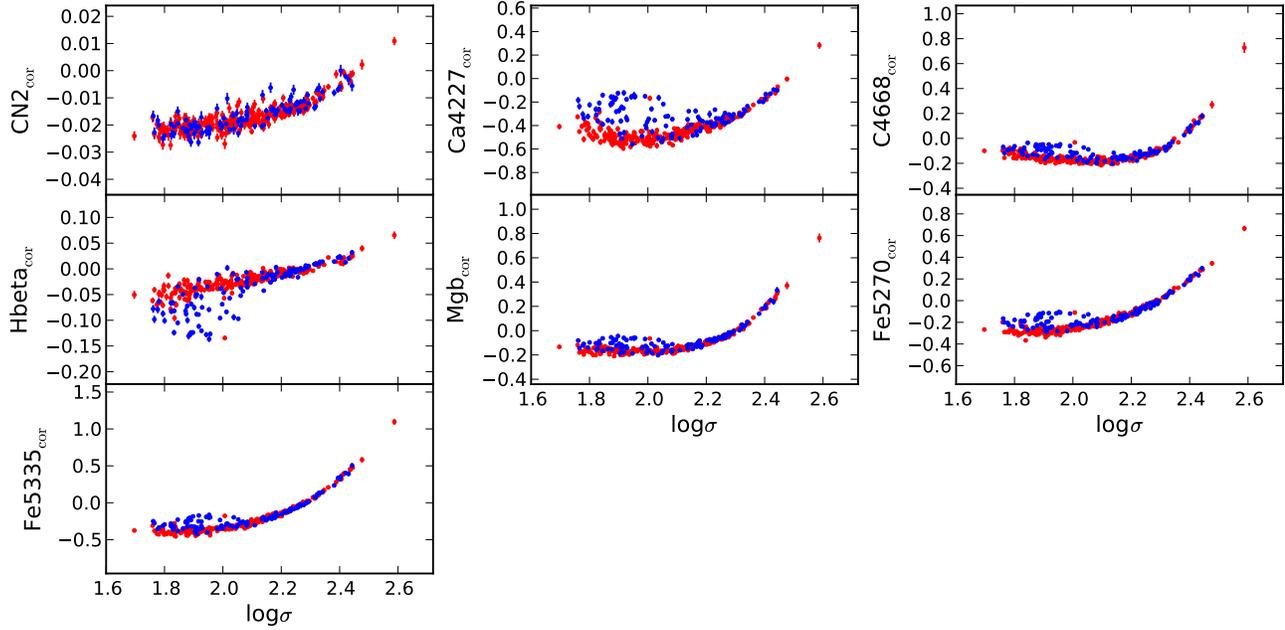}} 
\caption{Derived velocity broadening corrections for the seven indices used in our stellar population analysis. Red and blue filled circles are passive and emission line galaxies respectively. Here the y-axis refers to the bracketed term in equation \eqref{indexcor}.}
\label{indcors}
\end{figure*} 

The SDSS spectroscopic pipeline measures velocity dispersions within the survey's 3$^{\prime\prime}$ diameter fibre for the majority of spectra which are typical of early-type galaxies, and have redshifts z $<$ 0.4, using a direct fitting method\footnote{http://www.sdss.org/dr7/algorithms/veldisp.html}. However, 15 (6\%) of our non-emission sample and 36 (33\%) of our galaxies found to have emission do not have measured velocity dispersions in DR7. As such we opt to use the penalised pixel fitting code \citep[{\sc ppxf},][]{cap&em04} to obtain the velocity dispersions for our entire sample in a homogeneous fashion.

Details of the technique used here can be found in \cite{price09}. Briefly the routine uses the same 85 stellar templates used by {\sc Gandalf}, modulated by multiplicative and additive Legendre polynomials of orders 6 and 2 respectively and broadened by a parametric line-of-sight velocity distribution, in this case a gaussian, to construct a model of each galaxy's spectrum. Prior to fitting, the templates are matched to the variable resolution of the SDSS spectra. Fortunately the SDSS pipeline itself records spectral resolution as a function of wavelength for each fibre and so we use this information to smooth the templates with a variable width gaussian, typically with $\sigma_{inst}$ $\sim$ 58-70 km s$^{-1}$ in the 4000-6000 \AA\ fitting range used here, to match our data. Taking advantage of fitting in the pixel space we mask bad pixels and the NaD line at 5892 \AA\ which can often be affected by interstellar absorption.

To assess the uncertainty on our velocity dispersion measurements ideally one would like to undertake a bootstrap procedure, using the error spectra to generate new pseudo-observations and then passing them through {\sc ppxf}, but as our sample size is large this approach would be time consuming. As a compromise we create a subsample of galaxies, 10 per S/N bin with a bin size of 10 between S/N = 10-50, randomly selected from the full sample. For each selected galaxy we generate 50 realisations, run them through {\sc ppxf} and take the standard deviation of the resulting distribution as the error on that galaxy's velocity dispersion. Finally we interpolate the $\sigma$-S/N-$\sigma_{error}$ distribution formed by our subsample into a surface using a bivariate spline. It is then a simple matter of inputing the S/N and $\sigma$ of each galaxy from the full sample to obtain an estimated $\sigma_{error}$.

Comparing our velocity dispersion measurements to those from the SDSS pipeline for galaxies with both we find a mean offset $<$SDSS-This work$>$ = -6 km s$^{-1}$, likely attributed to our different fitting range and masking procedures. Once this systematic shift is accounted for we find an RMS scatter of 5 km s$^{-1}$ with an intrinsic component of 0. Comparing our measurements with those from \cite{moore02} we find a mean offset of 2.8 km s$^{-1}$, an RMS scatter accounting for the offset of 8 km s$^{-1}$ with an intrinsic component of 4.7 km s$^{-1}$. Thus, overall, we find good compatibility between our measurements and those made by the SDSS pipeline and \cite{moore02}.

The final step necessary before using the measured velocity dispersions in our analysis is to aperture correct them to a common sampling in terms of each galaxy's half-light radius. Here we follow the prescription of \cite{jorgensen95},

\begin{equation*}
\log\frac{\sigma_{SDSS}}{\sigma_{R_{e}/2}} = -0.04\log\frac{1.5^{\prime\prime}}{R_{e}/2}
\end{equation*}

\noindent where $R_{e}$ is the psf-corrected half-light or effective radius, obtained by fitting a single S{\'e}rsic model using {\sc Galfit} \citep{peng02} to the SDSS r-band image of each galaxy. $\sigma_{SDSS}$ and $\sigma_{R_{e}/2}$ are the galaxies' velocity dispersions through the 1.5$^{\prime\prime}$ radius SDSS fibre and the desired $R_{e}/2$ aperture respectively. Corrections are in the range -0.067 to 0.017 dex with a mean of $\sigma_{cor}$ = -0.004 dex or $\sim$ 1 \%. 

\subsection{Absorption line indices}

We use the {\sc Lick\_Ew} code which is provided as part of the {\sc Ez-Ages} package\footnote{http://www.ucolick.org/$\sim$graves/EZ\_Ages.html} \citep{graves08} to measure the Lick indices of our sample. The full set of indices defined in Table 1 of \cite{worthey94} and Table 1 of \cite{worthey&otta97} are measured where possible, although a small fraction of the galaxies have regions of bad pixels which prevent access to all indices. The routine uses the error spectra and the equations of \cite{cardiel98} to compute error estimates for the indices.

Having obtained the raw indices for our sample it is then necessary to correct them to the resolution of the Schiavon models, which are themselves based on stellar spectra smoothed to the wavelength-dependent Lick/IDS resolution. Effectively one wants to know the value of each observed index for a given galaxy at the model's resolution but without any doppler broadening from that galaxy's stellar velocity dispersion. One route is to employ multiplicative and additive corrections derived from artificially broadened stars observed with the same instrument as the galaxy spectra. More recent works have moved toward using stellar population model templates which permit a better match in terms of spectral type to observed galaxy spectra, while still maintaining the zero velocity dispersion requirement. 

\begin{figure*}
\flushleft
\scalebox{0.7}[0.7]{\includegraphics{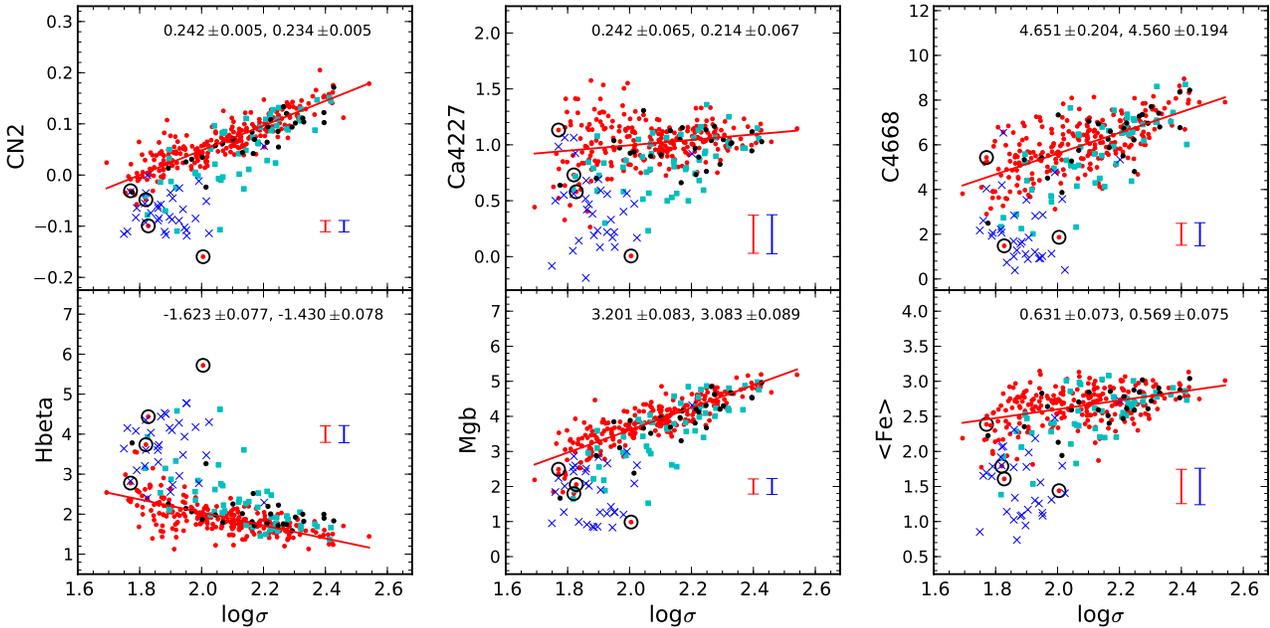}} 
\caption{Lick indices used in our stellar population analysis, plotted against velocity dispersion for our Coma galaxies. Red filled circles are our passive galaxy subsample, with the four blue passive galaxies highlighted by black circles. Our emission-line sample is denoted based on emission driver as determined in Fig. \ref{bpt} with cyan filled squares representing AGN/LINERs, blue crosses for star-forming systems and black points where the source is ambiguous. Red and blue error bars show typical (median) index errors for passive and emission galaxies respectively. The slope of the best fit trend line to the passive galaxy data is displayed in each panel where the first value includes the blue subset and the second value does not. The latter fit is also plotted as a red line in each panel.}
\label{indsig}
\end{figure*}

Here we opt to follow a similar method to that of \cite{kelson06} and also take advantage of the functionality of {\sc ppxf} in deriving our velocity broadening corrections. First, {\sc ppxf} is used to output the best fit combination of our stellar templates at the resolution of our data for each galaxy with and without smoothing by the velocity dispersion of that galaxy. Next {\sc Lick\_Ew} is executed three times; first on the observed spectrum (I$_{obs}$), then on the best fit spectrum with velocity dispersion broadening (I$_{T}$) and finally on the broadening free spectrum smoothed to the Lick resolution (I$_{lick}$) for each index as defined by Table 1 of \cite{schiavon07}. The corrected indices for each galaxy (I$_{cor}$) can then be obtained via:

\begin{equation}
\label{indexcor}
\operatorname{I}_{cor} = \operatorname{I}_{obs} + ( \operatorname{I}_{lick} - \operatorname{I}_{T} )
\end{equation}

\noindent In Fig. \ref{indcors} we present the index corrections, that is the value of the bracketed term on the right hand side of equation \eqref{indexcor}, for the indices which were used in our stellar population analysis (see Section \ref{sppsection}) as a function of galaxy velocity dispersion for both emission and non-emission line samples. Uncertainties on these corrections are obtained using a similar approach to that used previously for the velocity dispersion measurements but in practice are $\leq$ 10\% of the typical index errors and so are not propagated further.

From Fig. \ref{indcors} it is evident the corrections for all indices behave broadly as expected, that is negative corrections when a galaxy's velocity dispersion is sub-Lick resolution for a given index moving to positive corrections at super-Lick resolution. Just by inspection there appears to be a significantly larger scatter in the corrections of the emission line galaxies relative to their non-emission counterparts. This is caused by the corrections also being a function of index strength and, therefore, both samples in fact have comparable scatter. However, an increased scatter is noted at $\sigma \le$ 150 km s$^{-1}$ which can likely be attributed to a greater variation in fitted spectral type for galaxies with lower velocity dispersion.

As the Schiavon models are themselves based on a flux calibrated stellar library, with no transformation to the Lick/IDS system, we also make no attempt to calibrate our data on to the Lick system.

In Appendix \ref{appendixa} we present a comparison of our index measurements with two other literature studies and show that our data is generally in good agreement with both datasets. In Appendix \ref{appendixd} we tabulate our index and velocity dispersion data.

\subsection{Absorption line Analysis}

\begin{figure*}
\flushleft
\scalebox{0.7}[0.7]{\includegraphics{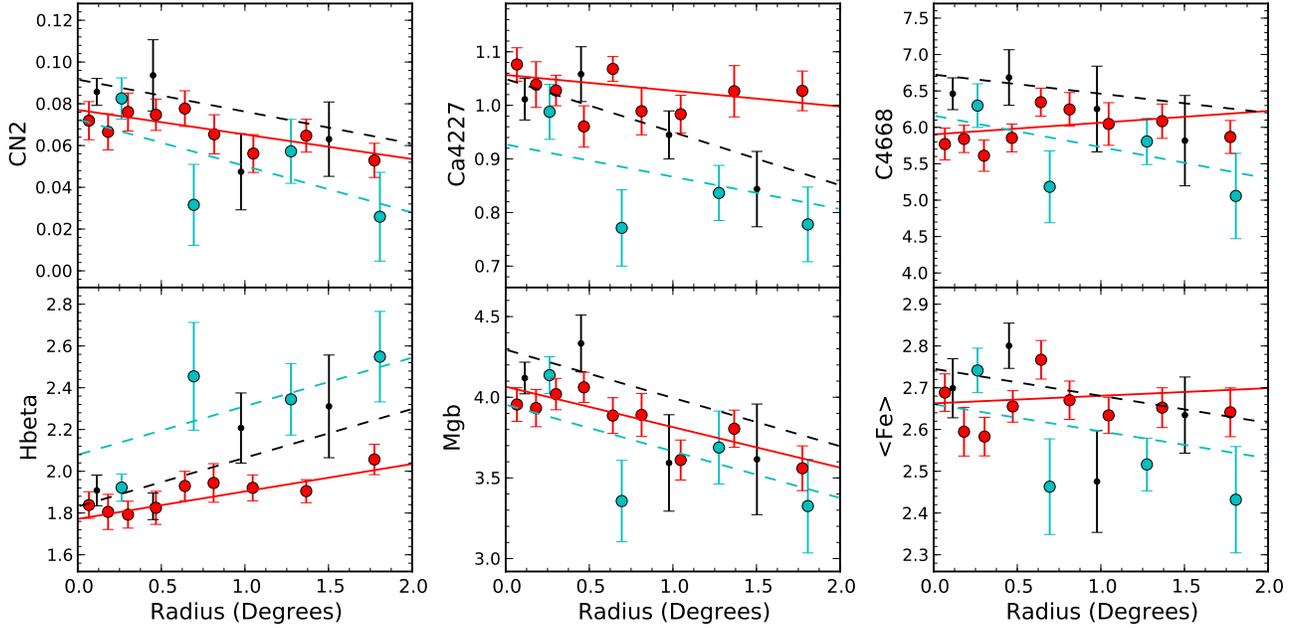}} 
\caption{Lick indices against cluster-centric radius. Data are binned with roughly the same number of galaxies per bin. Red ($\sim$ 27 galaxies per bin) and cyan ($\sim$ 13 galaxies per bin) filled circles denote passive and AGN/LINER galaxies respectively. The black points ($\sim$ 7 galaxies per bin) are galaxies whose emission driver could not be determined. Trend lines are fit to the unbinned data. The star forming subsample is not plotted for clarity (see text).}
\label{indrad}
\end{figure*}

\begin{figure*}
\flushleft
\scalebox{0.7}[0.7]{\includegraphics{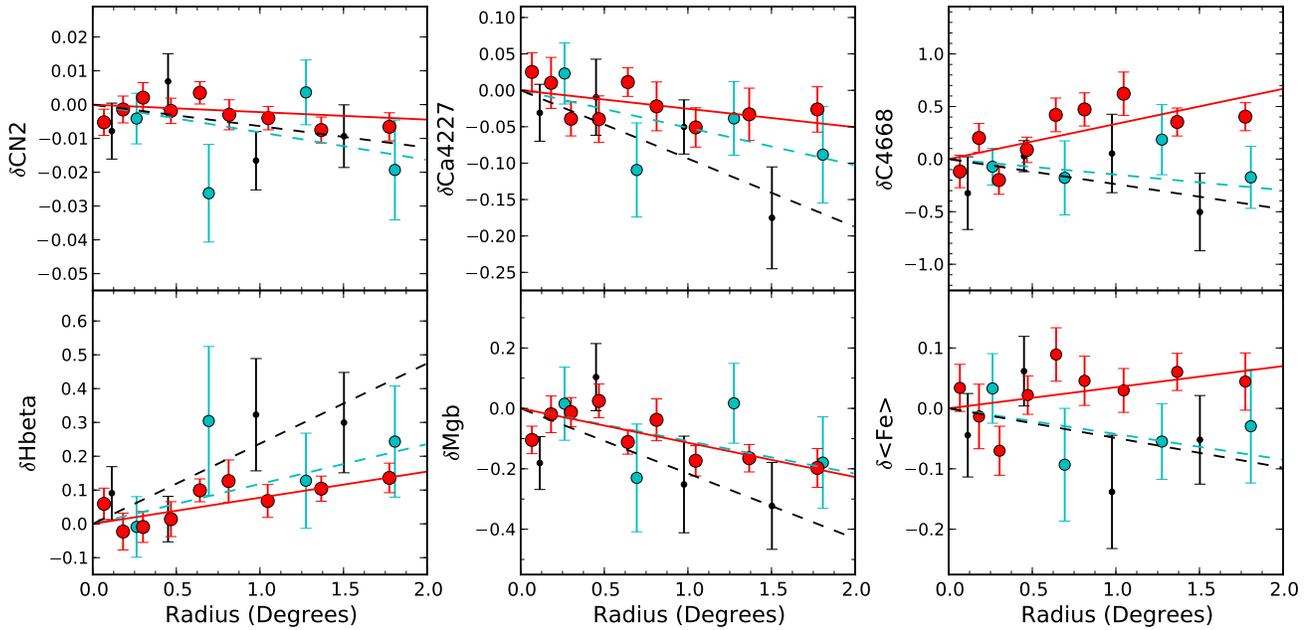}} 
\caption{Index-radius diagrams plotted in the form of a residual index $\delta I$ where $\delta I = I - (b\log\sigma + c\log(1.5^{\prime\prime}/R_{e}) + d)$. The gradient of the best fit lines corresponds to the coefficient $a$ in a planar fit of the form $I = aR_{cc} + b\log\sigma + c\log(1.5^{\prime\prime}/R_{e}) + d$ . Thus these diagrams compare the data at fixed velocity dispersion. Symbols and binning are as in Fig. \ref{indrad} and once more the star forming subsample is not plotted for clarity.}
\label{indrad_p}
\end{figure*} 

\begin{table*}
\begin{minipage}{165mm}
\centering
\caption{Cluster-centric index gradients in units of \AA, or mags for CN2, per degree obtained from the fits in Fig. \ref{indrad} and Fig. \ref{indrad_p}. For our passive galaxies two slopes are given, the first including the blue subset outlined earlier and the second masking them. The 2nd column indicates the number of galaxies in each respective passive galaxy fit out of a potential of 246. The number of galaxies in each emission driver delineated fit is shown in the header row and remains constant from one index to another. We express the significance of each fit, in units of standard error, in parentheses. }
\label{indradtab}
\begin{tabular}{ccr@{$\pm$}lr@{$\pm$}lr@{$\pm$}lr@{$\pm$}lr@{$\pm$}l}
\hline
Index & N & \multicolumn{4}{c}{Passive} & \multicolumn{2}{c}{AGN/LINER (51)} & \multicolumn{2}{c}{SF (32)} & \multicolumn{2}{c}{AED (27)} \\
\hline
\multicolumn{12}{c}{I -- R$_{cc}$ (Fig. \ref{indrad})} \\
\hline 
CN2&244/240&-0.009&0.001(6.3)&-0.012&0.001(8.1)&-0.022&0.003(8.8)&0.009&0.005(2.0)&-0.015&0.003(5.1) \\ 
Ca4227&246/242&-0.025&0.019(1.3)&-0.029&0.019(1.5)&-0.061&0.036(1.7)&0.058&0.078(0.7)&-0.100&0.050(2.0) \\ 
C4668&245/242&0.183&0.065(2.8)&0.161&0.061(2.6)&-0.432&0.109(4.0)&0.766&0.250(3.1)&-0.263&0.140(1.9) \\ 
Hbeta&246/242&0.104&0.026(4.1)&0.132&0.025(5.3)&0.233&0.046(5.1)&-0.344&0.107(3.2)&0.234&0.061(3.8) \\ 
Mgb&246/242&-0.223&0.029(7.7)&-0.251&0.028(8.9)&-0.289&0.044(6.5)&0.465&0.114(4.1)&-0.299&0.067(4.4) \\ 
$<$Fe$>$&245/241&0.023&0.025(0.9)&0.018&0.023(0.8)&-0.064&0.039(1.6)&0.212&0.096(2.2)&-0.064&0.058(1.1) \\  
\hline
\multicolumn{12}{c}{$\delta$I -- R$_{cc}$ (Fig. \ref{indrad_p})} \\
\hline 
CN2&244/240&-0.001&0.001(0.4)&-0.002&0.001(1.6)&-0.008&0.002(3.4)&0.008&0.005(1.5)&-0.006&0.004(1.6) \\ 
Ca4227&246/242&-0.022&0.021(1.0)&-0.025&0.020(1.3)&-0.052&0.034(1.5)&0.016&0.084(0.2)&-0.095&0.059(1.6) \\ 
C4668&245/242&0.345&0.060(5.8)&0.334&0.058(5.7)&-0.148&0.107(1.4)&0.692&0.253(2.7)&-0.239&0.188(1.3) \\ 
Hbeta&246/242&0.044&0.026(1.7)&0.077&0.025(3.1)&0.118&0.045(2.6)&-0.327&0.114(2.9)&0.237&0.083(2.9) \\ 
Mgb&246/242&-0.099&0.027(3.7)&-0.114&0.027(4.3)&-0.108&0.046(2.4)&0.467&0.116(4.0)&-0.216&0.082(2.6) \\ 
$<$Fe$>$&246/242&0.041&0.029(1.4)&0.035&0.023(1.5)&-0.042&0.056(0.7)&0.203&0.129(1.6)&-0.048&0.093(0.5) \\      
\hline
\end{tabular}
\end{minipage}
\end{table*}

In this section we shall study the stellar populations of our sample galaxies in a model independent way, using their absorption line indices. We will look for correlations between these measurements and galaxy velocity dispersion and local environment, parameterised by cluster-centric radial distance. The same analysis techniques are also applied in Sections \ref{sppsigsec} and \ref{sppradsec} to the model dependent stellar populations parameters.

In Fig. \ref{indsig} we plot the indices used in our analysis (where $<$Fe$> = $(Fe5270$+$Fe5335)/2) against velocity dispersion and in Fig. \ref{indrad} against cluster-centric radius. In the former figure the legend of each panel displays the results of a linear least-squares fit to the passive galaxy data weighted by the index uncertainties and iteratively taking into account intrinsic scatter. Slope uncertainties are computed by creating 500 realisations of the index data using their errors and taking the standard deviation of the resulting distribution of slopes. The first reported slope is computed including the blue subset identified earlier while the latter slope is derived having masked them out, resulting in a fit which is more comparable to previous works based on colour-selected samples. For the index-radius panels the index data are binned for clarity due to substantial intrinsic scatter and the emission-line sample is split by emission driver. The passive sample has bins derived excluding the four blue quiescent outliers and contains $\sim$ 27 galaxies per bin while the emission line bins contain $\sim$ 13 and $\sim$ 7 galaxies per bin for the AGN/LINER and AED samples respectively. The fits to the star forming sample are not plotted in Fig. \ref{indrad} for clarity. This is because they often have trends with strongly offset intercepts or slopes in the opposite sense to the other subsets and thus act to distract the eye from the passive sample which is more so the focus of this work. The results of the index-radius fits, including the star formers, are displayed in Table \ref{indradtab}. We stress that in all cases it is the unbinned data that are fit and the binning is for display purposes only. 

The relations for the non-emission line galaxies presented in Fig. \ref{indsig} are broadly consistent with previous studies on red sequence galaxies in Coma \citep{matkovic09}, and other low redshift clusters \citep{smith07}. Both of these works have spectra derived from comparable physical scales to this work, the former observing with 3.1$^{\prime\prime}$ diameter fibres and the latter with 2$^{\prime\prime}$ diameter fibres at z = 0.048, giving a spatial coverage of $\sim$ 1.8 kpc. Strong positive correlations are particularly evident for CN2, C4668 and Mgb while a significant anticorrelation is present for H$\beta$. There is evidence, at least by eye, for increased scatter about the fitted relations with decreasing $\sigma$ for all indices except perhaps CN2.

\begin{table*}
\begin{minipage}{170mm}
\centering
\caption{Comparison of the index gradients derived by this work (from the planar fitting technique, second section Table \ref{indradtab}) with those reported in the literature for Coma by C02 and S08 and for a large number of nearby clusters by S06. For the C02 results their Mg$_{2}$ gradient is converted to an equivalent gradient in Mgb. The results of S06 are converted to gradients in degrees assuming R$_{200}$ = 1.59$^{\circ}$ for Coma. ``Core-SW'' denotes our sample restricted to these areas (see text).}
\label{comptab}
\begin{tabular}{cccccc}
\hline
Index & This work & Core-SW & C02 & S08 & S06 \\
\hline
CN2 & --0.002$\pm$0.002 & --0.006$\pm$0.003 & 	- 			& --0.025$\pm$0.007 	& - \\
Ca4227 & --0.025$\pm$0.020 & --0.060$\pm$0.043 & 	- 			& --0.136$\pm$0.072 	& - \\
C4668 & +0.334$\pm$0.058 & +0.622$\pm$0.150 &	- 			& --0.207$\pm$0.270 	& --0.102$\pm$0.046     \\
H$\beta$ & +0.077$\pm$0.025 & +0.119$\pm$0.056 &	+0.327$\pm$0.116 	& +0.539$\pm$0.104 	& +0.046$\pm$0.013  \\
Mgb & --0.114$\pm$0.027 & --0.070$\pm$0.058 &	--0.491$\pm$0.125 	& --0.490$\pm$0.145 	& --0.128$\pm$0.016 \\
$<$Fe$>$ & +0.035$\pm$0.023 & +0.095$\pm$0.052 &	--0.082$\pm$0.084 	& +0.031$\pm$0.083 	& +0.018$\pm$0.015 \\
\hline
\end{tabular}
\end{minipage}
\end{table*}

The emission-line galaxies generally have higher H$\beta$ and weaker metal lines at low $\sigma$ with respect to the quiescent galaxies while at high $\sigma$ both subsamples have similar index strengths. As displayed in Fig. \ref{indsig} this trend is clearly driven by a transition from AGN/LINER dominated systems at high $\sigma$, with stellar populations at least comparable to quiescent bright early-type galaxies, to star-forming galaxies at low $\sigma$. Over the range in velocity dispersion sampled here the transition is fairly smooth, with a crossover at $\sim$ 100 km s$^{-1}$ and a marked overlap between the two emission drivers, a feature that acts to reiterate Fig. \ref{bpt} middle panel.

In Fig. \ref{indrad} significant radial trends are seen for the passive galaxies in CN2 (6.3$\sigma$), C4668 (2.8$\sigma$), H$\beta$ (4.1$\sigma$) and Mgb (7.7$\sigma$) which typically increase in significance when masking the blue subset. For H$\beta$ and C4668 the correlations are positive with larger index values at greater cluster-centric radii whereas CN2 and Mgb are observed to decrease with radius. Strong correlations are also identified in these indices for the AGN/LINER galaxies in the same sense as the passive sample, with the exception of C4668. The indices of star-forming galaxies behave somewhat differently with H$\beta$ decreasing and Mgb increasing with radius (see Table \ref{indradtab}), although the robustness of these trends given the small size of the subsample is questionable. 

The trends presented in Fig. \ref{indsig} and \ref{indrad} may be affected by correlations between radius and velocity dispersion and thus, for instance, if the outer parts of the cluster are preferentially inhabited by lower $\sigma$ galaxies the derived index-radius fits will be biased. A further factor we seek to address stems from the fact that the survey's 3$^{\prime\prime}$ fibres sample a different fraction of each galaxy based on their physical size. As such, if we assume some typical internal radial gradient in their stellar populations, and therefore index strengths, larger galaxies will generally have index strengths biased higher, in the case of negative gradients, or lower, in the case of positive internal trends, relative to smaller galaxies. Indeed, on average, negative gradients have been detected in the metallic indices while H$\beta$ gradients may be positive or negative for individual galaxies, but are typically nearly flat \citep[e.g.][]{davies93,gonzalez93,rawle08}.

To factor out the issues discussed above and derive index relations at fixed velocity dispersion and radius that are independent of aperture effects we follow \citet[][hereafter S06]{smith06} and S08 and fit our data with a multi-dimensional plane that incorporates all three terms,

\begin{equation}
\label{plane}
I = aR_{cc} + b\log\sigma + c\log\frac{1.5^{\prime\prime}}{R_{e}} + d
\end{equation} 

\noindent where 1.5$^{\prime\prime}$ and $R_{e}$ are again the fibre radius and effective radius of each galaxy respectively. Here R$_{cc}$ is the projected cluster-centric radius. The coefficients of the first two terms give the desired index trends. Note, while S08 use luminosity in their plane fits, we almost always find stronger index correlations with $\sigma$ and therefore opt to employ the latter in our fits \citep[see also][]{bernardi05,smith09b}. In order to display the results of the planar fits in a comparable manner to Fig. \ref{indrad} we plot the data in the form of a residual index $\delta I$, 

\begin{equation*}
\delta I = I - (b\log\sigma + c\log\frac{1.5^{\prime\prime}}{R_{e}} + d)
\end{equation*} 

\noindent We choose not to plot the equivalent $\delta I$ for $b\log\sigma$ since the slopes typically change by less than one standard error relative to those shown in Fig. \ref{indsig}. However, the effect on the radius relations is generally more substantial and so in Fig. \ref{indrad_p} we present $\delta I$-R$_{cc}$ for our sample galaxies with fit results again tabulated in Table \ref{indradtab}. The robustness of our fitting procedure is examined in Appendix \ref{appendixc}.

For the passive galaxies significant correlations are still found for C4668 (5.8$\sigma$), Mgb (3.7$\sigma$) and H$\beta$ (3.1$\sigma$), when the blue subset is masked, while radial trends in CN2 become insignificant. On the other hand, our sample of AGN/LINERs now generally have much weaker trends relative to their linear fits, with significant correlations still found for CN2 (3.4$\sigma$), H$\beta$ (2.6$\sigma$) and Mgb (2.4$\sigma$). This sharp reduction in the computed gradients stems from the fact that velocity dispersion correlates mildly with radius for this sample in the sense that lower dispersions are found at larger radii. Finally, the index trends for star-forming galaxies remain comparatively unchanged.

A valid question here is to what extent do the outermost points drive the significant radial trends recovered by both the linear and planar fits for our passive sample. To test this we have repeated the fitting procedure with the outermost points, those in the final bin with $R_{cc} > 1.55^{\circ}$, and the blue subset masked. There is some change to the fitted slopes reported in Table \ref{indradtab}. However, the three indices with significant radial coefficients in the planar fits remain C4668 (6$\sigma$), H$\beta$ (2.5$\sigma$) and Mgb (3.1$\sigma$) and with the same sign as given in Table \ref{indradtab}. Indeed, the same is true for the linear fits with C4668 (4.4$\sigma$), H$\beta$ (3.3$\sigma$) and Mgb (5.7$\sigma$).

In Table \ref{comptab} we compare the radial gradients for our passive sample to those reported in C02 and S08 for Coma passive galaxies and more generally in a sample of low redshift clusters from S06. The sign of our significantly detected slopes are broadly consistent with those found elsewhere, with the exception of C4668, and indeed we also find at most a very weak correlation for $<$Fe$>$. Interestingly C02 and S08 find much stronger gradients in H$\beta$ and Mgb relative to both this work and S06. This result may either be driven by the luminosity range covered here compared to the previous works, C02 analysed both dwarf and giant galaxy regimes while S08 only include dwarf galaxies, or the spatial coverage of the cluster provided by both previous studies which was limited to core and south-west regions, or some combination of both factors. To try to assess the impact of limited spatial coverage we create a subset of our passive sample with a highly comparable cluster footprint to that of S08 and refit our planar relations. Our Core-SW subset has 116 galaxies, 3 of which are classified here as blue passive galaxies. The results of these fits, with the blue galaxies masked, are presented in Table \ref{comptab} and do not display a significant increase in H$\beta$ or Mgb gradient, although we do note that when including the blue subset in the fit the derived H$\beta$ slope increases to 0.194$\pm$0.061 \AA\ deg$^{-1}$. This finding seems to imply that the weaker slopes found here for H$\beta$ and Mgb compared to S08 are driven by our luminosity range being constrained to bright Coma galaxies not by our cluster coverage. Similarly, C02 including a sample of dwarfs in their data likely results in their slopes being an intermediate between this work and S08.

\section{Stellar Population Parameters}
\label{sppsection}

Having measured the absorption indices for our passive and emission-line samples we are now able to convert these into estimates of their stellar population parameters using the Schiavon models and {\sc Ez-Ages} fitting code. As we plan to determine these parameters for both of our subsamples it is important to preface this section with a word of warning. 

It has been known for some time that the presence of composite stellar populations (CSP) resulting from multiple star-forming bursts significantly affects how the stellar age returned by SSP models should be interpreted. Recently \cite{serra07} conducted extensive simulations to fully assess the impact of trying to apply SSP models to CSPs, which they constructed based on a two burst scenario with a mass dominating old population and a frosting of younger stars. They found that rather than obtaining a luminosity-weighted age, the SSP-equivalent age is often skewed substantially lower by a relatively small mass fraction of younger stars, an outcome attributed to the non-linear response of the Balmer lines to stellar age \citep[see also][for a more detailed discussion]{trager09}. However, they report that SSP-equivalent [Fe/H] and [$\alpha$/Fe] track quite closely the luminosity-weighted parameters, being driven typically by the older population except in the most extreme cases of a very young and massive frosting. We therefore stress that, particularly in the case of the star-forming galaxies in our emission-line sample, the derived ages are Balmer-weighted SSP-equivalent parameters and may not be indicative of the global stellar populations in these galaxies.

\subsection{The Model and Parameter Fitting}
\label{paramerrors}

The Schiavon models are based on the \cite{Jones99} empirical stellar library and span a range of 1 to 17.7 Gyr and -1.3 to +0.2 [Fe/H]. They enable a determination of abundance ratios in terms of Mg, C, N and Ca by incorporating their stellar atmospheric effects. As the models are based on scaled-solar isochrones the evolutionary tracks and stellar atmospheres are not treated in a self-consistent manner in a similar vein to the widely used \citet[][hereafter TMBK]{thomas03,thomas04} models. Here we employ a Salpeter initial mass function.

We derive the SSP-equivalent parameters for our galaxies using the {\sc Ez-Ages} code of \cite{graves08} which, using the Schiavon models, seeks to return consistent stellar population parameters across a number of index-index diagrams. The routine begins by making an initial estimate of age and [Fe/H] by inverting a H$\beta$ -- $<$Fe$>$ grid, specifically chosen to strongly break the age-metallicity degeneracy while being largely insensitive to non-solar abundance patterns \citep{schiavon07}. Next the code moves on to determining [Mg/Fe] from a H$\beta$ -- Mgb grid. To do this it first inverts the new grid, compares the age and [Fe/H] measured from this grid to the fiducial estimates and if they do not agree assumes a non-solar [Mg/Fe] is present and recomputes the model with a new [Mg/Fe]. Further iterations occur until the derived age and [Fe/H] converge to the fiducial values within some tolerance. The code then goes on to repeat this iterative approach replacing Mg by C, N and Ca in turn with the final abundance pattern used to compute a new model and derive a new fiducial age and [Fe/H], with the entire fitting process repeating until the fit does not improve. The standard index set used for the fitting is H$\beta$ and $<$Fe$>$ for the initial age and [Fe/H] estimates, Mgb for [Mg/Fe], C4668 for [C/Fe], CN2 for [N/Fe] and Ca4227 for [Ca/Fe].

We could not obtain fits for 11\% of our entire sample as they fell off the fiducial H$\beta$ -- $<$Fe$>$ grid. This may be because they either have SSP parameters, perhaps only marginally, outside the models coverage or else were carried off the grid by measurement errors. As a work around we fit these galaxies using Fe5270 or Fe5335 only instead of their mean. Typically the results obtained from the three different approaches are consistent, within errors, for those galaxies which can be fit by all three.

\begin{figure}
\flushleft
\scalebox{0.44}[0.44]{\includegraphics{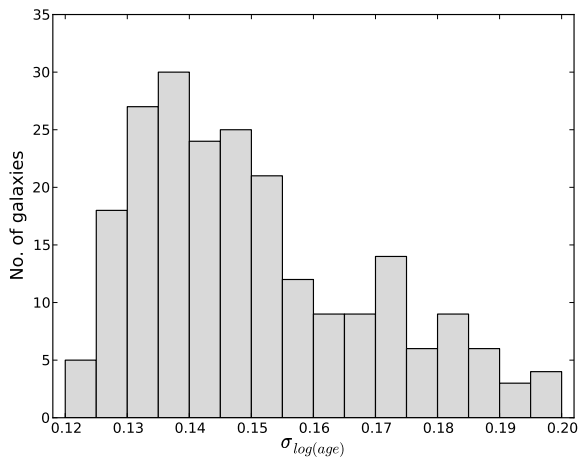}} 
\caption{Histogram showing the distribution of uncertainties on our derived SSP-equivalent ages for galaxies in our quiescent sample.}
\label{ageerr}
\end{figure}

To compute uncertainties on the stellar population parameters we select five quiescent galaxies from our original sample of 417 with median S/N per \AA\ $\sim$ 15, 25, 35, 45, 55 which have index errors representative of the entire sample in their respective S/N bin. Next for each S/N step we generate 300 new realisations of the index data by randomly perturbing the initial values by their errors and run these through {\sc Ez-Ages}. At each S/N we then assign the typical uncertainties for age, [Fe/H] and the abundance ratios as the standard deviation of the distribution of parameters obtained from our simulations. Finally we linearly fit to these points a relation of the form $P_{error} = m(S/N)^{-1} + c$, where $P_{error}$ is the standard error on P and P = $\log$(age), [Fe/H], [Mg/Fe], [C/Fe], [N/Fe] and [Ca/Fe] in turn. Therefore using this error model and a galaxy's S/N per \AA\ in the wavelength range 4000-6000\AA\ we are able to estimate the statistical uncertainty on all its stellar populations parameters. This approach also acts to assess the error correlations between fitted parameters. As an example, in Fig. \ref{ageerr} we show the distribution of the standard error on $\log$(age) for our quiescent sample. This figure demonstrates that our S/N cut results in at most a $\sim$ 50\% uncertainty on our age estimates, with a median error of $\sim$ 40\%. Note here we assume that the index error distributions for the quiescent and emission-line sample are comparable and as such do not attempt to quantify any systematics generated by our emission cleaning technique.

\subsection{Identifying Composite Stellar Populations}

As outlined earlier, the measured stellar age reported by our analysis can be significantly biased by a frosting of hot young stars and in such cases is far from a luminosity-weighted parameter. Fortunately tools are available that allow the detection of particularly young ($\lesssim$ 1 Gyr) star bursts, thus placing a lower limit on the cessation of star formation in a given galaxy. This is particularly applicable to our passive sample since we wish to test that if a relation exists between age and velocity dispersion it is not driven by, for instance, the ages of lower $\sigma$ quiescent galaxies being biased younger by recent star formation.

\begin{figure}
\flushleft
\scalebox{0.45}[0.45]{\includegraphics{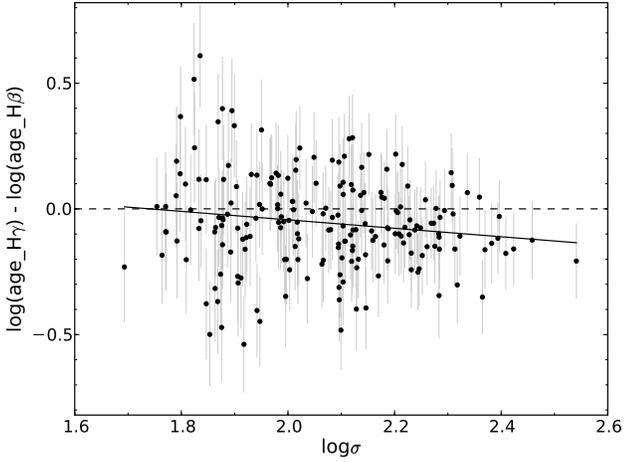}} 
\caption{The difference between SSP-equivalent ages derived using H$\gamma$ and H$\beta$ indices plotted against galaxy velocity dispersion for our passive sample. The solid line represents a simple linear fit to the data, the slope of which is only mildly significant ($\sim$ 2$\sigma$). The panel indicates that it is in fact high $\sigma$ galaxies which typically have a weak discrepancy between their Balmer line ages rather than galaxies with low velocity dispersions.}
\label{frosting1}
\end{figure}

\begin{figure}
\flushleft
\scalebox{0.45}[0.45]{\includegraphics{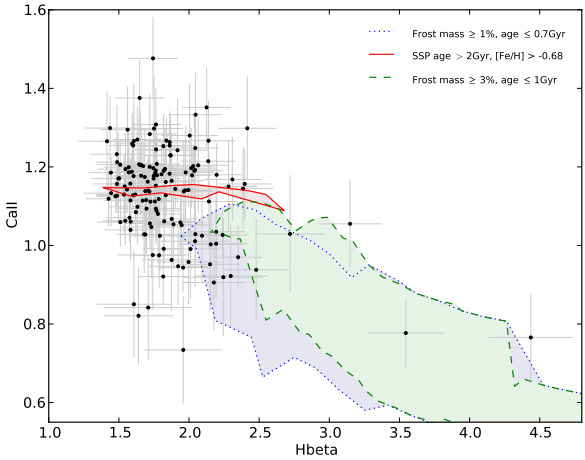}} 
\caption{Rose CaII index against H$\beta$ for passive sample galaxies with errors in CaII $\leq$ 0.15. The parameter coverage of simulated CSPs with frosting mass $\geq$ 1\% and age $\leq$ 0.7 Gyr and frosting mass $\geq$ 3\% and age $\leq$ 1.0 Gyr are denoted by blue and green shaded polygons respectively. The region spanned by genuine SSPs with 2 Gyr $<$ age $\leq$ 14 Gyr is highlighted by a red shaded polygon. Despite sizable scatter in CaII, the majority of our passive sample show little evidence for episodes of star formation within the past $\sim$ 1 Gyr. }
\label{frosting2}
\end{figure}

One such method involves the comparison of SSP-equivalent ages derived using different Balmer indices, relying on the fact that the higher order Balmer lines are more strongly affected by the presence of a young, blue subpopulation. Therefore, using H$\delta$ or H$\gamma$ rather than H$\beta$ as our age sensitive index should result in younger ages being obtained if a sufficiently young frosting is present. Here we employ H$\gamma$ derived ages which are measured by {\sc Ez-Ages} following the conclusion of abundance pattern fitting. In Fig. \ref{frosting1} we plot the difference between the H$\gamma$ and H$\beta$ based age estimates for the 205 passive galaxies with both, as a function of their velocity dispersion.  The diagram demonstrates that the ages derived for low $\sigma$ galaxies are consistent, despite sizable scatter, and that it is in fact the high velocity dispersion galaxies which show a mild discrepancy, having H$\gamma$ ages at most some $\sim$ 25\% younger than those from H$\beta$. Furthermore, the distribution of age differences about zero is dominated by a scatter due to measurement errors of 0.18, with a small intrinsic scatter of 0.07.

Following \cite{smith09b}, a second technique to search for the presence of recent star bursts in our passive sample makes use of the \cite{rose84} CaII index. This index is computed as the ratio of the residual flux at the core of the Ca K and H lines and is approximately constant for stars of later type than F2, but changes strongly for hotter stars. Thus, when combined with Balmer line indices, this index can be used to distinguish near SSP-like galaxies from those with a small frosting of younger stars plus a mass-dominant underlying old population \citep{leonardi&rose96}. Indeed, such a test also places limits on the contribution from any other A-type stars, such as hot horizontal branch stars \citep[e.g.][]{maraston00}.

\begin{table}
\centering
\caption{Measured stellar population parameters for all subsamples, including those for early-type galaxies in the passive sample based on morphologies taken from NED. Columns 2-5 give the parameter means ($\mu$), total scatter about the mean ($\sigma_{tot}$), scatter due to errors ($\sigma_{errs}$) and intrinsic scatter in the distribution ($\sigma_{int}$) respectively. The number of galaxies in each subset is given in parentheses.}
\label{spptab}
\begin{tabular}{cr@{$\pm$}lccc}
\hline
Parameter & \multicolumn{2}{c}{$\mu$} & $\sigma_{tot}$ & $\sigma_{errs}$ & $\sigma_{int}$ \\
\hline
Passive (222) \\
\hline
log(age)&0.85&0.02&0.25&0.15&0.20 \\
$[$Fe/H$]$&-0.08&0.01&0.17&0.13&0.11 \\
$[$Mg/Fe$]$&0.15&0.01&0.11&0.11&0.00 \\
$[$C/Fe$]$&0.15&0.01&0.15&0.10&0.11 \\
$[$N/Fe$]$&0.04&0.01&0.19&0.12&0.14 \\
$[$Ca/Fe$]$&0.00&0.01&0.12&0.14&0.00 \\
\hline
Passive Early-type (160) \\
\hline
log(age)&0.88&0.02&0.25&0.15&0.20 \\
$[$Fe/H$]$&-0.08&0.01&0.17&0.12&0.11 \\
$[$Mg/Fe$]$&0.17&0.01&0.10&0.11&0.00 \\
$[$C/Fe$]$&0.15&0.01&0.16&0.10&0.12 \\
$[$N/Fe$]$&0.05&0.01&0.19&0.12&0.15 \\
$[$Ca/Fe$]$&0.00&0.01&0.12&0.14&0.00 \\
\hline
AGN/LINER (44) \\
\hline
log(age)&0.70&0.05&0.31&0.15&0.27 \\
$[$Fe/H$]$&-0.03&0.03&0.18&0.12&0.13 \\
$[$Mg/Fe$]$&0.16&0.02&0.10&0.10&0.02 \\
$[$C/Fe$]$&0.14&0.03&0.17&0.10&0.15 \\
$[$N/Fe$]$&0.06&0.02&0.16&0.11&0.12 \\
$[$Ca/Fe$]$&-0.04&0.02&0.13&0.13&0.04 \\
\hline
SF (24) \\
\hline
log(age)&0.26&0.03&0.14&0.17&0.00 \\
$[$Fe/H$]$&-0.52&0.08&0.41&0.16&0.37 \\
$[$Mg/Fe$]$&0.13&0.04&0.19&0.15&0.12 \\
$[$C/Fe$]$&-0.05&0.04&0.18&0.13&0.12 \\
$[$N/Fe$]$&-0.17&0.06&0.31&0.16&0.26 \\
$[$Ca/Fe$]$&-0.08&0.05&0.24&0.20&0.14 \\
\hline
AED (26) \\
\hline
log(age)&0.78&0.04&0.21&0.14&0.16 \\
$[$Fe/H$]$&-0.03&0.03&0.14&0.11&0.08 \\
$[$Mg/Fe$]$&0.19&0.02&0.09&0.10&0.00 \\
$[$C/Fe$]$&0.21&0.03&0.15&0.09&0.12 \\
$[$N/Fe$]$&0.08&0.04&0.19&0.11&0.15 \\
$[$Ca/Fe$]$&0.00&0.02&0.09&0.12&0.00 \\
\hline
\end{tabular}
\end{table}

Taking this approach, in Fig. \ref{frosting2} we plot CaII against H$\beta$ for the 149 (67\%) of our passive sample fit by Ez-Ages (see next section) with uncertainties in their CaII index $\leq$ 0.15. Next we simulate a set of CSPs using the SSP models of \cite{vazdekis10} with a base population of age $>$ 5 Gyr and [Fe/H] $>$ -0.68 and frostings of age $\leq$ 1 Gyr and [Fe/H] $>$ -0.68. The frostings have mass fractions of 1\%, 3\% and 5\%. The measured CaII and H$\beta$ indices for all combinations of frosting mass $\geq$ 1\% and age $\leq$ 0.7 Gyr and frosting mass $\geq$ 3\% and age $\leq$ 1.0 Gyr are plotted on Fig. \ref{frosting2}. For clarity the regions in the parameter space spanned by these models are represented by blue and green shaded polygons respectively. Finally we also plot the coverage predicted for genuine SSPs with 2 Gyr $<$ age $\leq$ 14 Gyr and  -0.68 $<$ [Fe/H] $\leq$ 0.2, roughly comparable to the parameters observed for our Coma passive sample (see Fig. \ref{sppsig}), in red. Of course here we note the mixing of two model sets, Vazdekis and Schiavon, although we expect the effects to be limited.

Fig. \ref{frosting2} shows that, despite sizable scatter in CaII, the majority of the 149 galaxies tested do not show evidence for small amounts of recent star formation. Indeed, by resampling the data given the measurement uncertainties we confirm that 88$\pm$2\% have CaII $\geq$ 1.1 or H$\beta$ $\leq$ 2 and thus are outside the parameter space covered by the explored frosting options. Furthermore, these galaxies have $<$CaII$>$ = 1.150 $\pm$ 0.01 which may be compared to $<$CaII$>$ = 1.140 for the plotted SSP models. That said, $\sim$ 15\% of the sample have 2.0 $\leq$ H$\beta$ $\leq$ 2.5 and could reach the blue and green polygons within their 1$\sigma$ errors. However, it is important to note that, while frosting may be present to some extent in these systems, many of them are also consistent with SSPs within their uncertainties. For reference, the most extreme galaxy on this plot, H$\beta$ $\sim$ 4.5 and CaII $\sim$ 0.7, is the k+a galaxy GMP 3892.

To summarise, we have found little evidence that the majority of the passive sample are having their SSP-equivalant ages biased low by small mass fractions of young stars from recent star formation episodes. We cannot, however, exclude this scenario for a small fraction of the sample.

\subsection{Results}
\label{sppresults} 

\begin{figure}
\scalebox{0.45}[0.45]{\includegraphics{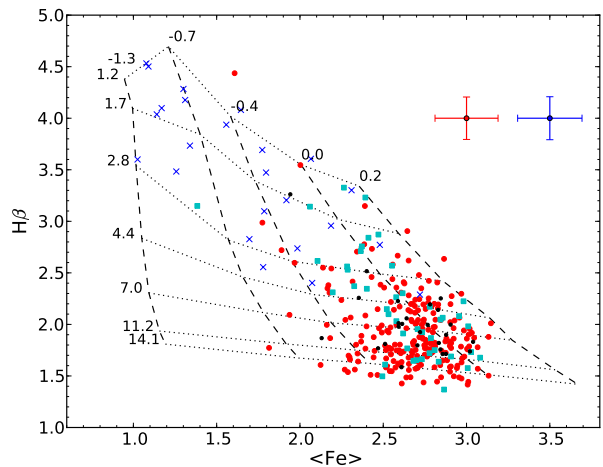}} 
\caption{H$\beta$ and $<$Fe$>$ indices for the 316 (89\%) of sample galaxies which were fit by {\sc Ez-Ages} overplotted on a Schiavon model grid with a solar abundance pattern. Red filled circles, Cyan filled squares and blue crosses are passive, AGN/LINER and star forming galaxies respectively. Black points represent galaxies whose emission driver could not be determined. Red and blue error bars show median index errors ($\sigma_{H\beta} = 0.21$ and $\sigma_{<Fe>} = 0.19$ in both cases) for passive and emission line galaxies respectively. Due to the relative insensitivity of this particular index pair to abundance pattern variations the grid is fairly stable in index space.}
\label{hbfe}
\end{figure}

In Fig. \ref{hbfe} we plot the index data for both our passive and emission-line samples, again separated by emission driver, on an example H$\beta$ -- $<$Fe$>$ model grid. This diagram contains the 222 (90\%) passive and 94 (85\%) emission-line galaxies for which {\sc Ez-Ages} yielded results. The vast majority of the failed fits continue to be caused by galaxies falling off the fiducial Balmer grid, regardless of which of the three Iron indices (Fe5270, Fe5335 or the mean of the two) were used. Also, we were unable to measure one or more of the required indices for a full abundance pattern fit for four of our passive sample due to bad pixels in the respective index bandpasses. We note that the model grid presented in Fig. \ref{hbfe} is for a solar-abundance model only but, as this index pair is specifically selected for its weak sensitivity to non-solar elemental abundances, this particular grid is relatively stable in index space. 

Simple inspection of Fig. \ref{hbfe} indicates that quiescent and AGN/LINER dominated galaxies have a comparable spread in age and [Fe/H] which, from visual assessment alone, typically spans from age$\geq$ 3 Gyr and [Fe/H] $>$ -0.7 to the edge of the model grid. On the other hand the star-forming subsample are generally younger, age$\leq$ 3 Gyr, with a spread in metallicity that covers the entire grid, -1.3 $\leq$ [Fe/H] $\leq$ +0.2. In Table \ref{spptab} we compare the subsamples in a more quantitative manner, presenting the mean, rms scatter ($\sigma_{tot}$), scatter due to errors ($\sigma_{errs}$) and intrinsic scatter ($\sigma_{int}$) of the six stellar population parameters measured here. The latter is computed as $\sigma_{int}^{2}$ = $\sigma_{tot}^{2}$ - $\sigma_{errs}^{2}$. A number of noteworthy points emerge from this table.

Firstly, our passive galaxy sample shows a large age spread of $\pm$0.19 dex (3.1 Gyr) even after factoring out scatter due to statistical error. While not directly comparable to the recent work of T08 who selected their sample based on morphology, this large intrinsic scatter in the ages of Coma passive galaxies, the vast majority of which occupy the cluster's red sequence, does not support their finding of age uniformity across the range in luminosity and $\sigma$ studied here. They also conclude that differing sample selection techniques in terms of morphology (early-types) vs colour (red sequence) are unlikely to be the main driver behind this discrepancy. To test this we obtained morphologies from NED for as many galaxies in our passive sample as possible and find that $\sim$ 72\% of the 222 galaxies are of types E-S0. The mean stellar population parameters for the passive early-types are presented in Table \ref{spptab}. Since the two samples are not independent, we use a resampling method to assess the significance of the differences between them. Only the mean log(age) and [Mg/Fe] are found to be different at the $\sim$ 3$\sigma$ level. This test also confirms that the observed intrinsic age scatter is not driven by our sample including emission-free spiral galaxies.

Secondly, the AGN/LINER systems have a younger mean age than the passive galaxies yet similar mean [Fe/H] and elemental abundances, in agreement with the findings of \cite{graves07}. Of course \cite{graves07} apply a colour constraint when selecting their sample and so, as our analysis is based typically on red sequence quiescent galaxies and AGN/LINERs which display a range of colours, a direct comparison may be inappropriate. To test this we apply the same colour cut used to identify the blue passive subset and find that, for the 40 galaxies that meet this criteria, their mean stellar population parameters remain effectively constant with $\mu_{\mathrm{log(age)}}$ changing by less than one standard error ($\mu_{\mathrm{log(age)}}$ = 0.72 $\pm$ 0.05 dex). Thus AGN/LINER galaxies that are on the cluster's red sequence are $\sim$ 15-20 \% younger than our passive sample. One further caveat here, however, is that the two subsamples may have different velocity dispersion distributions which perhaps introduces a further bias. This issue will be addressed in the next section.

\begin{figure*}
\flushleft
\scalebox{0.7}[0.7]{\includegraphics{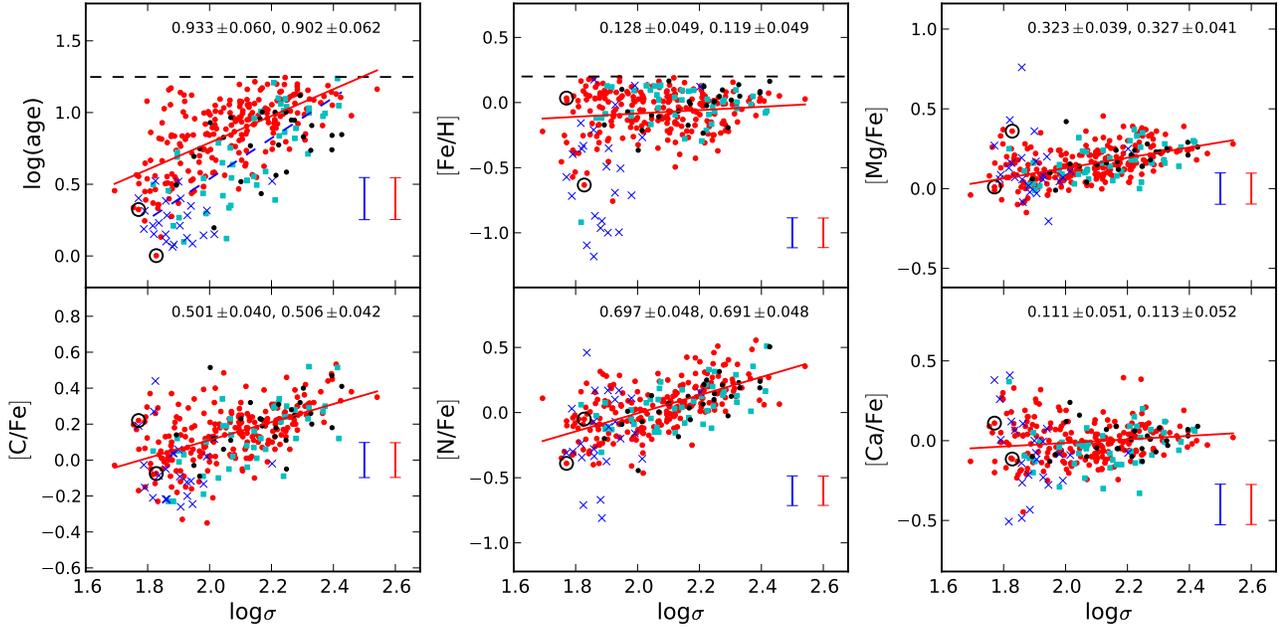}} 
\caption{The six derived stellar population parameters as a function of galaxy velocity dispersion for the 222 passive and 94 emission line galaxies fit by {\sc Ez-Ages}. Symbols are the same as in Fig. \ref{indsig}. The legend again contains the slope of a fit to the passive sample, first not masking and then masking the blue subset. The latter fit is also plotted as a red line in each panel. The blue line is a fit to the red AGN/LINER sample identified in the Section \ref{sppresults}. Error bars are median parameter errors for the passive, red, and emission, blue, samples. The black dashed lines in the first two panels denoted the age (17.7 Gyr) and [Fe/H] (+0.2) limit of the Schiavon models.}
\label{sppsig}
\end{figure*}

Thirdly, the 26 Coma galaxies with unclassified emission have average stellar population parameters similar to those of the passive and AGN/LINER samples. Indeed they are found to have a mean age between that of the former and latter. This likely rules out the presence of weak ongoing star formation in these galaxies and points toward low-level AGN/LINER activity. This prediction is further reinforced by the results of \cite{graves07} who find that weak AGN/LINERs display ages closer to those of passive galaxies (see their Fig. 12, top row of panels) and that, while not plotted in Fig. \ref{bpt} due to low line A/N, the AED galaxies all have [NII]/H$\alpha$ and [OIII]/H$\beta$ ratios compatible with AGN/LINER activity.

\subsection{Trends with Velocity Dispersion}
\label{sppsigsec}

\begin{figure*}
\flushleft
\scalebox{0.7}[0.7]{\includegraphics{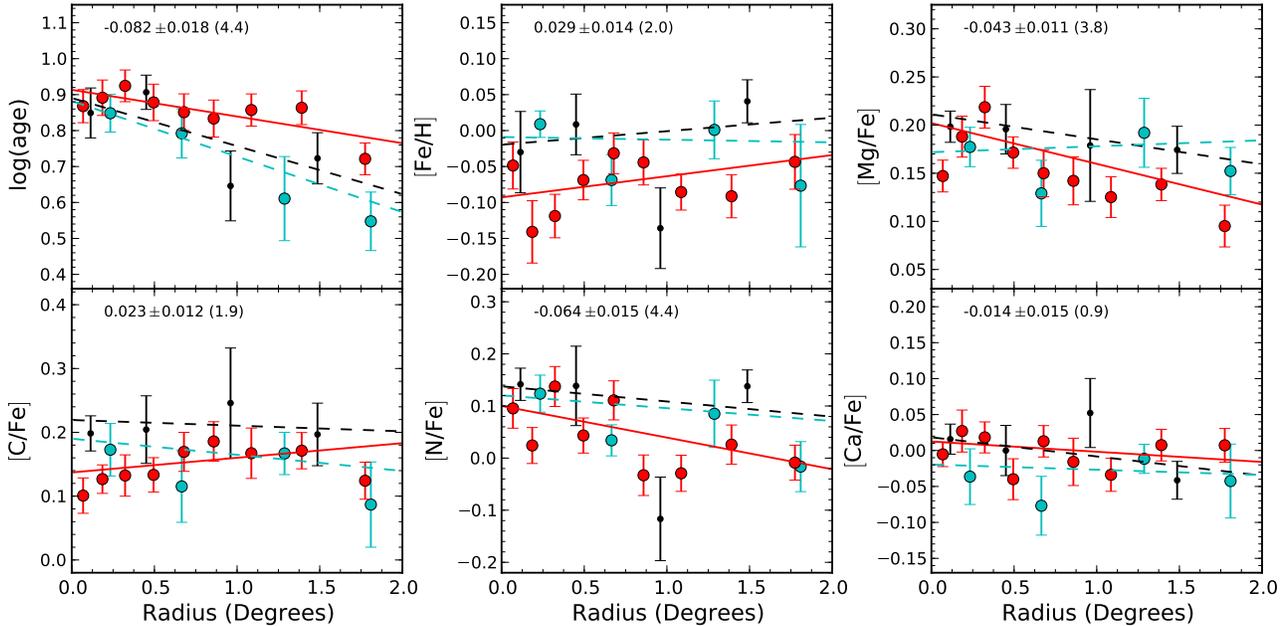}} 
\caption{Stellar population parameters as a function of projected cluster-centric radius for the passive and emission line galaxies that were fit by {\sc Ez-Ages}. Symbols are as in Fig. \ref{indrad}. Trend lines are fit to the raw data which are then binned with $\sim$ 27, $\sim$ 11 and $\sim$ 7 galaxies per bin for red, cyan and black bins respectively. Star forming galaxies are not plotted for clarity. The legend contains the slope of the passive galaxy fit, excluding the blue subset, and its significance.}
\label{spprad}
\end{figure*}

\begin{figure*}
\flushleft
\scalebox{0.7}[0.7]{\includegraphics{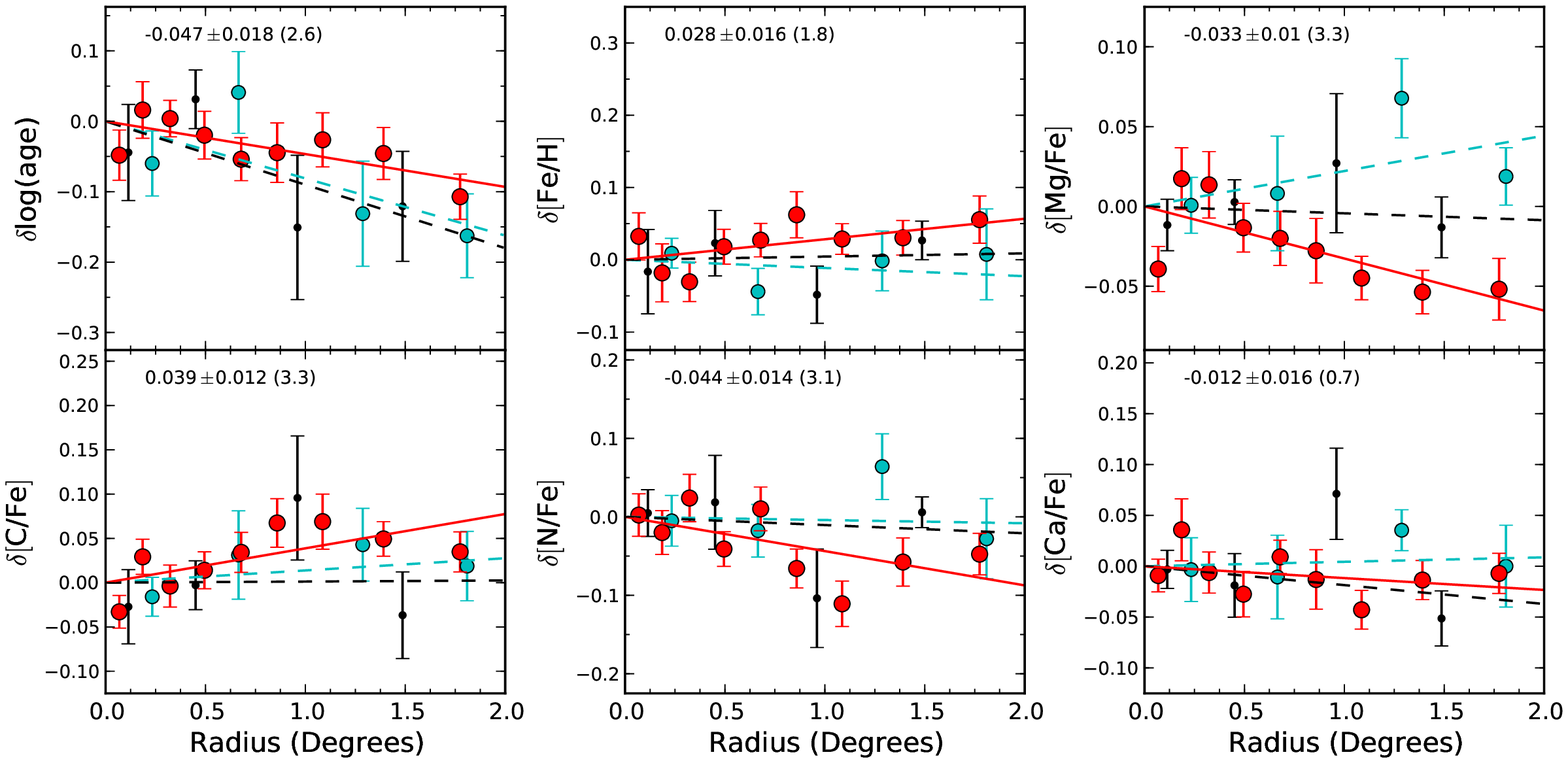}} 
\caption{Stellar population parameter residuals as a function of cluster-centric radius. Symbols as in Fig. \ref{indrad} and binning as in Fig. \ref{spprad}. Trend lines are fit to the raw data. The legend contains the slope of the passive galaxy fit, excluding the blue subset, and its significance. Star forming galaxies are not plotted for clarity. }
\label{spprad_p}
\end{figure*}

In this and the following section we seek to identify if the stellar population parameters of our sample galaxies display correlations with velocity dispersion and cluster-centric radius. To this end we will conduct a similar analysis to that carried out for the absorption line indices, fitting for both linear and planar relations.

In Fig. \ref{sppsig} we present the six measured stellar population parameters as a function of galaxy velocity dispersion for both passive and emission line samples. The legend of each panel again displays the results of a weighted linear fit to, firstly, the 222 galaxies in the quiescent sample and, secondly, having masked the two blue passive galaxies that were fit by {\sc Ez-Ages}. Once more these fits are obtained taking into account intrinsic scatter with the slope uncertainties being computed by resampling the stellar population parameter data using their errors. As with the index-$\sigma$ diagrams presented earlier, we find that the velocity dispersion coefficient in a planar fit is always well within one standard error of the slope derived from its respective linear fit. We therefore choose to only plot the linear relations but note that the discussion that follows also applies to trends corrected to fixed radius and taking into account aperture effects.

\begin{table*}
\centering
\caption{Stellar population parameter gradients per degree, from the linear fits presented in Fig. \ref{spprad} and the planar fits displayed in Fig. \ref{spprad_p}. The number of galaxies in each respective fit is displayed in parentheses in the table header and the significance of each fit in parentheses in the body of the table. For the passive sample the results in the first column include the blue subset while second column do not.}
\label{sppradtab}
\begin{tabular}{cr@{$\pm$}lr@{$\pm$}lr@{$\pm$}lr@{$\pm$}lr@{$\pm$}l}
\hline
Parameter & \multicolumn{4}{c}{Passive (222/220)} & \multicolumn{2}{c}{AGN/LINER (44)} & \multicolumn{2}{c}{SF (24)} & \multicolumn{2}{c}{AED (26)} \\ 
\hline
\multicolumn{11}{c}{Parameter -- R$_{cc}$ (Fig. \ref{spprad})} \\
\hline 
log(age)&-0.074&0.019(3.8)&-0.082&0.018(4.4)&-0.154&0.033(4.6)&0.079&0.067(1.2)&-0.134&0.046(2.9) \\ 
$[$Fe/H$]$&0.029&0.014(2.1)&0.029&0.014(2.0)&-0.004&0.025(0.2)&0.097&0.061(1.6)&0.019&0.034(0.6) \\ 
$[$Mg/Fe$]$&-0.042&0.012(3.5)&-0.043&0.011(3.8)&0.006&0.019(0.3)&0.061&0.056(1.1)&-0.026&0.028(0.9) \\ 
$[$C/Fe$]$&0.023&0.012(1.9)&0.023&0.012(1.9)&-0.025&0.023(1.1)&0.069&0.055(1.3)&-0.009&0.030(0.3) \\ 
$[$N/Fe$]$&-0.061&0.014(4.4)&-0.064&0.015(4.4)&-0.025&0.024(1.0)&-0.269&0.060(4.5)&-0.029&0.034(0.9) \\ 
$[$Ca/Fe$]$&-0.014&0.015(1.0)&-0.014&0.015(0.9)&-0.007&0.026(0.3)&-0.116&0.076(1.5)&-0.027&0.035(0.8) \\   
\hline
\multicolumn{11}{c}{$\delta$Parameter -- R$_{cc}$ (Fig. \ref{spprad_p})} \\
\hline 
log(age)&-0.041&0.018(2.3)&-0.047&0.018(2.6)&-0.081&0.036(2.2)&0.085&0.079(1.1)&-0.090&0.053(1.7) \\ 
$[$Fe/H$]$&0.029&0.015(1.9)&0.028&0.016(1.8)&-0.011&0.025(0.5)&0.065&0.061(1.1)&0.004&0.042(0.1) \\ 
$[$Mg/Fe$]$&-0.033&0.011(2.9)&-0.033&0.010(3.3)&0.022&0.022(1.0)&0.076&0.057(1.3)&-0.004&0.034(0.1) \\ 
$[$C/Fe$]$&0.037&0.012(3.1)&0.039&0.012(3.3)&0.014&0.024(0.6)&0.070&0.052(1.3)&0.001&0.033(0.0) \\ 
$[$N/Fe$]$&-0.042&0.014(3.0)&-0.044&0.014(3.1)&-0.004&0.026(0.2)&-0.229&0.068(3.4)&-0.010&0.041(0.3) \\ 
$[$Ca/Fe$]$&-0.012&0.015(0.8)&-0.012&0.016(0.7)&0.004&0.027(0.2)&-0.130&0.079(1.6)&-0.019&0.049(0.4) \\
\hline 
\end{tabular}
\end{table*}

The first panel of Fig. \ref{sppsig} indicates a strong positive correlation between SSP-equivalent age and velocity dispersion for Coma passive galaxies. From visual inspection it is apparent that the scatter about the fitted relation increases for lower $\sigma$ galaxies to the extent that a fraction of these systems have ages comparable to galaxies with the highest velocity dispersions in the cluster. Indeed, low $\sigma$ galaxies are seen to span the full range of ages provided by the Schiavon models. Extensive scatter in the ages of Coma low $\sigma$ galaxies has also been reported by T08, based on index data from \cite{moore02}, and by \cite{smith09a} for their dwarf sample. This implies that passive galaxies with low velocity dispersions in Coma show a wider range of star formation histories than do high $\sigma$ galaxies.
 
This result is unlikely to be driven by an increased incidence of star bursts with ages $\lesssim$ 1 Gyr in low $\sigma$ systems as confirmed earlier. As before, fitting separately to the early-type subsample does not significantly affect the observed relation, in fact we find a slope that is highly consistent with that obtained for the entire passive sample. Furthermore, the trend is altered by less than one standard error when derived from a planar fit. 

One additional factor deserves further comment. \cite{kelson06} pointed out that, at fixed $\sigma$, the stellar population parameters of a magnitude-limited sample will be biased in the sense that, for example, older galaxies are fainter and so may be excluded by the limit. This issue will of course be more prevalent at lower velocity dispersions where an increasingly larger fraction of old systems are too faint and fall out of the sample. In the first panel of Fig. \ref{sppsig} this would then result in the fitted slope being biased too steep. However, it has been demonstrated by both \cite{smith09b} and \cite{allanson09} that such a selection effect does not strongly influence simple two-parameter fits and so it is unlikely to significantly alter the recovered relation.

Having considered all the points discussed above, we must conclude that the data presented here support a correlation between galaxy SSP-equivalent age and velocity dispersion, or in other words archaeological downsizing, for passive galaxies in the Coma cluster across the studied $\sigma$ range.

In this panel we also fit separately to the red AGN/LINER galaxies identified previously, denoted by the blue dashed line in Fig. \ref{sppsig}, the result of which clearly demonstrates that at fixed velocity dispersion these systems are, on average, younger than their passive counterparts. 

The second panel of Fig. \ref{sppsig} displays a much weaker positive correlation, relative to the age-$\sigma$ relation, between [Fe/H] and $\sigma$ for the passive galaxies. There is some evidence for an asymmetric scatter toward lower [Fe/H] at the low $\sigma$ end of the relation, although we note that the upper bound is inherently limited by the models coverage which is capped at [Fe/H] = 0.2. AGN/LINER systems appear to be distributed in a similar manner to the passive galaxies in this parameter space while the star forming sample shows only a limited overlap, with a sizeable fraction having [Fe/H] $\leq$ -0.5. In fact a number of the latter sample have [Fe/H] $\lesssim$ -1, a regime where {\sc Ez-Ages}, and presumably the Schiavon models, may well be unsuitable as cautioned by the authors \citep{graves08}.

In the third panel of Fig. \ref{sppsig} we see that higher $\sigma$ quiescent galaxies tend to have higher [Mg/Fe], and therefore shorter star formation timescales, than lower $\sigma$ galaxies. The fitted slope in this panel is highly comparable to that derived by \cite{smith09a}, an interesting result given the different velocity dispersion coverage of the two samples, and \cite{graves07} for giant galaxies based on stacked SDSS spectra. The majority of the AGN/LINER and star forming galaxies scatter across a similar range in [Mg/Fe] to the passive galaxies with only one sizeable outlier, likely explained by this galaxy also having [Fe/H] = -1.18.

The final three panels of Fig. \ref{sppsig} show relatively strong positive correlations for [C/Fe]-$\sigma$ and [N/Fe]-$\sigma$ while [Ca/Fe] is seen to only weakly correlate with velocity dispersion over the range studied here. Once more, the emission line galaxies occupy similar regions of each respective parameter space to the passive galaxies with only a few star forming outliers, typically with [Fe/H] $\sim$ -1. The fitted slopes for [C/Fe], [N/Fe] and [Ca/Fe] are all in agreement with those found by \cite{graves07} within the joint parameter uncertainties. They are somewhat larger than those reported by \cite{smith09a} for their simple two-parameter fits to Coma dwarf galaxies. However, those fits may well be poorly constrained given the narrow $\sigma$ range covered by their dwarf sample.

\subsection{Trends with Radius}
\label{sppradsec}

\begin{table*}
\centering
\caption{Stellar population parameter radial gradients obtained by fitting a plane to each passive galaxy subset as defined by the first column. Column 2 gives the number of galaxies in each fit. Significant gradients ($>$ 2$\sigma$) are denoted by bold text. ETG and LTG refer to early-type and late-type galaxies respectively.}
\label{select}
\begin{tabular}{ccr@{$\pm$}lr@{$\pm$}lr@{$\pm$}lr@{$\pm$}lr@{$\pm$}lr@{$\pm$}l}
\hline
Criteria & Ngal & \multicolumn{2}{c}{log(age)} & \multicolumn{2}{c}{$[$Fe/H$]$} & \multicolumn{2}{c}{$[$Mg/Fe$]$} & \multicolumn{2}{c}{$[$C/Fe$]$} & \multicolumn{2}{c}{$[$N/Fe$]$} & \multicolumn{2}{c}{$[$Ca/Fe$]$} \\
\hline
All & 220 & {\bf -0.047}&{\bf 0.018} & 0.028& 0.016 & {\bf -0.033} & {\bf 0.010} & {\bf 0.039}&{\bf 0.012} & {\bf -0.044} & {\bf 0.014} & -0.012&0.016 \\
$\sigma <$ 125 km s$^{-1}$ & 119&-0.006&0.025 &  -0.006&0.023 &  -0.033&0.019 &  0.032&0.019 &  -0.032&0.021 &  0.007&0.025 \\ 
$\sigma >$ 125 km s$^{-1}$ & 101&{\bf -0.084} &{\bf 0.025}  &  {\bf 0.067} &{\bf 0.019}  &  {\bf -0.036} &{\bf 0.014}  &  {\bf 0.047} &{\bf 0.016}  &  {\bf -0.072} &{\bf 0.018}  &  -0.025&0.019 \\ 
ETG & 158&-0.039&0.024 &  0.039&0.020 &  {\bf -0.032} &{\bf 0.016}  &  {\bf 0.048} &{\bf 0.017}  &  {\bf -0.092} &{\bf 0.020}  &  -0.018&0.019 \\ 
ETG $\sigma <$ 125 km s$^{-1}$ & 79&0.046&0.038 &  0.016&0.036 &  -0.050&0.032 &  {\bf 0.069} &{\bf 0.029}  &  {\bf -0.105} &{\bf 0.035}  &  -0.005&0.042 \\ 
ETG $\sigma >$ 125 km s$^{-1}$ & 79&{\bf -0.092} &{\bf 0.033}  &  {\bf 0.061} &{\bf 0.023}  &  -0.035&0.018 &  0.034&0.020 &  {\bf -0.099} &{\bf 0.022}  &  -0.029&0.024 \\ 
LTG & 20&-0.086&0.067 &  -0.029&0.053 &  -0.024&0.045 &  -0.005&0.044 &  0.011&0.050 &  -0.022&0.055 \\   
\hline
\end{tabular}
\end{table*}

In Fig. \ref{spprad} we plot the measured stellar population parameters against cluster-centric radius for three out of four subsamples, only excluding the star forming systems again for clarity. The trend lines are linear least-square fits to the raw data, weighted by their uncertainties, which are then binned for clarity. Each bin contains $\sim$ 28, $\sim$ 11 and $\sim$ 7 galaxies per bin for the passive, AGN/LINER and AED samples respectively. The results of these fits, including those to the star forming galaxies, are presented in Table \ref{sppradtab}, first section. 

For the quiescent galaxies significant correlations are obtained for log(age) (3.8$\sigma$), [Mg/Fe] (3.5$\sigma$) and [N/Fe] (4.4$\sigma$). These trends typically get marginally stronger when masking the blue subset. No significant trend is seen for [Fe/H]. This implies passive galaxies on the outskirts of the cluster are, on average, 0.15 dex ($\sim$ 2 Gyr) younger and have more extended star formation histories than those in the cluster core. Both AGN/LINER galaxies and those with unclassified emission display significant age trends with comparable slopes. It is only the small sample of star forming galaxies which do not possess a correlation between SSP-equivalent age and radius.

As demonstrated for the index-radius relations in Fig. \ref{indrad_p}, controlling for variations in velocity dispersion with radius can have a substantial effect on the significance of the observed radial trends. Therefore in Fig. \ref{spprad_p} we plot the residual stellar population parameters, again derived by fitting a multivariate plane to the data, against radius and as such track the parameter variations at fixed velocity dispersion. The results of these fits are tabulated in the second section of Table \ref{sppradtab}. 

For the passive galaxies, significant, although typically weaker, trends are still found for log(age) (2.6$\sigma$), [Mg/Fe] (3.3$\sigma$), [C/Fe] (3.3$\sigma$) and [N/Fe] (3.1$\sigma$). Thus bright quiescent galaxies on the outskirts of the cluster are on average $\sim$ 20\% younger with lower [Mg/Fe] and [N/Fe] but higher [C/Fe] than those in the core at fixed velocity dispersion. 

By contrast the only remaining significant trends for the emission line galaxies are log(age) (2.2$\sigma$) for AGN/LINER systems and [N/Fe] (3.4$\sigma$) for star forming galaxies. This may well stem from the small size of these subsamples. In fact we find that if the AGN/LINER and AED galaxies are fit together, an approach that is not wholly unjustifiable given the identified similarities between the two samples, a log(age) correlation of $-0.101\pm$0.028 dex deg$^{-1}$ is recovered. Repeating this for [Fe/H] and [Mg/Fe] we find no significant trends for the combined sample.

Again, a valid question here is to what extent are the relatively weak but significant radial trends obtained for the red passive sample driven by the outermost data points. Once more we have repeated the fits with red passive galaxies beyond R$_{cc}$ = 1.55$^{\circ}$ masked. For the planar fits the results for [Mg/Fe] (2.5$\sigma$), [C/Fe] (3.7$\sigma$) and [N/Fe] (3.3$\sigma$) remain comparatively unchanged in terms of strength and sign. The same is also true for the linear fits, although [C/Fe] now becomes significant at the 3.4$\sigma$ level. The radial trend in log(age), however, becomes insignificant in the linear (1.7$\sigma$) and planar (1.2$\sigma$) fits yet the signs remain the same as those presented in Table \ref{sppradtab}. Thus it would seem the trend toward younger ages with radius begins to become significantly defined at R$_{cc} \gtrsim 1.5^{\circ}$ for the red passive galaxies, although the present
data cannot trace the form of the relation in any detail.

Next we seek to check the correlations found for the passive galaxies against other possible methods of subdivision such as morphology and $\sigma$. The former is of particular interest since, as no morphological criteria were imposed, the morphology-density relation may well play a key role in driving the observed radial stellar population gradients for this sample. Thus in Table \ref{select} we apply a variety of further cuts to the quiescent galaxies excluding the two particularly blue early-type galaxies. 

Firstly, the sample is split into subsets based on their velocity dispersion into low ($\sigma <$ 125 km s$^{-1}$) and high ($\sigma >$ 125 km s$^{-1}$) velocity dispersion bins. Here we see that, most notably, significant trends in log(age) (3.2$\sigma$) and [Fe/H] (3.3$\sigma$) are recovered for the high $\sigma$ sample. Next we separate the sample into early-type and late-type galaxies based on their NED morphologies. Here we find that the former does not have a significant radial age gradient, though it is consistent with that found for the passive galaxies as a whole taking into account the fact that these two samples are not independent. For the latter we find a sizable but insignificant negative trend. At face value one might now be tempted to conclude that the gradient found for the entire sample (-0.047$\pm$0.018 dex deg$^{-1}$) may well be driven by a combination of a very weak trend in the early-type galaxies modulated by a larger negative gradient in the late-type galaxies. To test this we subdivide the early-type galaxies into low and high $\sigma$ bins. Interestingly we find that the higher $\sigma$ early-type galaxies possess a strong negative age gradient, comparable in magnitude to the late-types, while the low $\sigma$ early-types show no significant gradient. This perhaps implies that the radial variation of the stellar populations of early-type galaxies in Coma may be more complex than the simple planar model applied here. Nevertheless we confirm that the morphology-density relation does not drive the log(age)-radius correlation seen for the passive sample.

\begin{table}
\centering
\caption{Comparison between the results obtained for this work's passive galaxies, both full sample and core-South-West, and those in the literature. S06 gradients are again converted to per degree assuming R$_{200}$ = 1.59$^{\circ}$ for Coma and we assume their [$\alpha$/Fe] gradient is comparable to our [Mg/Fe] gradient.}
\label{sppcomp}
\begin{tabular}{lr@{$\pm$}lr@{$\pm$}l}
\hline
Source & \multicolumn{2}{c}{log(age)} & \multicolumn{2}{c}{$[$Mg/Fe$]$}  \\
\hline
This Work: All Passive 	& -0.047&0.018 	& -0.033&0.010 \\
This Work: Passive Core-SW		& -0.041&0.048	& -0.047&0.028 \\	
S08 (Coma dwarfs)		& -0.369&0.069	& -0.128&0.039 \\
S06 		& -0.045&0.011	& -0.030&0.006 \\
\hline
\end{tabular}
\end{table}

\begin{figure*}
\scalebox{0.43}[0.43]{\includegraphics{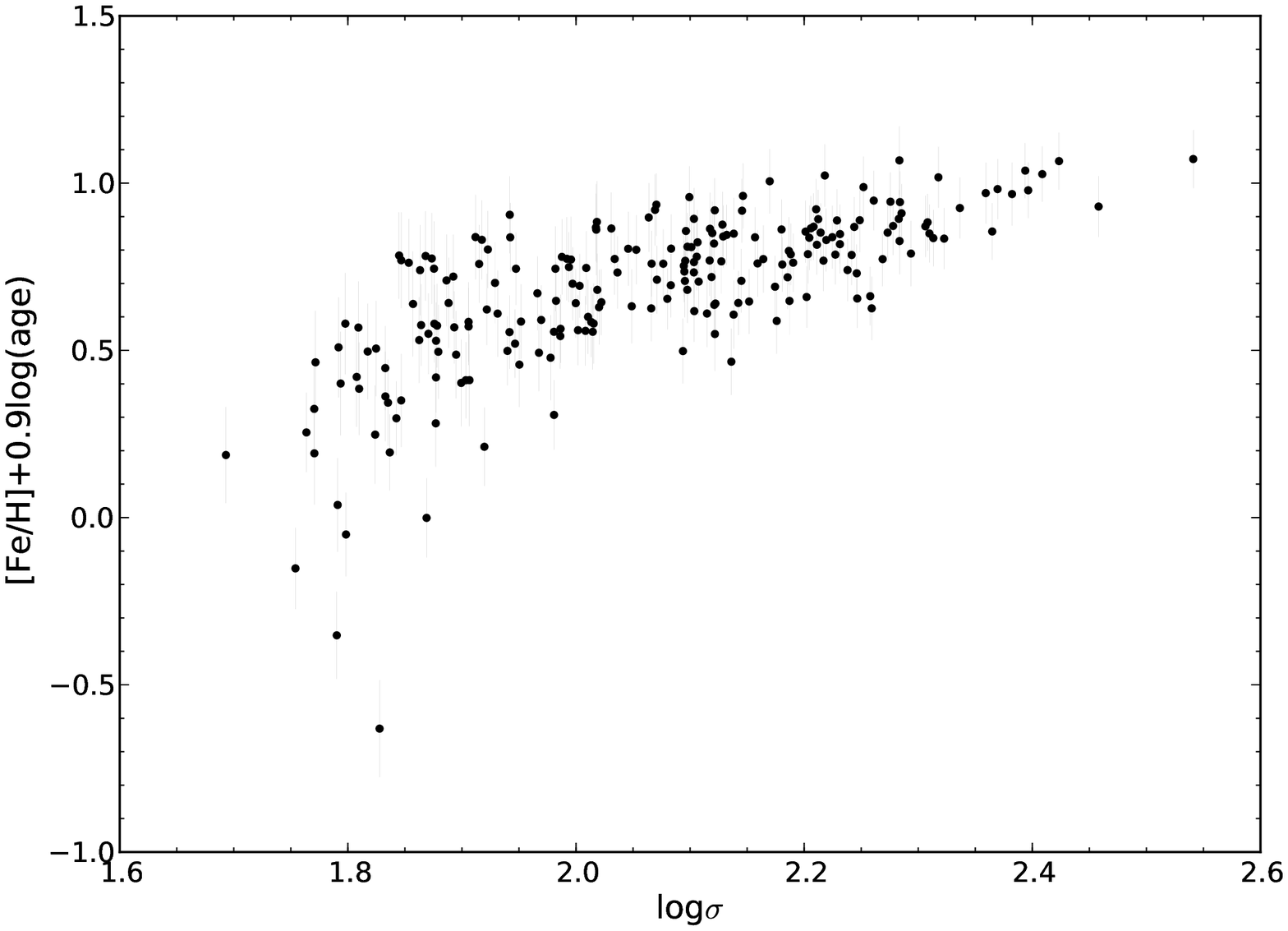}} \scalebox{0.43}[0.43]{\includegraphics{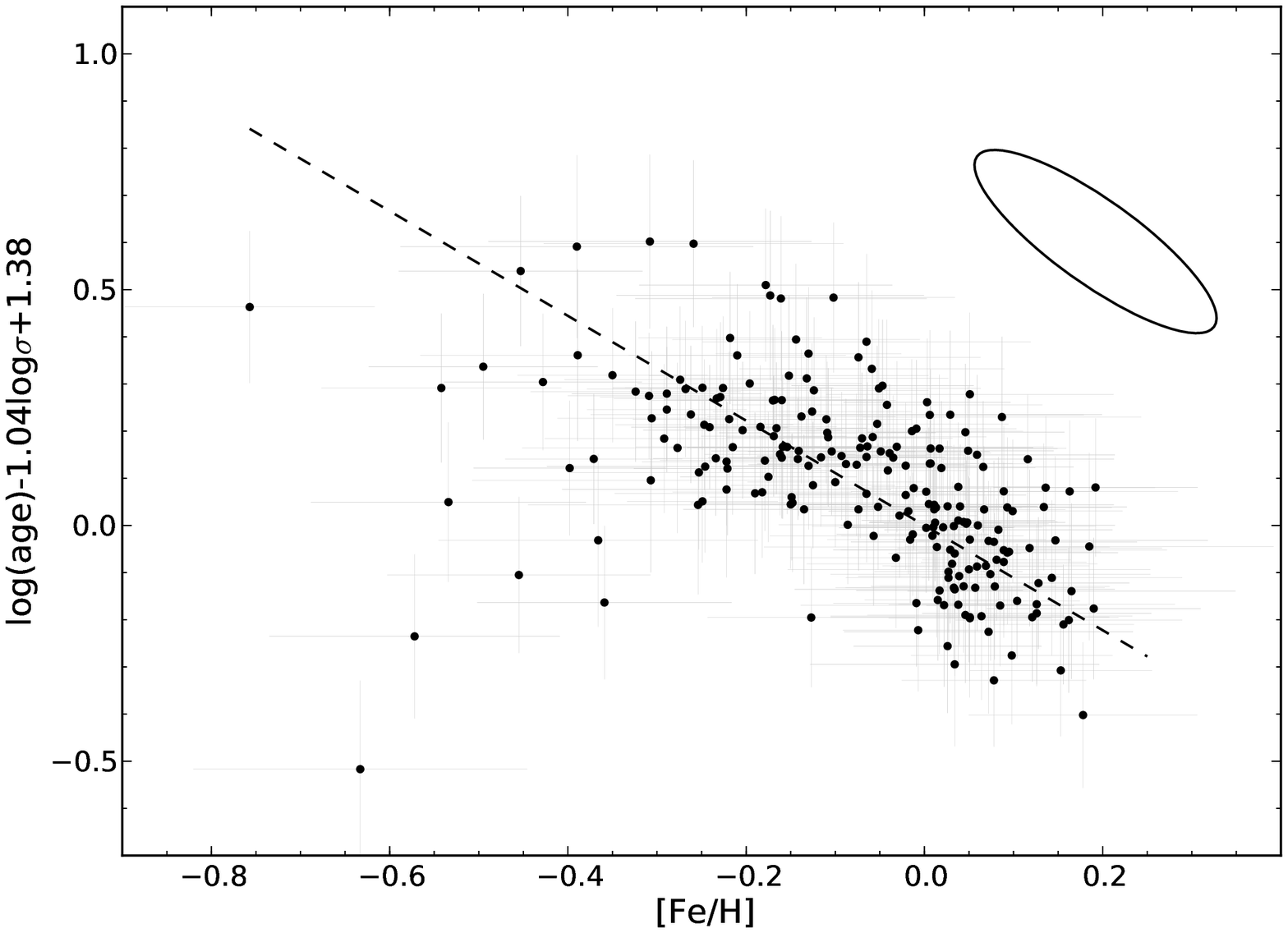}} 
\caption{Two edge on views of our version of the \citet{trager00b} Z-plane, plotted here as an Fe-plane, for our quiescent sample. In the right panel we plot a typical 68\% confidence error ellipse which, as discussed in the text, is nearly parallel to the observed age-metallicity anti-correlation. The dashed line in this panel shows the best fit relation.}
\label{zplane}
\end{figure*}

In Table \ref{sppcomp} we present a comparison of the significant results obtained in this work for the passive galaxies to those studies featured previously in Table \ref{comptab}, excluding C02 who did not derive log(age) or [Mg/Fe] gradients. Again we fit separately to the Core-SW region subsample. We consider the [$\alpha$/Fe] result of S06 comparable to our [Mg/Fe] gradient. From this table it is immediately apparent that highly consistent age and abundance ratio gradients are found by this work and S06, despite each using different SSP models. It also indicates that S08 derived much stronger trends for their dwarf sample. We confirm that for bright Coma galaxies a limited azimuthal coverage does not result in significantly stronger gradients, again taking into account that the overall and Core-SW samples are not independent.

\subsection{The Z-plane}
\label{zplanesec}

It was \cite{trager00b} who first highlighted an anti-correlation between age and metallicity at fixed velocity dispersion in early-type galaxies, terming it the Z-plane. In physical terms this relation implies that younger galaxies, in an SSP-equivalent sense, have played host to a larger number of stellar generations allowing for greater metallic enrichment relative to older galaxies. To test for such a trend in our data we fit our quiescent sample with an Fe-plane of the form,

\begin{equation*}
\mathrm{[Fe/H]} = a\log\sigma + b\log\mathrm{age} + c
\end{equation*}

\noindent where the coefficients $a$ and $b$ give the dependence of [Fe/H] on $\sigma$ at fixed age and on age at fixed $\sigma$ respectively. The fit is weighted by errors in [Fe/H] and age but no attempt to account for correlated errors is made at this stage (but see below). We recover $a$ = 0.93$\pm$0.10, $b$ = -0.90$\pm$0.08 and $c$ = -1.24$\pm$0.17 with an rms scatter of 0.17 and in Fig. \ref{zplane} plot two projections of the fitted plane. The slope of [Fe/H] with $\sigma$ is steeper than that found by \cite{trager00b}, $a$ = 0.48$\pm$0.12, and by \cite{smith09a} for their Coma dwarf sample, $a$ = 0.51$\pm$0.2. However, the age-metallicity correlation we derive is comparable within the joint errors to that obtained by both works, $b$ = -0.74$\pm$0.09 from the former and $b$ = -0.74$\pm$0.14 from the latter. We estimate an intrinsic scatter of 0.13 in [Fe/H] and in the left panel of Fig. \ref{zplane}, just by eye, note an increased scatter for lower $\sigma$ galaxies, as also found by \cite{smith09a}. Quantitatively, the intrinsic scatter increases from 0.04 for $\log\sigma >$ 2.1 to 0.17 for $\log\sigma <$ 2.1.

\begin{figure}
\scalebox{0.43}[0.43]{\includegraphics{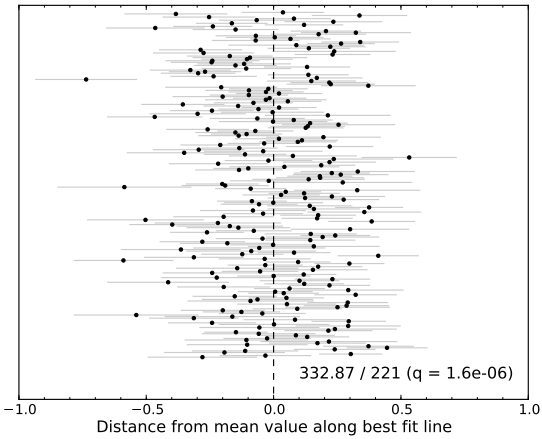}}
\caption{Distribution of distances from the mean displacement following each point in the right hand panel of Fig. \ref{zplane} being deprojected onto the best fitting line. The one-dimensional $\chi^{2}$, degrees of freedom and q, the probability that the observed distribution could be attributed solely to correlated errors, are shown.}
\label{ptest}
\end{figure}

Correlated errors between age and [Fe/H] can be a serious problem when trying to establish whether an age-metallicity anti-correlation exists at fixed velocity dispersion \citep[e.g.][]{trager00b}. Indeed it may be the case, given the fact that the orientation of the error ellipse in the right hand panel of Fig. \ref{zplane} is parallel to the slope of the dashed line, that the observed anti-correlation can be completely accounted for by uncertainties in the data. To test whether this is the case we employ a similar method to \cite{rawle08} and deproject the data onto the fitted plane (the trend line in Fig. \ref{zplane}, second panel). Next we compute the one-dimensional $\chi^{2}$ for the data along the line about a constant value, taking the one-dimensional error to be the semi-major axis of the 1$\sigma$ (40\% joint confidence in two-dimensions, 68\% in one-dimension along the major axis) error ellipse obtained from our error model simulations rescaled to the S/N of each data point. See Appendix \ref{appendixb} for an examination of the robustness of this choice. In Fig. \ref{ptest} we plot the distribution of one-dimensional residuals against an arbitrary y-axis, together with the one-dimensional $\chi^{2}$, degrees of freedom and Q, the probability that the observed distribution could be attributed solely to correlated errors. This test demonstrates that the Z-plane has a high probability of being resolved and would require an increase in errors in log(age) and [Fe/H] of $>$20\% to become unresolved, which, given the analysis conducted in Appendix \ref{appendixb}, is very unlikely.

It is also possible that the correlated errors may bias $b$, the slope of the age-metallicity trend at fixed $\sigma$. To test for this effect we use simple simulations that take as their input the observed ages and velocity dispersions and new artificial [Fe/H] values which are assigned using the Z-plane with coefficients as stated above. We generate new realisations of this dataset by resampling the ages and metallicities using the error ellipse, again rescaled to the S/N of each data point. To each new realisation we fit the Z-plane and obtain a new $b$ coefficient. Repeating this process 500 times we find that, for an anti-correlation as strong as the one we observe, $b$ is not significantly influenced by the correlated errors between age and metallicity.

\section{Discussion}

\subsection{Downsizing in the Coma Cluster}

Evidence of downsizing, referred to here as a correlation between galaxy velocity dispersion and SSP-equivalent age, has been identified by many recent studies on the stellar populations of quiescent galaxies in the local universe \citep[e.g.][]{caldwell03,nelan05,smith06,graves07,smith07,thomas09,graves09}. This relation can be considered the archaeological equivalent of the downsizing trend observed by authors such as \cite{cowie96} and more recently by \cite{perezgonzalez08}, who find that the total luminosity of galaxies which are undergoing rapid star formation declines as a function of redshift. 

The slope of the age-$\sigma$ relation reported by those works listed above are between 0.35 \citep{graves07} and 0.72 (S06) across approximately the same $\sigma$ range as studied here. In fact \cite{bernardi06} find a slope of 1.1 based on stacked early-type galaxy spectra from the SDSS. This sizable spread can, at some level, likely be attributed to a host of systematics such as sample selection (morphology, colour or both), data processing techniques and the SSP model of choice for each study. Nevertheless a picture is developing such that age does correlate with velocity dispersion in the sense that higher $\sigma$ quiescent galaxies formed their stars earlier than lower $\sigma$ galaxies. 

Previously \cite{sanchezblaz06b} and T08 have reported that early-type galaxies in the core of the cluster show no signal of downsizing. T08 go on to highlight that the Coma sample used in \cite{thomas05}, 35 early-type galaxies from \cite{mehlert00}, also does not support downsizing and that the age-$\sigma$ relation they find for the 97 Coma red-sequence galaxies in \citet[][hereafter NFPS]{nelan05} is the result of no Balmer line emission infill correction being applied. On the contrary \cite{matkovic09} do find, albeit limited, evidence for downsizing in their Coma galaxies. Given the wealth of studies on the quiescent galaxy population in the local volume that support a significant age-$\sigma$ relation, one must ask why evidence for its signature in Coma seems inconclusive at best.

With this in mind, in this section we seek to test the robustness of our observed age-$\sigma$ correlation against a number of systematic effects. These are the level at which we claim a galaxy to be quiescent, the approach used to correct the observed indices for each galaxy's velocity dispersion and the choice of SSP model employed.

As detailed earlier, our analysis requires that a galaxy must not have emission in any of H$\beta$, [OIII]5007, H$\alpha$ or [NII]6585 detected by {\sc Gandalf} with a A/N $\geq$ 4 in order to be considered quiescent. This detection threshold was set in accordance with the authors of the routine but we note that T08 opt for a much stricter cut, cleaning any emission observed at the A/N $\geq$ 2 level. In part this more severe criteria likely helps compensate for their study not having access to the, typically, stronger H$\alpha$ and [NII]6585 lines. That said the S/N of their spectra, which is much higher than that used here, will result in more sensitivity to emission and so allow the detection of fainter lines at fixed A/N. To test what effect cutting at this level, which amounts to a detection threshold of EW(H$\beta$) $\sim$ 0.13 \AA\ in our SDSS spectra, has on our observed downsizing trend we re-fit the 158 of the 222 passive galaxies that meet this criterion and have measured stellar population parameters. We find age $\propto \sigma^{0.96 \pm 0.07}$, consistent with the slope obtained when fitting the entire sample. Thus a stricter emission line cut appears not to affect the downsizing trend we recover.

While perhaps only a minor effect, \cite{kelson06} find some evidence for the particular scheme used to velocity broadening correct their index data influencing their measured parameter correlations. Here we opt to test two further correction techniques, that employed by {\sc Lick\_Ew} and the more standard multiplicative method of \cite{trager98}. The former smooths the observed spectra by $\sigma_{smooth}=\sqrt{\sigma_{lick}^2-\sigma_{res}^2-\sigma_{gal}^2}$ where $\sigma_{lick}$ is the Lick resolution of the particular index being measured \citep[taken from Table 1 of ][]{schiavon07}, $\sigma_{res}$ is the SDSS instrumental resolution and $\sigma_{gal}$ the galaxy velocity dispersion. Should the sign of the term under the square root turn negative the code uses multiplicative corrections based on smoothed SSP spectra from \cite{schiavon07}. We take the velocity broadening polynomial coefficients for the \cite{trager98} multiplicative method from their Table 5. The technique used by {\sc Lick\_Ew} yields age $\propto \sigma^{0.83 \pm 0.07}$ while the multiplicative route gives age $\propto \sigma^{0.78 \pm 0.06}$. Thus, while the latter is a shallower trend, our correction scheme and both of those tested produce consistent results within the joint errors. 

To test the model dependence of our observed downsizing trend we make use of the empirically derived transformations of \cite{smith09a} to convert our stellar population parameters to the TMBK model system. Fitting the age-$\sigma$ relation to the transformed data we find age $\propto \sigma^{0.83 \pm 0.06}$. Again this is consistent, within the joint uncertainties, to the result obtained when the Schiavon models are employed.

In summary, following the above tests, and those conducted earlier, we conclude that our SDSS data does robustly support downsizing in both early-type and passive galaxy populations in the Coma cluster. As demonstrated in Appendix \ref{appendixa}, we have found good consistency between this work and NFPS in terms of derived age-$\sigma$ relations. We speculate that small sample size may well explain the absence of downsizing seen by T08 using their Keck/LRIS data, given the sizable scatter in the first panel of Fig. \ref{sppsig}. Furthermore we have identified that, for galaxies which are in both studies, the \citet[][hereafter M02]{moore02} dataset results in a significantly flatter H$\beta$-$\sigma$ slope than this work. This issue, with as yet unknown origin, may well drive T08 to conclude the M02 data displays no age-$\sigma$ correlation.

\subsection{Environment}

The Coma cluster is an ideal place to study how environment affects the stellar populations of galaxies across a wide range of local densities. Its core represents one of the highest density regions that is readily observable for line strength studies while the cluster's outskirts provide excellent insight into galaxies that are, perhaps for the first time, just starting to feel the effects of a cluster environment. Indeed evidence has been mounting for some time that galaxies in the cluster's core display systematically different line indices, and by inference stellar populations, than those in the outer regions.

The first study to highlight the spatial dependence of the stellar populations of Coma members using line indices was that of \cite{guzman92}. They observed 51 Coma elliptical galaxies located in the cluster's core and ``halo'', defined by them having a projected radius of 1$^{\circ}$ $<$ R $<$ 5$^{\circ}$, with $\sigma >$ 100 km s$^{-1}$. As such their sample is comparable to the high $\sigma$ end of this work. They found that their Mg$_{2}$-$\sigma$ relation displayed a small environmental dependence such that, at fixed $\sigma$, galaxies in the core have higher index strengths than those in the cluster's halo. Employing early stellar population models, they went on to attribute this trend to variations in age with younger galaxies at larger cluster-centric radii.

Shortly after the work of Guzm{\'a}n et al., the aforementioned study of \cite{caldwell93} uncovered a sample of post-starburst early-type galaxies in the outer South-West region of the cluster. These galaxies were identified based on their strong high-order Balmer lines, indicative of a subpopulation of A stars with lifetimes $\sim$ 1 Gyr. Caldwell et al. could not comment further on whether this post-starburst population were localised to the NGC 4839 subgroup or spatially a more general phenomenon, as they were only able to observe one field on the clusters outskirts \citep[although see][]{caldwell97}. That said, very recently \cite{mahajan10} found that k+a galaxies in Coma are in fact uniformly distributed about the cluster, only avoiding regions with substantial X-ray emission such as the core.

C02 once more addressed the effect of environment on the stellar populations of Coma members, this time spanning a wide range in total magnitude from dwarf to giants, using Lick indices. They found evidence for a significant radial gradient in Mg$_{2}$, even after factoring out the index's correlation with luminosity. As previously mentioned, based on the weaker trends they recovered for H$\beta$, they interpreted this as a metallicity gradient. More recently, S08 have concluded that the radial index trends in a sample of red-sequence Coma dwarf galaxies in fact support a strong gradient in their SSP-equivalent ages, with older dwarfs preferentially found in the cluster's core. Again neither of these studies were able to comment on the generality of this gradient due to limited azimuthal coverage.

The results for our quiescent sample, presented in Table \ref{sppradtab}, provide further support for an interpretation of the observed radial index trends in terms of variations in galaxy age rather than metallicity. We also find evidence for significant gradients in [Mg/Fe], [C/Fe] and [N/Fe]. These findings indicate that passive galaxies on the outskirts of the cluster are younger, formed their stars over a longer period of time and have higher [C/Fe] but lower [N/Fe] than galaxies located in the cluster's core. Furthermore, when considering our core-South-West subsample we find no significant increase in recovered age or [Mg/Fe] gradients, providing the first hint that, at least for bright galaxies, the South-West field is not unique.

In Tables \ref{comptab} and \ref{sppcomp} we have shown that the samples of S08 and C02 typically display much stronger cluster-centric radial gradients in terms of line indices and stellar population parameters than derived for our sample. We have proposed that this finding can be attributed to the differing luminosity coverage of the three samples rather than a bias being introduced by S08 and C02 only observing the cluster's South-West region. Taken together this implies that either; brighter galaxies had finished forming the majority of their stars before entering the cluster, star formation within these systems was less affected by the cluster environment, or some mix of the two. The former relies on such galaxies having undergone substantial pre-processing in groups \citep[e.g][]{kodama01,mcgee09} prior to being accreted into the cluster while the latter can likely be put down to their deeper potential wells being better able to retain gas for star formation. Irrespective of the precise scenario, our results indicate that the stellar populations of bright galaxies are less influenced by entry into the cluster environment than those of dwarf galaxies. Furthermore, this picture fits well with that discussed by \cite{gavazzi10}. They studied galaxies in the Coma supercluster and report that the colours and luminosity distributions of high luminosity galaxies display little environmental dependence relative to those of low luminosity galaxies. Thus they propose a scenario where the former were in place at earlier times while the latter are currently, cosmologically speaking, undergoing environmentally driven transformations as they enter denser structures.

We have also compared our findings to those from S06 who studied quiescent galaxies in a host of nearby clusters across a comparable $\sigma$ range to this work. Using a similar planar analysis technique to that used here but employing different SSP models, they found average cluster-centric radial gradients that are highly consistent with those reported here.

In terms of other recent more general studies, \cite{bernardi06} and \cite{clemens09} both used large samples of early-type galaxies selected from the SDSS to test the environmental dependence of their stellar populations. Using differing techniques, they split galaxies into high and low local galaxy density subsamples. Here we note that the outskirts of Coma are unlikely to be comparable to low density or field regions due to the pre-processing of in-falling galaxies. Nevertheless, both studies report that early-types in low density regions are systematically younger than those in high density environments, in agreement with our results. By contrast, \cite{clemens09} find marginal evidence that [Z/H] is lower in dense regions but that [$\alpha$/Fe] does not depend on environment, while \cite{bernardi06} identify no [Z/H] correlation but conclude that $\alpha$-enhancement is greater in high density regions. Thus it would appear the latter study fits better with our findings.

Finally, we are able to compare our results with the predictions from the semi-analytic galaxy formation models of \cite{delucia06}. Their Fig. 8a indicates how galaxy median luminosity-weighted age varies, on average, as a function of physical distance out from cluster centres. While direct comparison is complicated by deprojection effects, mass segregation and conversion from luminosity-weighted to SSP-equivalent age \citep{trager09}, this figure predicts a typical change of $\sim$ 13\% between 0.1R$_{200}$ and R$_{200}$. This trend is consistent with our finding of 19\% $\pm$ 8\% within the same radial range in Coma. In addition, our results qualitatively agree with those from the N-body simulations of \cite{gao04}. They find that subhaloes located near the centre of cluster-sized parent haloes were accreted earlier. Thus a combination of this correlation with any mechanism that acts to inhibit star formation in galaxies as they enter the cluster environment will lead to a cluster-centric radial gradient in their stellar populations.

\section{Conclusions}

In this paper we have studied the stellar populations of bright Coma cluster galaxies using SDSS archival spectroscopy and up-to-date SSP models. The key findings of this work can be summarised as follows:

\begin{enumerate}
\item Of the 110 galaxies (31\% of the entire sample) with detectable emission we find 46\% show AGN/LINER like emission, 29\% have ongoing star formation and 25\% have an ambiguous emission driver based on this data. The fraction of AGN/LINER systems is seen to increase with increasing galaxy luminosity while the star forming fraction decreases.

\item For the 246 passive galaxies that meet our selection criteria we find strong correlations between absorption line index strength and velocity dispersion for CN2, C4668, Mgb and H$\beta$. Particularly in the case of H$\beta$, scatter about these relations is seen to increase with decreasing velocity dispersion. At low $\sigma$ the emission line galaxies typically have stronger H$\beta$ and weaker metal lines than quiescent galaxies, while at high $\sigma$ both subsets have comparable index strengths. This trend is driven by the aforementioned transition from star forming to AGN/LINER galaxies with increasing luminosity, and therefore $\sigma$.

\item Using a planar analysis technique that factors out index correlations with $\sigma$, we find significant cluster-centric radial gradients in H$\beta$, Mgb and C4668 for passive galaxies. Comparing our results with those of other recent studies that typically show stronger radial index gradients, we have shown that this difference can be attributed to the luminosity range covered by each study and not the uniqueness of the NGC 4839 in-fall region.

\item We have demonstrated that the vast majority of our passive sample do not play host to a frosting of young ($\lesssim$ 1 Gyr) stars on top of an underlying, mass dominant old population.

\item For 222 passive galaxies in Coma we find a strong positive correlation between their SSP-equivalent age and velocity dispersion. The scatter about this relation increases toward lower $\sigma$. This trend is found to be robust against variations in sample selection criteria (early-type vs quiescent), emission line detection thresholds, index velocity broadening corrections and the particular SSP model employed. Weaker positive correlations are recovered between $\sigma$ and [Fe/H], [Mg/Fe], [C/Fe] and [N/Fe]. Thus higher $\sigma$ quiescent galaxies in Coma are found to have formed their stars earlier, more rapidly and with higher [Fe/H], [C/Fe] and [N/Fe] than cluster members with lower velocity dispersions.

\item Galaxies which show AGN/LINER emission are typically found to have younger SSP-equivalent ages than quiescent galaxies at fixed velocity dispersion. 

\item Significant cluster-centric radial stellar population gradients are found for our passive sample in terms of their SSP-equivalent age, [Mg/Fe], [C/Fe] and [N/Fe]. These trends are in the sense that, at fixed $\sigma$, galaxies on the cluster's outskirts are younger, formed stars over a longer period of time and have higher [C/Fe] but lower [N/Fe]. Again we attribute fainter luminosity coverage rather than the uniqueness of the NGC 4839 group as the driver for the stronger radial gradients found by other recent studies. We find that the morphology-density relation does not drive the recovered age-radius relation.

\item Interpreting galaxies with barely detectable emission as weak AGN/LINER systems, galaxies with this emission driver are found to have a significant cluster-centric radial gradient in their SSP-equivalent ages. This trend is more than a factor of two stronger than that seen for quiescent galaxies.

\item Finally, for our passive sample we have found an age-metallicity anti-correlation that is consistent with previous works. We have shown that this anti-correlation is resolved, and so is not the result of correlated errors between age and metalicity, and that correlated errors do not significantly bias the measured slope.

\end{enumerate}

We have demonstrated that the stellar populations of bright quiescent galaxies in Coma depend strongly on their $\sigma$ and, relative to dwarf cluster members in the South-West, less so on their local environment. We have also shown that the stellar populations of such galaxies in the NGC 4839 in-fall region are not unique with respect to galaxies elsewhere in the outskirts of the cluster. In a forthcoming paper, we will explore these two related issues using an enlarged sample of faint galaxies. Firstly, to determine whether the strong environmental trends seen for dwarf galaxies in the South-West are localised there or a more general effect in the outskirts of the cluster. Secondly, to identify the transition $\sigma$, or range of velocity dispersions, at which the stellar populations of a galaxy start to become significantly more affected by its local environment.

\section*{Acknowledgements}
JP acknowledges support from the UK Science and Technology Facilities Council. SP and AH acknowledge the support of a grant from The Leverhulme Trust to the University of Bristol. RJS was supported for this work by STFC Rolling Grant PP/C501568/1 ``Extragalactic Astronomy and Cosmology at Durham 2008--2013''.

We thank Mark Taylor for writing {\it Tool for Operations on Catalogues And Tables} (http://www.star.bris.ac.uk/$\sim$mbt/topcat/) and providing support on its use.

Funding for the SDSS and SDSS-II has been provided by the Alfred P. Sloan Foundation, the Participating Institutions, the National Science Foundation, the U.S. Department of Energy, the National Aeronautics and Space Administration, the Japanese Monbukagakusho, the Max Planck Society, and the Higher Education Funding Council for England. The SDSS Web Site is http://www.sdss.org/.

The SDSS is managed by the Astrophysical Research Consortium for the Participating Institutions. The Participating Institutions are the American Museum of Natural History, Astrophysical Institute Potsdam, University of Basel, University of Cambridge, Case Western Reserve University, University of Chicago, Drexel University, Fermilab, the Institute for Advanced Study, the Japan Participation Group, Johns Hopkins University, the Joint Institute for Nuclear Astrophysics, the Kavli Institute for Particle Astrophysics and Cosmology, the Korean Scientist Group, the Chinese Academy of Sciences (LAMOST), Los Alamos National Laboratory, the Max-Planck-Institute for Astronomy (MPIA), the Max-Planck-Institute for Astrophysics (MPA), New Mexico State University, Ohio State University, University of Pittsburgh, University of Portsmouth, Princeton University, the United States Naval Observatory, and the University of Washington.


\begin{thebibliography}{}

\bibitem[\protect\citeauthoryear{{Aaronson}, {Persson} \& {Frogel}}{{Aaronson}
  et~al.}{1981}]{aaronson81}
{Aaronson} M.,  {Persson} S.~E.,    {Frogel} J.~A.,  1981, \apj, 245, 18

\bibitem[Abazajian et al.(2009)]{abazajian09} Abazajian K.~N. et 
al.\ 2009, \apjs, 182, 543 

\bibitem[Allanson et al.(2009)]{allanson09} Allanson S.~P., 
Hudson, M.~J., Smith, R.~J., \& Lucey, J.~R.\ 2009, \apj, 702, 1275

\bibitem[\protect\citeauthoryear{{Baldwin}, {Phillips} \&
  {Terlevich}}{{Baldwin} et~al.}{1981}]{bpt81}
{Baldwin} J.~A.,  {Phillips} M.~M.,    {Terlevich} R.,  1981, \pasp, 93, 5

\bibitem[\protect\citeauthoryear{{Bender}, {Burstein} \& {Faber}}{{Bender}
  et~al.}{1992}]{bender92}
{Bender} R.,  {Burstein} D.,    {Faber} S.~M.,  1992, \apj, 399, 462

\bibitem[Bernardi et al.(2005)]{bernardi05} Bernardi, M., Sheth, 
R.~K., Nichol, R.~C., Schneider, D.~P., 
\& Brinkmann, J.\ 2005, \aj, 129, 61 

\bibitem[\protect\citeauthoryear{{Bernardi}, {Nichol}, {Sheth}, {Miller} \&
  {Brinkmann}}{{Bernardi} et~al.}{2006}]{bernardi06}
{Bernardi} M.,  {Nichol} R.~C.,  {Sheth} R.~K.,  {Miller} C.~J.,    {Brinkmann}
  J.,  2006, \aj, 131, 1288

\bibitem[\protect\citeauthoryear{{Bower}, {Lucey} \& {Ellis}}{{Bower}
  et~al.}{1992}]{bower92}
{Bower} R.~G.,  {Lucey} J.~R.,    {Ellis} R.~S.,  1992, \mnras, 254, 601

\bibitem[\protect\citeauthoryear{{Briel}, {Henry} \& {Boehringer}}{{Briel}
  et~al.}{1992}]{briel92}
{Briel} U.~G.,  {Henry} J.~P.,    {Boehringer} H.,  1992, \aap, 259, L31

\bibitem[\protect\citeauthoryear{{Burstein}, {Faber}, {Gaskell} \&
  {Krumm}}{{Burstein} et~al.}{1984}]{burstein84}
{Burstein} D.,  {Faber} S.~M.,  {Gaskell} C.~M.,    {Krumm} N.,  1984, \apj,
  287, 586

\bibitem[\protect\citeauthoryear{{Caldwell}, {Rose}, {Sharples}, {Ellis} \&
  {Bower}}{{Caldwell} et~al.}{1993}]{caldwell93}
{Caldwell} N.,  {Rose} J.~A.,  {Sharples} R.~M.,  {Ellis} R.~S.,    {Bower}
  R.~G.,  1993, \aj, 106, 473
  
\bibitem[Caldwell 
\& Rose(1997)]{caldwell97} Caldwell N., \& Rose, J.~A.\ 1997, \aj, 113, 492
  
\bibitem[Caldwell et al.(2003)]{caldwell03} Caldwell N., Rose, 
J.~A., \& Concannon, K.~D.\ 2003, \aj, 125, 2891 


\bibitem[\protect\citeauthoryear{{Cappellari} \& {Emsellem}}{{Cappellari} \&
  {Emsellem}}{2004}]{cap&em04}
{Cappellari} M.,  {Emsellem} E.,  2004, \pasp, 116, 138

\bibitem[\protect\citeauthoryear{{Cardiel}, {Gorgas}, {Cenarro} \&
  {Gonzalez}}{{Cardiel} et~al.}{1998}]{cardiel98}
{Cardiel} N.,  {Gorgas} J.,  {Cenarro} J.,    {Gonzalez} J.~J.,  1998, \aaps,
  127, 597
  
\bibitem[Carter et al.(2002)]{carter02} Carter D. et al.\ 
2002, \apj, 567, 772 

\bibitem[Carter et al.(2009)]{carter09} Carter D. et al.\ 
2009, \mnras, 397, 695 

\bibitem[\protect\citeauthoryear{{Clemens}, {Bressan}, {Nikolic} \&
  {Rampazzo}}{{Clemens} et~al.}{2009}]{clemens09}
{Clemens} M.~S.,  {Bressan} A.,  {Nikolic} B.,    {Rampazzo} R.,  2009, \mnras,
  392, L35

\bibitem[\protect\citeauthoryear{{Colless} \& {Dunn}}{{Colless} \&
  {Dunn}}{1996}]{colless96}
{Colless} M.,  {Dunn} A.~M.,  1996, \apj, 458, 435

\bibitem[\protect\citeauthoryear{{Cowie}, {Songaila}, {Hu} \& {Cohen}}{{Cowie}
  et~al.}{1996}]{cowie96}
{Cowie} L.~L.,  {Songaila} A.,  {Hu} E.~M.,    {Cohen} J.~G.,  1996, \aj, 112,
  839

\bibitem[\protect\citeauthoryear{{Davies}, {Sadler} \& {Peletier}}{{Davies}
  et~al.}{1993}]{davies93}
{Davies} R.~L.,  {Sadler} E.~M.,    {Peletier} R.~F.,  1993, \mnras, 262, 650

\bibitem[\protect\citeauthoryear{{De Lucia}, {Springel}, {White}, {Croton} \&
  {Kauffmann}}{{De Lucia} et~al.}{2006}]{delucia06}
{De Lucia} G.,  {Springel} V.,  {White} S.~D.~M.,  {Croton} D.,    {Kauffmann}
  G.,  2006, \mnras, 366, 499
  
\bibitem[Dotter et al.(2007)]{dotter07} Dotter A., Chaboyer, 
B., Ferguson, J.~W., Lee, H.-c., Worthey, G., Jevremovi{\'c}, D., 
\& Baron, E.\ 2007, \apj, 666, 403 


\bibitem[\protect\citeauthoryear{{Djorgovski} \& {Davis}}{{Djorgovski} \&
  {Davis}}{1987}]{djorgovski87}
{Djorgovski} S.,  {Davis} M.,  1987, \apj, 313, 59

\bibitem[\protect\citeauthoryear{{Dressler}}{{Dressler}}{1980}]{dressler80}
{Dressler} A.,  1980, \apj, 236, 351

\bibitem[Dressler et al.(1987)]{dressler87} Dressler A., 
Lynden-Bell, D., Burstein, D., Davies, R.~L., Faber, S.~M., Terlevich, R., 
\& Wegner, G.\ 1987, \apj, 313, 42 

\bibitem[\protect\citeauthoryear{{Faber}, {Friel}, {Burstein} \&
  {Gaskell}}{{Faber} et~al.}{1985}]{faber85}
{Faber} S.~M.,  {Friel} E.~D.,  {Burstein} D.,    {Gaskell} C.~M.,  1985,
  \apjs, 57, 711

\bibitem[\protect\citeauthoryear{{Gao}, {White}, {Jenkins}, {Stoehr} \&
  {Springel}}{{Gao} et~al.}{2004}]{gao04}
{Gao} L.,  {White} S.~D.~M.,  {Jenkins} A.,  {Stoehr} F.,    {Springel} V.,
  2004, \mnras, 355, 819

\bibitem[\protect\citeauthoryear{{Gavazzi}, {Fumagalli}, {Cucciati} \&
  {Boselli}}{{Gavazzi} et~al.}{2010}]{gavazzi10}
{Gavazzi} G.,  {Fumagalli} M.,  {Cucciati} O.,    {Boselli} A.,  2010, ArXiv
  e-prints
  
\bibitem[Godwin et al.(1983)]{gmp83} Godwin J.~G., Metcalfe, 
N., \& Peach, J.~V.\ 1983, \mnras, 202, 113

\bibitem[Gonz{\'a}lez(1993)]{gonzalez93} Gonz{\'a}lez J.~J.\ 
1993, Ph.D.~Thesis, University of California

\bibitem[\protect\citeauthoryear{{Goudfrooij} \& {Emsellem}}{{Goudfrooij} \&
  {Emsellem}}{1996}]{g&e96}
{Goudfrooij} P.,  {Emsellem} E.,  1996, \aap, 306, L45

\bibitem[\protect\citeauthoryear{{Graves}, {Faber}, {Schiavon} \&
  {Yan}}{{Graves} et~al.}{2007}]{graves07}
{Graves} G.~J.,  {Faber} S.~M.,  {Schiavon} R.~P.,    {Yan} R.,  2007, \apj,
  671, 243

\bibitem[\protect\citeauthoryear{{Graves} \& {Schiavon}}{{Graves} \&
  {Schiavon}}{2008}]{graves08}
{Graves} G.~J.,  {Schiavon} R.~P.,  2008, \apjs, 177, 446

\bibitem[Graves et al.(2009)]{graves09} Graves G.~J., Faber, 
S.~M., \& Schiavon, R.~P.\ 2009, \apj, 693, 486 

\bibitem[\protect\citeauthoryear{{Guzman}, {Lucey}, {Carter} \&
  {Terlevich}}{{Guzman} et~al.}{1992}]{guzman92}
{Guzman} R.,  {Lucey} J.~R.,  {Carter} D.,    {Terlevich} R.~J.,  1992, \mnras,
  257, 187

\bibitem[\protect\citeauthoryear{{Hinshaw}, {Weiland} \& {Hill}}{{Hinshaw}
  et~al.}{2009}]{hinshaw09}
{Hinshaw} G.,  {Weiland} J.~L.,    {Hill} R.~S.,  2009, \apjs, 180, 225

\bibitem[\protect\citeauthoryear{{James}, {Salaris}, {Davies}, {Phillipps} \&
  {Cassisi}}{{James} et~al.}{2006}]{james06}
{James} P.~A.,  {Salaris} M.,  {Davies} J.~I.,  {Phillipps} S.,    {Cassisi}
  S.,  2006, \mnras, 367, 339

\bibitem[\protect\citeauthoryear{{Jones}}{{Jones}}{1999}]{Jones99}
{Jones} L.~A.,  1999, PhD thesis, University of North Carolina

\bibitem[\protect\citeauthoryear{{J{\o}rgensen}}{{J{\o}rgensen}}{1999}]{jorgensen99}
{J{\o}rgensen} I.,  1999, \mnras, 306, 607

\bibitem[\protect\citeauthoryear{{J{\o}rgensen}, {Franx} \&
  {Kjaergaard}}{{J{\o}rgensen} et~al.}{1995}]{jorgensen95}
{J{\o}rgensen} I.,  {Franx} M.,    {Kjaergaard} P.,  1995, \mnras, 276, 1341

\bibitem[Kauffmann et al.(2003)]{kauffmann03} Kauffmann G. et 
al.\ 2003, \mnras, 346, 1055

\bibitem[\protect\citeauthoryear{{Kelson}, {Illingworth}, {Franx} \& {van
  Dokkum}}{{Kelson} et~al.}{2006}]{kelson06}
{Kelson} D.~D.,  {Illingworth} G.~D.,  {Franx} M.,    {van Dokkum} P.~G.,
  2006, \apj, 653, 159

\bibitem[\protect\citeauthoryear{{Kodama} \& {Smail}}{{Kodama} \&
  {Smail}}{2001}]{kodama01}
{Kodama} T.,  {Smail} I.,  2001, \mnras, 326, 637

\bibitem[\protect\citeauthoryear{{Kormendy}}{{Kormendy}}{1977}]{kormendy77}
{Kormendy} J.,  1977, \apj, 218, 333

\bibitem[\protect\citeauthoryear{{Kuntschner}, {Lucey}, {Smith}, {Hudson} \&
  {Davies}}{{Kuntschner} et~al.}{2001}]{kuntschner01}
{Kuntschner} H.,  {Lucey} J.~R.,  {Smith} R.~J.,  {Hudson} M.~J.,    {Davies}
  R.~L.,  2001, \mnras, 323, 615
  
\bibitem[Leonardi \& Rose(1996)]{leonardi&rose96} Leonardi A.~J., \& Rose, J.~A.\ 1996, \aj, 111, 182 

\bibitem[\protect\citeauthoryear{{Mahajan}, {Haines} \&
  {Raychaudhury}}{{Mahajan} et~al.}{2010}]{mahajan10}
{Mahajan} S.,  {Haines} C.~P.,    {Raychaudhury} S.,  2010, ArXiv e-prints

\bibitem[Maraston 
\& Thomas(2000)]{maraston00} Maraston C., \& Thomas, D.\ 2000, \apj, 541, 126

\bibitem[\protect\citeauthoryear{{Matkovi{\'c}}, {Guzm{\'a}n},
  {S{\'a}nchez-Bl{\'a}zquez}, {Gorgas}, {Cardiel} \& {Gruel}}{{Matkovi{\'c}}
  et~al.}{2009}]{matkovic09}
{Matkovi{\'c}} A.,  {Guzm{\'a}n} R.,  {S{\'a}nchez-Bl{\'a}zquez} P.,  {Gorgas}
  J.,  {Cardiel} N.,    {Gruel} N.,  2009, \apj, 691, 1862
  
\bibitem[McGee et al.(2009)]{mcgee09} McGee S.~L., Balogh, 
M.~L., Bower, R.~G., Font, A.~S., 
\& McCarthy, I.~G.\ 2009, \mnras, 400, 937

\bibitem[\protect\citeauthoryear{{Mehlert}, {Saglia}, {Bender} \&
  {Wegner}}{{Mehlert} et~al.}{2000}]{mehlert00}
{Mehlert} D.,  {Saglia} R.~P.,  {Bender} R.,    {Wegner} G.,  2000, \aaps, 141,
  449

\bibitem[\protect\citeauthoryear{{Moore}}{{Moore}}{2001}]{moore01}
{Moore} S.~A.~W.,  2001, PhD thesis, University of Durham

\bibitem[\protect\citeauthoryear{{Moore}, {Lucey}, {Kuntschner} \&
  {Colless}}{{Moore} et~al.}{2002}]{moore02}
{Moore} S.~A.~W.,  {Lucey} J.~R.,  {Kuntschner} H.,    {Colless} M.,  2002,
  \mnras, 336, 382

\bibitem[\protect\citeauthoryear{{Nelan}, {Smith}, {Hudson}, {Wegner}, {Lucey},
  {Moore}, {Quinney} \& {Suntzeff}}{{Nelan} et~al.}{2005}]{nelan05}
{Nelan} J.~E.,  {Smith} R.~J.,  {Hudson} M.~J.,  {Wegner} G.~A.,  {Lucey}
  J.~R.,  {Moore} S.~A.~W.,  {Quinney} S.~J.,    {Suntzeff} N.~B.,  2005, \apj,
  632, 137

\bibitem[Osterbrock \& Ferland(2006)]{osterbrock05} Osterbrock D.~E., \& Ferland, G.~J.\ 2006, Astrophysics of Gaseous Nebulae and Active Galactic Nuclei, Second Edition, CA: University Science Books  

\bibitem[\protect\citeauthoryear{{Peletier}}{{Peletier}}{1989}]{peletier89}
{Peletier} R.~F.,  1989, PhD thesis, University of Groningen

\bibitem[\protect\citeauthoryear{{Peng}, {Ho}, {Impey} \& {Rix}}{{Peng}
  et~al.}{2002}]{peng02}
{Peng} C.~Y.,  {Ho} L.~C.,  {Impey} C.~D.,    {Rix} H.-W.,  2002, \aj, 124, 266

\bibitem[P{\'e}rez-Gonz{\'a}lez et al.(2008)]{perezgonzalez08} 
P{\'e}rez-Gonz{\'a}lez P.~G. et al.\ 2008, \apj, 675, 234 

\bibitem[\protect\citeauthoryear{{Poggianti}, {Bridges}, {Komiyama}, {Yagi},
  {Carter}, {Mobasher}, {Okamura} \& {Kashikawa}}{{Poggianti}
  et~al.}{2004}]{poggianti04}
{Poggianti} B.~M.,  {Bridges} T.~J.,  {Komiyama} Y.,  {Yagi} M.,  {Carter} D.,
  {Mobasher} B.,  {Okamura} S.,    {Kashikawa} N.,  2004, \apj, 601, 197
  
\bibitem[Poggianti et al.(2001)]{poggianti01} Poggianti B.~M. et 
al.\ 2001, \apj, 562, 689

\bibitem[Price et al.(2009)]{price09} Price J. et al.\ 2009, 
\mnras, 397, 1816 

\bibitem[\protect\citeauthoryear{{Proctor} \& {Sansom}}{{Proctor} \&
  {Sansom}}{2002}]{proctor02}
{Proctor} R.~N.,  {Sansom} A.~E.,  2002, \mnras, 333, 517

\bibitem[\protect\citeauthoryear{{Rawle}, {Smith}, {Lucey} \&
  {Swinbank}}{{Rawle} et~al.}{2008}]{rawle08}
{Rawle} T.~D.,  {Smith} R.~J.,  {Lucey} J.~R.,    {Swinbank} A.~M.,  2008,
  \mnras, 389, 1891

\bibitem[Rose(1984)]{rose84} Rose J.~A.\ 1984, \aj, 89, 1238 

\bibitem[\protect\citeauthoryear{{Rose}}{{Rose}}{1985}]{rose85}
{Rose} J.~A.,  1985, \aj, 90, 1927

\bibitem[\protect\citeauthoryear{{S{\'a}nchez-Bl{\'a}zquez} P.and~{Gorgas},
  {Cardiel} \& {Gonz{\'a}lez}}{{S{\'a}nchez-Bl{\'a}zquez}
  et~al.}{2006}]{sanchezblaz06b}
{S{\'a}nchez-Bl{\'a}zquez} P.and~{Gorgas} J.,  {Cardiel} N.,    {Gonz{\'a}lez}
  J.~J.,  2006, \aap, 457, 809
  
\bibitem[Sarzi et al.(2006)]{sarzi06} Sarzi M. et al.\ 2006, 
\mnras, 366, 1151 

\bibitem[\protect\citeauthoryear{{Schiavon}}{{Schiavon}}{2007}]{schiavon07}
{Schiavon} R.~P.,  2007, \apjs, 171, 146

\bibitem[\protect\citeauthoryear{{Serra} \& {Trager}}{{Serra} \&
  {Trager}}{2007}]{serra07}
{Serra} P.,  {Trager} S.~C.,  2007, \mnras, 374, 769

\bibitem[\protect\citeauthoryear{{Smith}, {Hudson}, {Lucey}, {Nelan} \&
  {Wegner}}{{Smith} et~al.}{2006}]{smith06}
{Smith} R.~J.,  {Hudson} M.~J.,  {Lucey} J.~R.,  {Nelan} J.~E.,    {Wegner}
  G.~A.,  2006, \mnras, 369, 1419

\bibitem[\protect\citeauthoryear{{Smith}, {Lucey} \& {Hudson}}{{Smith}
  et~al.}{2007}]{smith07}
{Smith} R.~J.,  {Lucey} J.~R.,    {Hudson} M.~J.,  2007, \mnras, 381, 1035

\bibitem[\protect\citeauthoryear{{Smith}, {Lucey}, {Hudson}, {Allanson},
  {Bridges}, {Hornschemeier}, {Marzke} \& {Miller}}{{Smith}
  et~al.}{2009a}]{smith09a}
{Smith} R.~J.,  {Lucey} J.~R.,  {Hudson} M.~J.,  {Allanson} S.~P.,  {Bridges}
  T.~J.,  {Hornschemeier} A.~E.,  {Marzke} R.~O.,    {Miller} N.~A.,  2009a,
  \mnras, 392, 1265

\bibitem[Smith et al.(2009b)]{smith09b} Smith R.~J., Lucey, 
J.~R., \& Hudson, M.~J.\ 2009b, \mnras, 400, 1690
   
\bibitem[Smith et al.(2008)]{smith08} Smith R.~J. et al.\ 
2008, \mnras, 386, L96 

\bibitem[\protect\citeauthoryear{Strateva et al.}{2001}]{strateva01}
{Strateva} I. et al.,  2001, \aj, 122, 1861

\bibitem[\protect\citeauthoryear{Strauss et al.}{2002}]{strauss02}
{Strauss} M.~A. et al., 2002, \aj, 124, 1810

\bibitem[\protect\citeauthoryear{{Thomas}, {Maraston} \& {Bender}}{{Thomas}
  et~al.}{2003}]{thomas03}
{Thomas} D.,  {Maraston} C.,    {Bender} R.,  2003, \mnras, 339, 897

\bibitem[\protect\citeauthoryear{{Thomas}, {Maraston}, {Bender} \& {Mendes de
  Oliveira}}{{Thomas} et~al.}{2005}]{thomas05}
{Thomas} D.,  {Maraston} C.,  {Bender} R.,    {Mendes de Oliveira} C.,  2005,
  \apj, 621, 673

\bibitem[\protect\citeauthoryear{{Thomas}, {Maraston} \& {Korn}}{{Thomas}
  et~al.}{2004}]{thomas04}
{Thomas} D.,  {Maraston} C.,    {Korn} A.,  2004, \mnras, 351, L19

\bibitem[\protect\citeauthoryear{{Thomas}, {Maraston}, {Schawinski}, {Sarzi} \&
  {Silk}}{{Thomas} et~al.}{2009}]{thomas09}
{Thomas} D.,  {Maraston} C.,  {Schawinski} K.,  {Sarzi} M.,    {Silk} J.,
  2009, ArXiv e-prints

\bibitem[\protect\citeauthoryear{{Trager}}{{Trager}}{1997}]{trager97}
{Trager} S.~C.,  1997, PhD thesis, University of California

\bibitem[Trager 
\& Somerville(2009)]{trager09} Trager S.~C., \& Somerville, R.~S.\ 2009, \mnras, 395, 608 

\bibitem[\protect\citeauthoryear{{Trager}, {Faber} \& {Dressler}}{{Trager}
  et~al.}{2008}]{trager08}
{Trager} S.~C.,  {Faber} S.~M.,    {Dressler} A.,  2008, \mnras, 386, 715

\bibitem[\protect\citeauthoryear{{Trager}, {Faber}, {Worthey} \&
  {Gonz{\'a}lez}}{{Trager} et~al.}{2000b}]{trager00b}
{Trager} S.~C.,  {Faber} S.~M.,  {Worthey} G.,    {Gonz{\'a}lez} J.~J.,  2000a,
  \aj, 120, 165

\bibitem[\protect\citeauthoryear{{Trager}, {Faber}, {Worthey} \&
  {Gonz{\'a}lez}}{{Trager} et~al.}{2000a}]{trager00a}
{Trager} S.~C.,  {Faber} S.~M.,  {Worthey} G.,    {Gonz{\'a}lez} J.~J.,  2000b,
  \aj, 119, 1645

\bibitem[\protect\citeauthoryear{{Trager}, {Worthey}, {Faber}, {Burstein} \&
  {Gonzalez}}{{Trager} et~al.}{1998}]{trager98}
{Trager} S.~C.,  {Worthey} G.,  {Faber} S.~M.,  {Burstein} D.,    {Gonzalez}
  J.~J.,  1998, \apjs, 116, 1

\bibitem[\protect\citeauthoryear{{Valdes}, {Gupta}, {Rose}, {Singh} \&
  {Bell}}{{Valdes} et~al.}{2004}]{valdes04}
{Valdes} F.,  {Gupta} R.,  {Rose} J.~A.,  {Singh} H.~P.,    {Bell} D.~J.,
  2004, \apjs, 152, 251

\bibitem[\protect\citeauthoryear{{Vazdekis}, {Casuso}, {Peletier} \&
  {Beckman}}{{Vazdekis} et~al.}{1996}]{vazdekis96}
{Vazdekis} A.,  {Casuso} E.,  {Peletier} R.~F.,    {Beckman} J.~E.,  1996,
  \apjs, 106, 307
  
\bibitem[Vazdekis et al.(2010)]{vazdekis10} Vazdekis A., 
S{\'a}nchez-Bl{\'a}zquez, P., Falc{\'o}n-Barroso, J., Cenarro, A.~J., 
Beasley, M.~A., Cardiel, N., Gorgas, J., 
\& Peletier, R.~F.\ 2010, arXiv:1004.4439 

\bibitem[\protect\citeauthoryear{{Worthey}}{{Worthey}}{1994}]{worthey94}
{Worthey} G.,  1994, \apjs, 95, 107

\bibitem[\protect\citeauthoryear{{Worthey}, {Faber} \& {Gonzalez}}{{Worthey}
  et~al.}{1992}]{worthey92}
{Worthey} G.,  {Faber} S.~M.,    {Gonzalez} J.~J.,  1992, \apj, 398, 69

\bibitem[\protect\citeauthoryear{{Worthey} \& {Ottaviani}}{{Worthey} \&
  {Ottaviani}}{1997}]{worthey&otta97}
{Worthey} G.,  {Ottaviani} D.~L.,  1997, \apjs, 111, 377

\bibitem[York et al.(2000)]{york00} York D.~G. et al.\ 2000, 
\aj, 120, 1579 

\end{thebibliography}

\appendix
\section{Index-Index Comparisons}
\label{appendixa}

In this section we compare our measured index data with other studies from the literature both focusing on and involving Coma cluster galaxies. Specifically we opt for two datasets which possess a sizable number of galaxies in common with this work, namely M02 and NFPS. The former has 78 galaxies in common with our passive sample while the latter has 75, permitting a suitable robustness check of our measurements. Here we make no attempt to match spectroscopic fibre sizes between datasets, noting that M02 observed with a 2.7$^{\prime\prime}$ diameter fibre and NFPS with a 2$^{\prime\prime}$ diameter fibre. In addition we note that both studies are based on flux calibrated spectra without explicit correction to the Lick/IDS system, in a similar vein to our data.

\begin{figure*}
\centering
\scalebox{0.44}[0.44]{\includegraphics{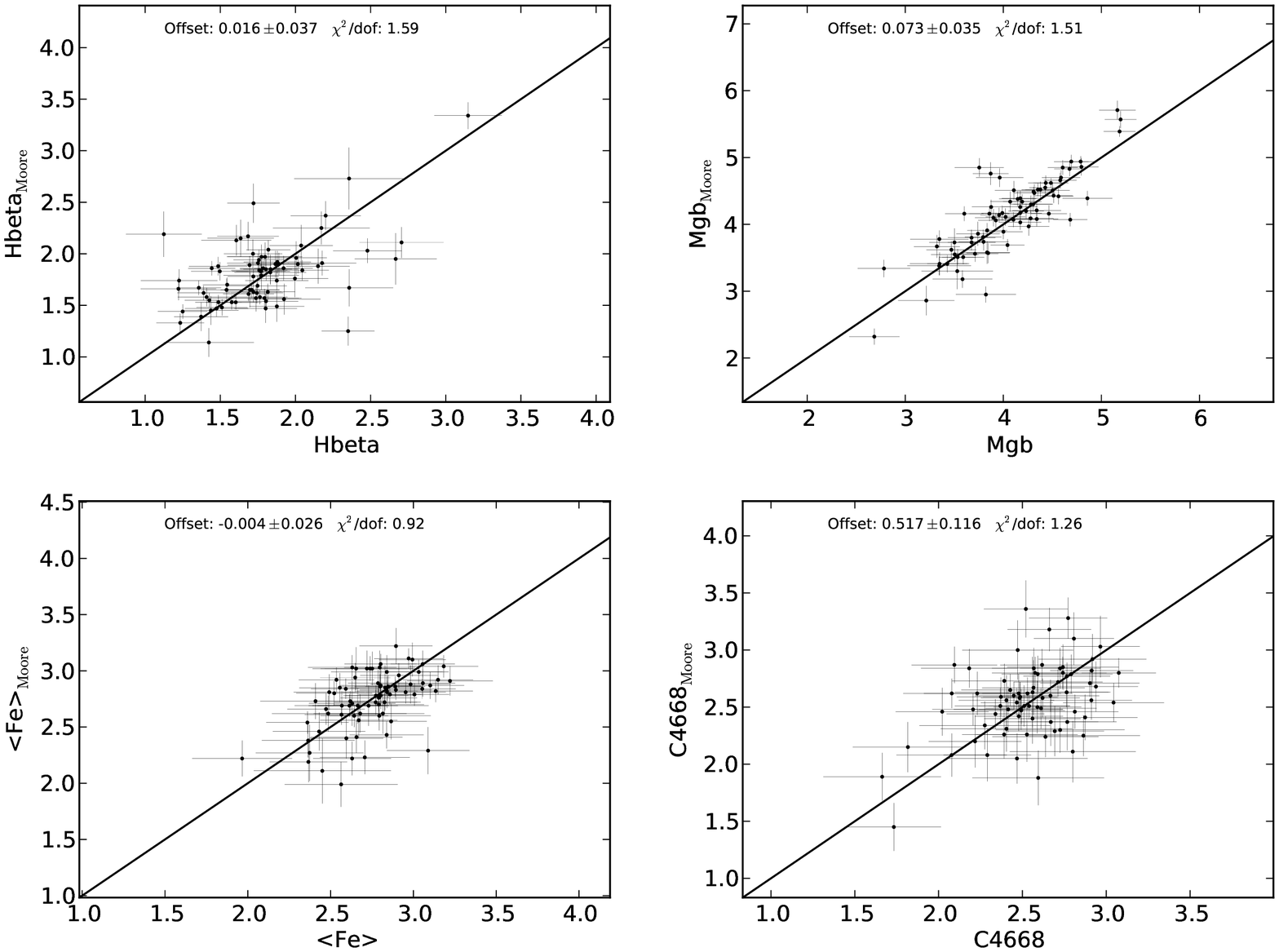}} 
\caption{Index-index comparisons between this work and data from \citet{moore02}. No aperture corrections are applied and the data is compared at the Lick resolution. The mean offset is computed as $<$Literature - This work$>$. The quoted reduced $\chi^{2}$ is about the equality line displayed in each panel unless the offset has a significance $\geq$ 3$\sigma$, in which case the model takes into account the mean offset. For clarity we restrict the axis range of the Hbeta and Mgb panels, excluding one k+a galaxy in common between the two samples. Note this galaxy is included in the offset and $\chi^{2}$ calculations.}
\label{m02}
\end{figure*}

\begin{figure*}
\centering
\scalebox{0.53}[0.53]{\includegraphics{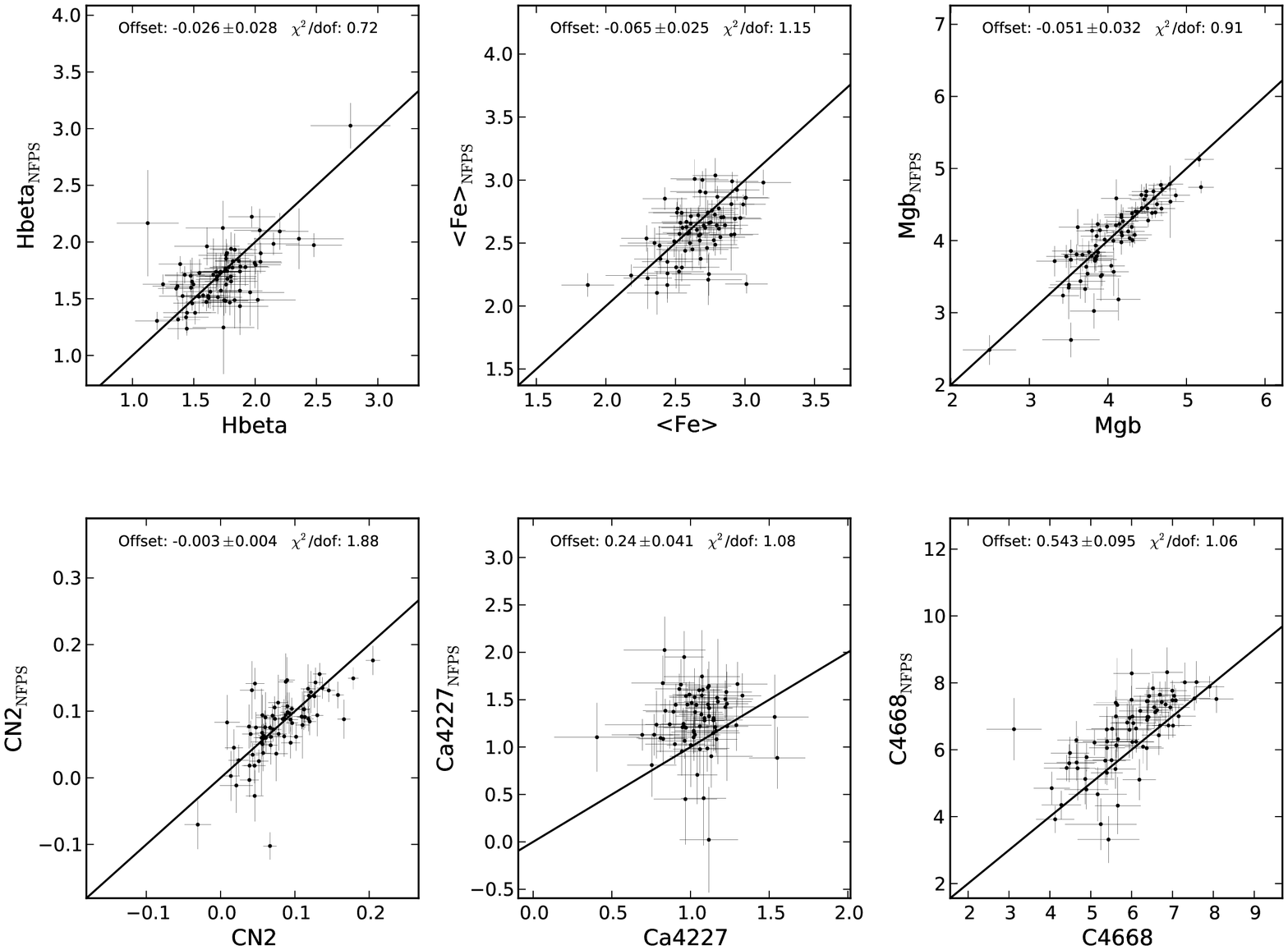}} 
\caption{Index-index comparisons between this work and data from \citet{nelan05}. No aperture corrections are applied and the data is compared at the Lick resolution. The mean offset is computed as $<$Literature - This work$>$. The quoted reduced $\chi^{2}$ is about the equality line displayed in each panel unless the offset has a significance $\geq$ 3$\sigma$, in which case the model takes into account the mean offset.}
\label{NFPS}
\end{figure*}

In Fig. \ref{m02} and Fig. \ref{NFPS} we present diagrams comparing the six indices used in our analysis with those from the literature studies. For the M02 sample not all indices are available due to the wavelength coverage of that work's spectra. Mean index offsets are computed as $<$Literature - This work$>$. The quoted reduced $\chi^{2}$ is computed for a model with no offset if the offsets significance is $<$ 3$\sigma$, otherwise it is about a model taking into account the mean offset. A one to one compatibility between the index data is also represented by the straight line in each panel. 

From Fig. \ref{m02} we see that only C4668 has a significant offset at the 4$\sigma$ level. H$\beta$ and Mgb have relatively high reduced $\chi^{2}$, indicating perhaps underestimated errors in this work or M02. In Fig. \ref{NFPS} significant offsets are seen for Ca4227 (6$\sigma$) and C4668 (5$\sigma$). When these systematic shifts are accounted for, however, both Ca4227 and C4668 from NFPS and this work are in agreement. High reduced $\chi^{2}$ is only obtained for CN2, again probably indicating underestimated errors. 

To summarise, generally speaking good agreement is seen between this work and the two literature overlap samples. There is, however, reasonable evidence for a systematic offset in our C4668 index measurements relative to those in both M02 and NFPS.

\begin{figure*}
\centering
\scalebox{0.55}[0.55]{\includegraphics{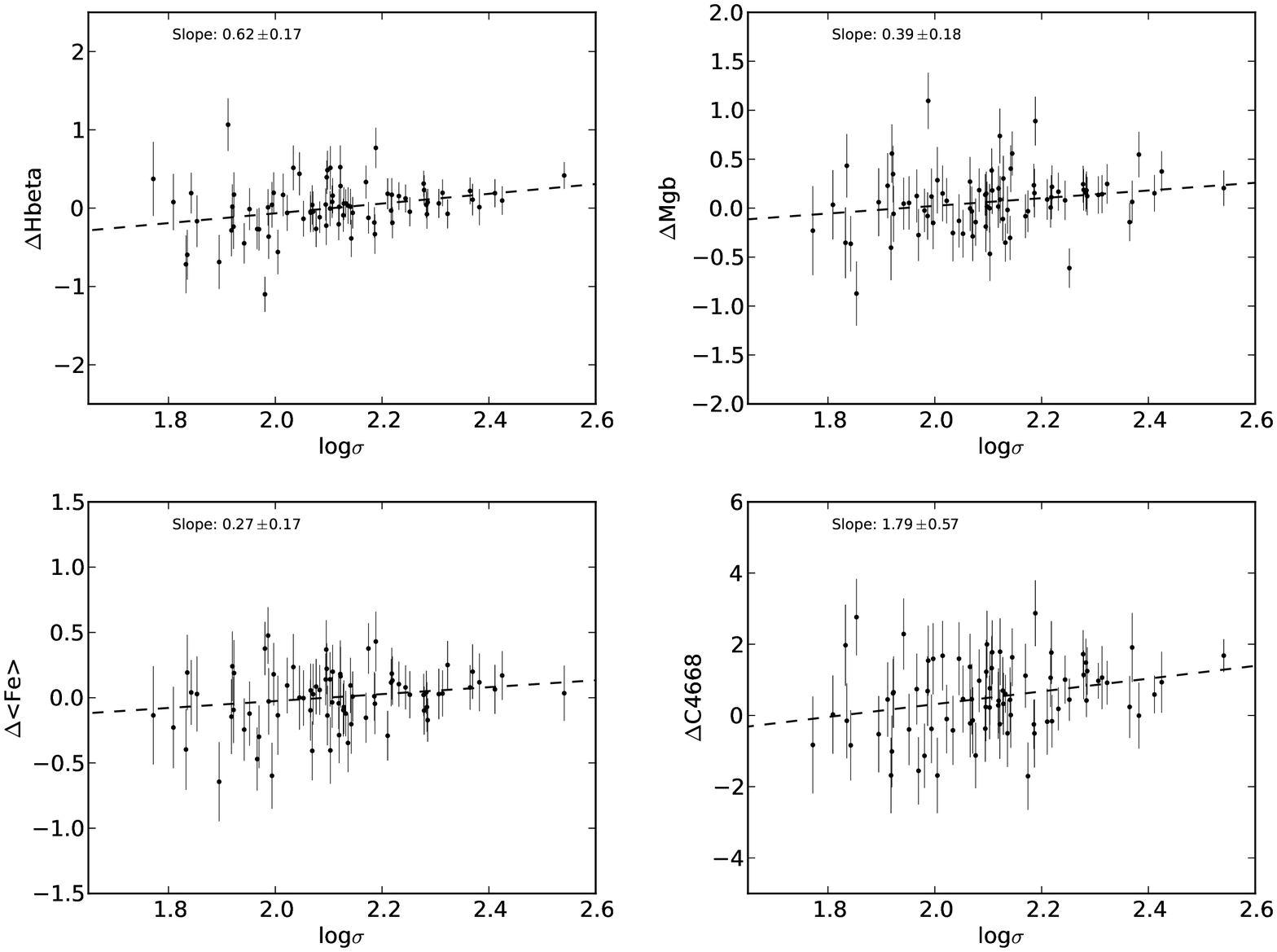}} 
\caption{$\Delta$index-$\sigma$ trends for the M02 overlap sample. $\Delta$index is computed as Literature index - This work and the velocity dispersion measurements are those from this work. }
\label{deltasigm}
\end{figure*}

\begin{figure*}
\centering
\scalebox{0.65}[0.65]{\includegraphics{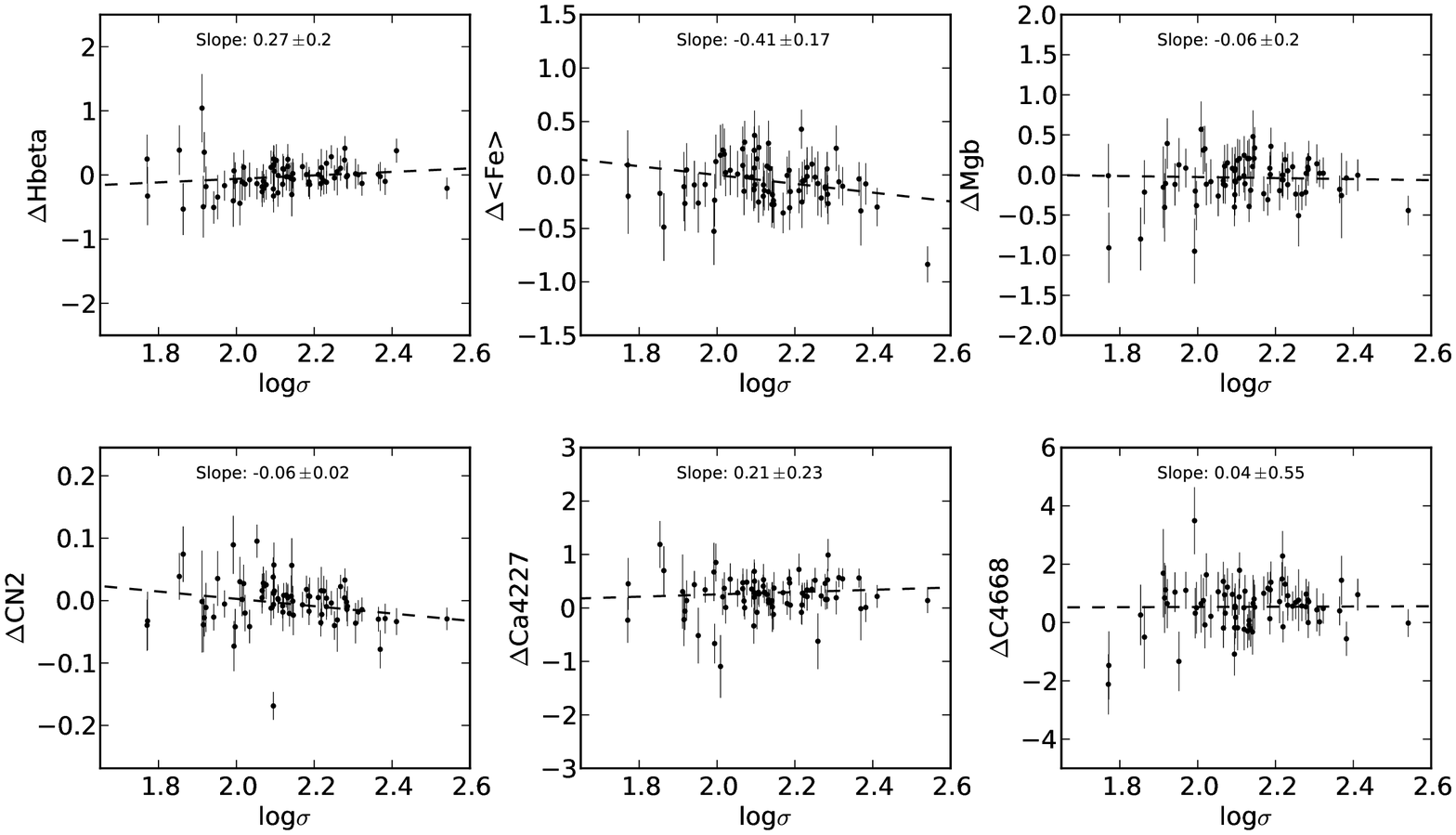}} 
\caption{$\Delta$index-$\sigma$ trends for the NFPS overlap sample. $\Delta$index is computed as Literature index - This work and the velocity dispersion measurements are those from this work.}
\label{deltasign}
\end{figure*}

Next we seek to identify what is driving the apparent lack of downsizing in the M02 data, as determined by T08, and a weaker signal of its presence in the NFPS data relative to this work. The logical place to search for the source of these differences is H$\beta$, the main SSP-equivalent age indicator in all three studies, as a function of velocity dispersion. With this in mind, in Fig. \ref{deltasigm} and Fig. \ref{deltasign} we plot $\Delta$index-$\sigma$ for both M02 and NFPS where $\Delta$index is computed as Literature index - This work and the velocity dispersion measurements are those from this work.

Fig. \ref{deltasigm} shows significant $\Delta$index-$\sigma$ correlations for H$\beta$ and C4668. Importantly this demonstrates M02 measure stronger H$\beta$ at high $\sigma$ and weaker H$\beta$ at low $\sigma$ than our SDSS data. This in turn implies, everything else being equal, they would recover younger ages for higher $\sigma$ galaxies and older ages for lower $\sigma$ galaxies than this work. Thus the age-$\sigma$ relation would become flattened, although it is unclear to what extent. By contrast, in Fig. \ref{deltasign} we see that the NFPS overlap displays no similarly significant $\Delta$H$\beta$-$\sigma$ correlation. Here we do note that, while consistent with zero, the NFPS slope is also consistent with the trend seen for the M02 sample. In this case we are able to take the comparison a step further and compare both H$\beta$-$\sigma$ and, using the Schiavon models, log(age)-$\sigma$ relations. For the former we obtain the slopes H$\beta_{nfps}$-$\sigma$ = -0.93 $\pm$ 0.09 and H$\beta_{sdss}$-$\sigma$ = -0.99 $\pm$ 0.16, highly consistent within the uncertainties. For the latter we are restricted to the 58 galaxies in common that {\sc Ez-Ages} is able to fit and recover slopes of log(age)$_{NFPS}$-$\sigma$ = 0.67 $\pm$ 0.13 and log(age)$_{sdss}$-$\sigma$ = 0.77 $\pm$ 0.14. Again these trends are consistent within the uncertainties.

In summary, we have demonstrated that NFPS and our SDSS data imply compatible downsizing trends where they overlap. However, for the M02 and SDSS overlap we find a significantly flatter H$\beta$-$\sigma$ relation from the former, which in turn may result in T08 obtaining no age-$\sigma$ correlation when employing the M02 dataset. Given the fact that we estimate velocity broadening differences introduce $\sim$ 15\% difference in H$\beta$-$\sigma$ slopes, it is unclear what drives this discrepancy.

\section{Error Model Robustness}
\label{appendixb}

\begin{table}
\centering
\caption{Relevant parameters for the galaxies used to test the error model. The first two columns give the galaxy ID, for comparison with Fig. \ref{emodgrid}, and signal-to-noise.  The final two columns show the correlation coefficient ($\rho$) between log(age) and [Fe/H] and the orientation of the resultant error ellipse ($\theta$), with respect to the log(age) axis, as computed from each galaxy's simulations. Galaxies e1-4 were used to construct the error model.}
\label{emodresult}
\begin{tabular}{cccc}
\hline
ID & S/N & $\rho$ & $\theta$ (deg) \\
\hline
1 & 33.9 & -0.71 & -44.34 \\
2 & 23.6 & -0.60 & -37.24 \\
3 & 49.3 & -0.67 & -28.13 \\
4 & 46.0 & -0.59 & -40.74 \\
5 & 38.3 & -0.65 & -33.74 \\
6 & 28.5 & -0.66 & -41.46 \\
e1 & 45.6 & -0.66 & -33.17 \\
e2 & 25.2 & -0.53 & -30.89 \\
e3 & 35.1 & -0.56 & -34.84 \\
e4 & 53.9 & -0.64 & -24.10 \\
\hline
\end{tabular}
\end{table}

\begin{figure*}
\centering
\scalebox{0.85}[0.85]{\includegraphics{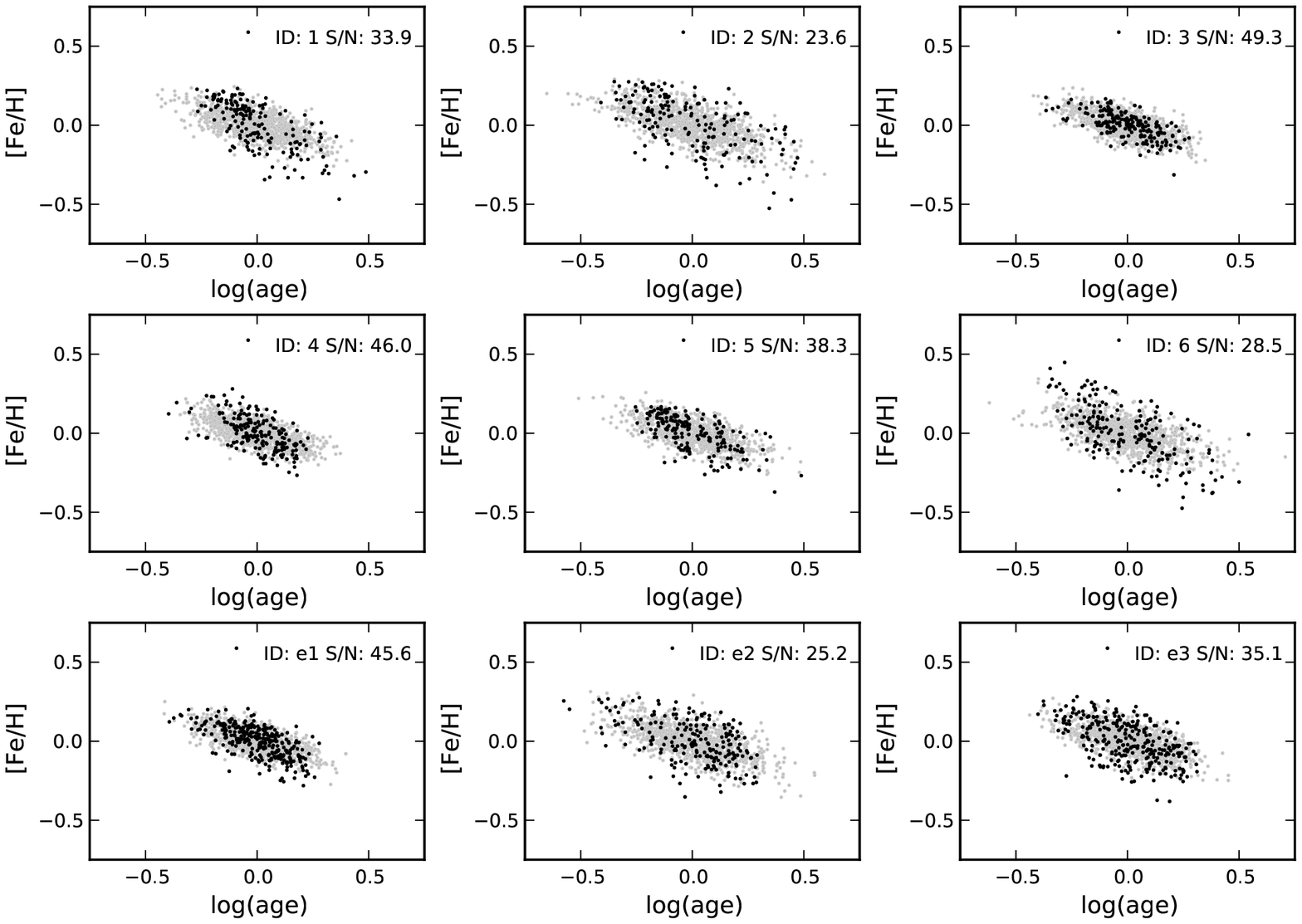}} 
\caption{Diagrams comparing the result of drawing 1000 samples from the rescaled error ellipse of e4, see Table \ref{emodresult}, (grey points) to the output of 200 realisations of each galaxy's index data which has been ran through {\sc Ez-Ages} (black points). As such this figure may be used to visually assess the suitability of the error model. The legend of each panel displays the galaxy ID from Table \ref{emodresult} and its S/N.}
\label{emodcompare}
\end{figure*}

\begin{figure}
\flushleft
\scalebox{0.45}[0.45]{\includegraphics{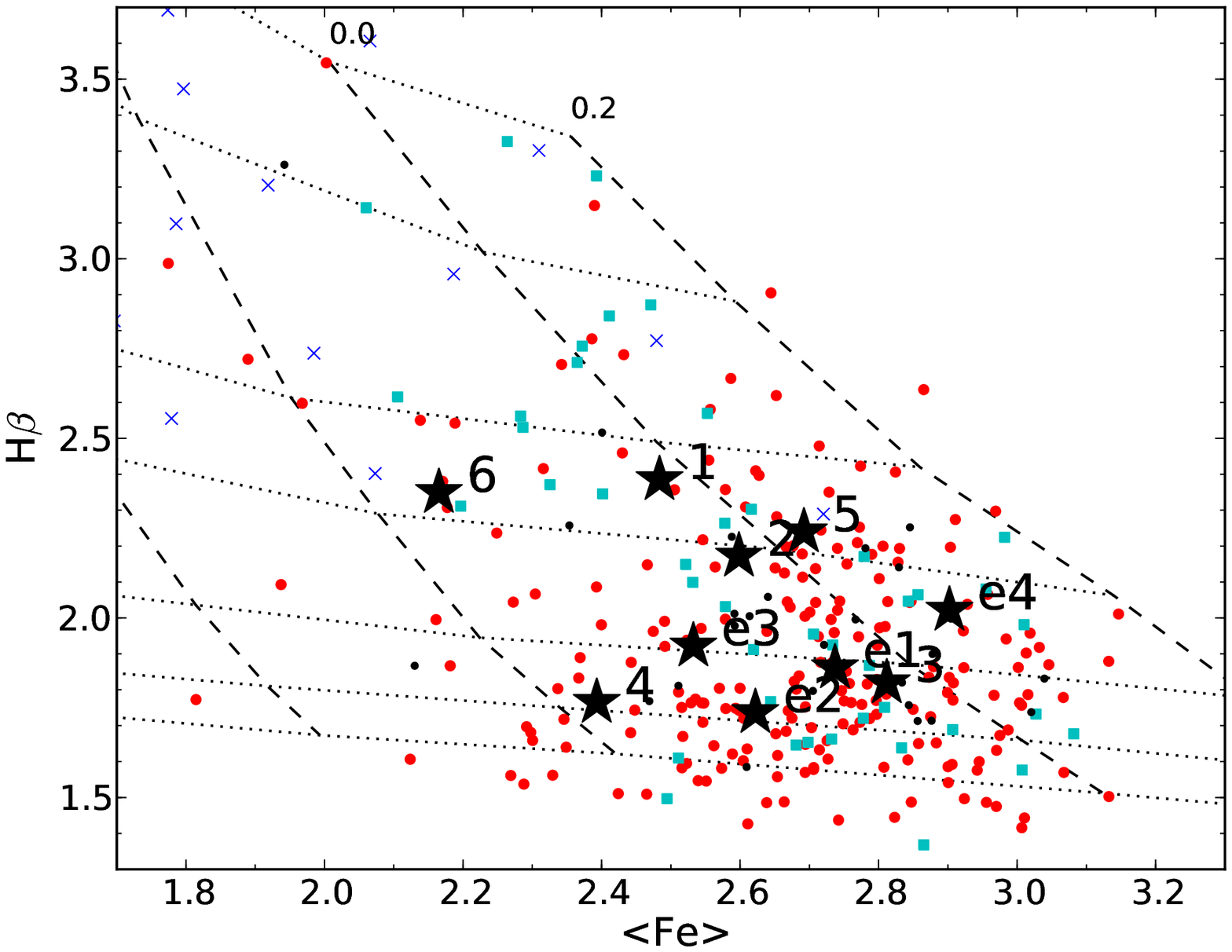}} 
\caption{H$\beta$--$<$Fe$>$ model grid with the galaxies used for the error model test highlighted by large black stars. For further details on the grid see Fig. \ref{hbfe}.}
\label{emodgrid}
\end{figure}

In this section we seek to test the robustness of the error model used to compute the stellar population parameter uncertainties both in terms of its ability to reproduce parameter error at a given S/N and suitably measure the correlated error between age and metallicity required for Section \ref{zplanesec}. To that end we select six galaxies of varying S/N and at different positions on the H$\beta$--$<$Fe$>$ grid (see Fig. \ref{emodgrid}) from our passive sample and generate 200 realisations of their index data using their respective uncertainties. Each realisation is ran through {\sc Ez-Ages} using the same approach as described for the real data. Next we use the S/N = 55 (e4, see Table \ref{emodresult}) simulation data, used to make the error model, to assess the age-metallicity covariance and construct the resultant error ellipse, which is in turn rescaled to the respective S/N of each of the six selected galaxies. Finally, we draw 1000 samples from the rescaled error ellipse and in Fig. \ref{emodcompare} compare the predicted and actual age-metallicity distributions for the six galaxies. We also include the S/N = 25, 35 and 45 galaxies used to build the error model since it is only the covariance of the S/N = 55 galaxy which is currently being tested. In addition in Table \ref{emodresult} we present the relevant parameters for the six randomly selected galaxies and four error model galaxies.

Fig. \ref{emodcompare} demonstrates that the rescaled error ellipse does a good job at reproducing the scatter from the realisations. Quantitatively the error model is confirmed to better than 10\% for these galaxies. Table \ref{emodresult} shows that $\rho$ is stable to within 10-15\% ($<\rho>$ = -0.62 $\pm$ 0.06) and $\theta$ varies by at most 30\% ($<\theta>$ = -34.42 $\pm$ 5.35$^{\circ}$). Indeed, by comparing Table \ref{emodresult} and Fig. \ref{emodgrid} we see that for the H$\beta$--$<$Fe$>$ parameter space occupied by the bulk of our sample galaxies $\theta$ is between -24$^{\circ}$ and -37$^{\circ}$. Given its associated $\rho$ and $\theta$ values, which are close to the means computed above, we choose to use the S/N = 45 (e1) as the base error ellipse for rescaling when testing the impact of correlated errors in Section \ref{zplanesec}.

\section{Radial Trends: Testing a Null Model}
\label{appendixc}

\begin{figure*}
\centering
\scalebox{0.67}[0.67]{\includegraphics{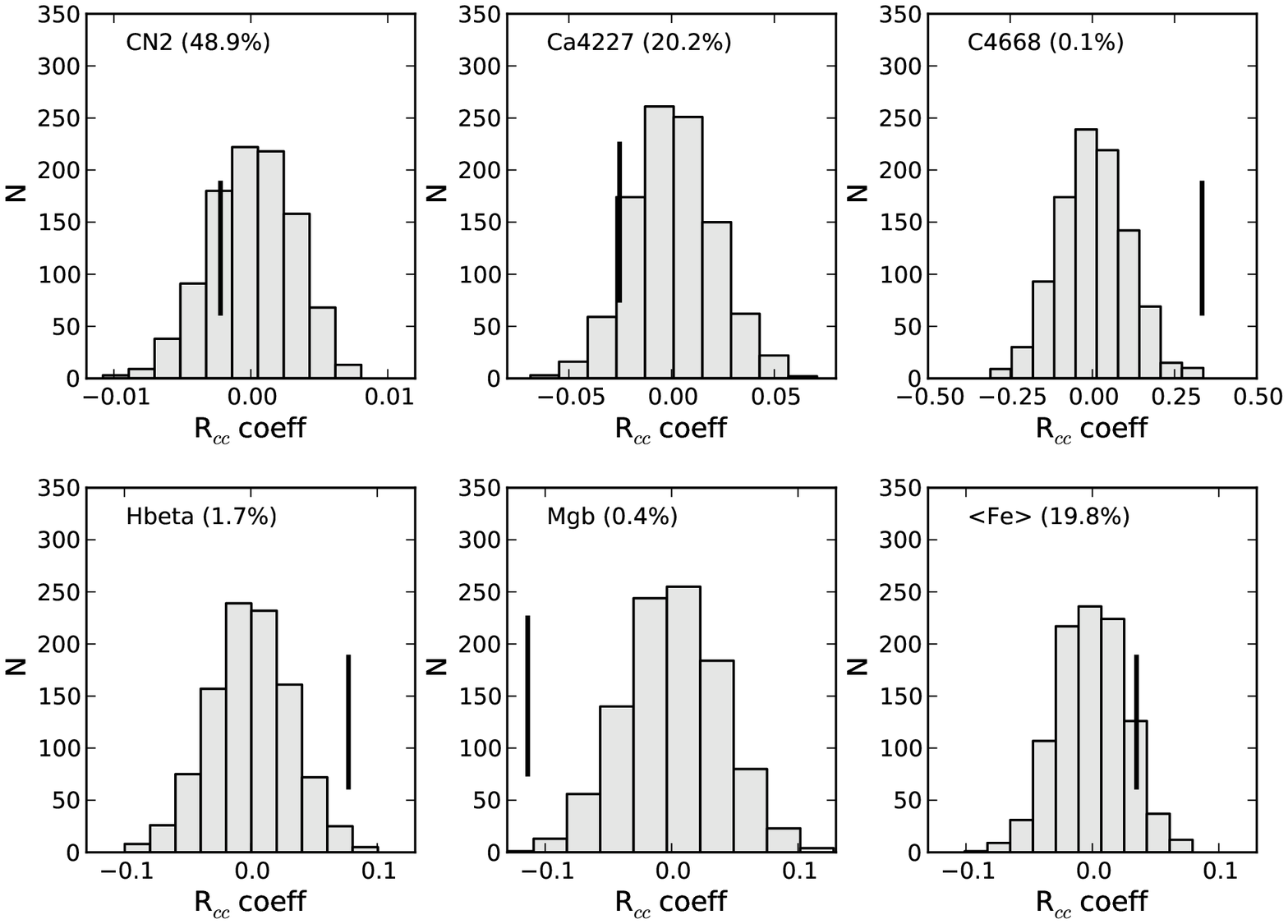}} 
\caption{Distribution of radial coefficients ($a$ in equation \ref{plane}) derived for the null model in which the indices of the passive sample are not correlated with projected cluster-centric radius. The black vertical bar denotes the value of the coefficient measured from the real data. The fraction of trials that yield an $|a|$ larger than the observed value is noted in parentheses.}
\label{hist_ind}
\end{figure*}

\begin{figure*}
\centering
\scalebox{0.67}[0.67]{\includegraphics{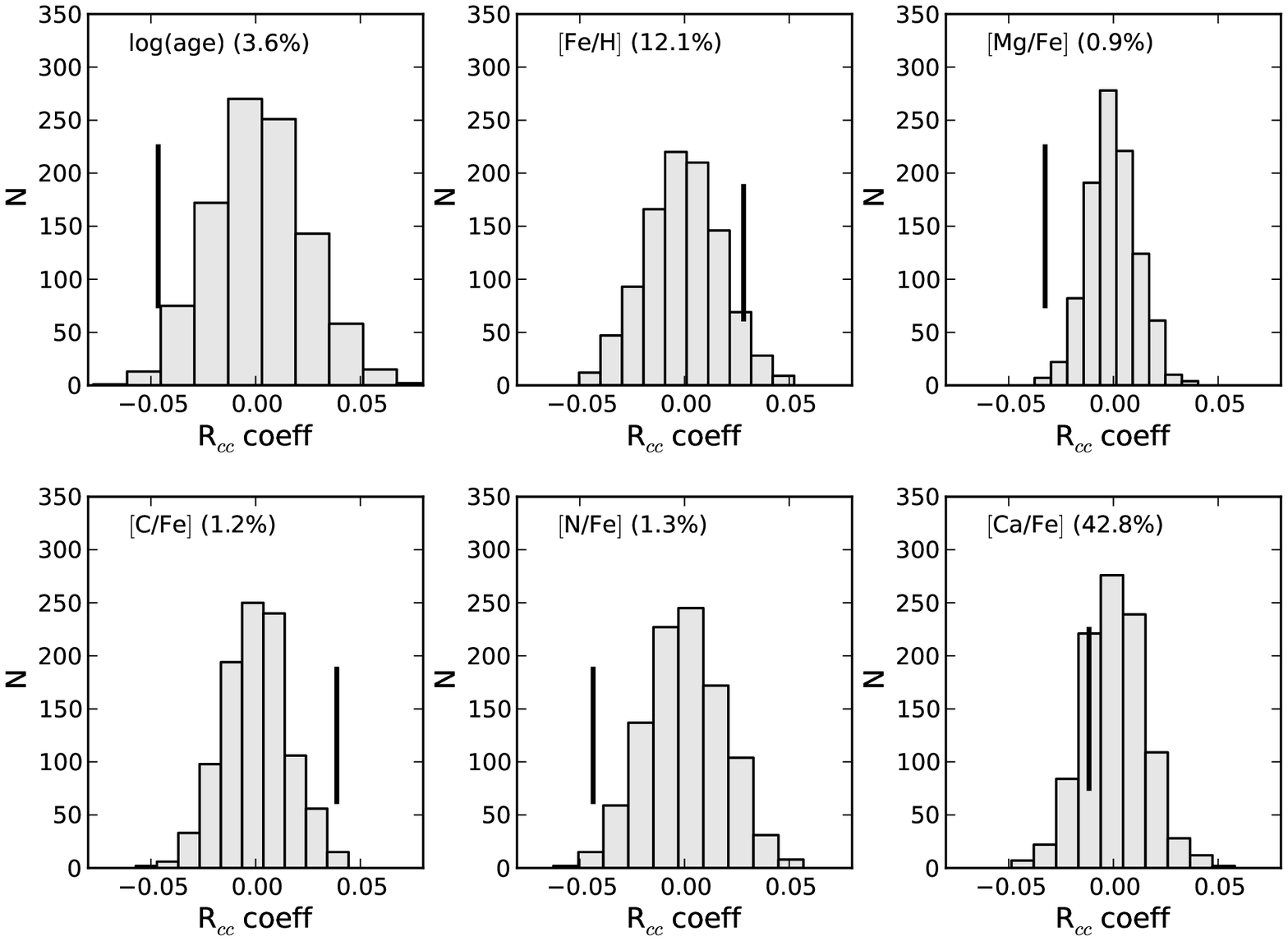}} 
\caption{Distribution of radial coefficients ($a$ in equation \ref{plane} but with $I$ set equal to each stellar population parameter in turn) derived for the null model in which the stellar population parameters of the passive sample are not correlated with projected cluster-centric radius. The black vertical bar denotes the value of the coefficient measured from the real data. The fraction of trials that yield an $|a|$ larger than the observed value is noted in parentheses.}
\label{hist_spp}
\end{figure*}

Given the significant, but relatively weak, radial trends detected for the passive sample in terms of their indices and stellar population parameters it is interesting to test the validity of these results, and indeed the fitting procedure, against a null model. As such, in this section we undertake simulations based on an artifical version of this sample, excluding the blue subset, in which there is no radial correlation. As input data for this test we use the velocity dispersions and half-light radii as measured and generate fake indices and stellar population parameters using the coefficients derived from our planar fits but with the radial component set equal to zero. Next, for each of the six indices and stellar population parameters we create 1000 realisations of the input data using the measurement errors and intrinsic scatter derived from the planar fits to the real data. Finally, each realisation is ran through the same planar fitting procedure used for the real dataset and in Fig. \ref{hist_ind} and Fig. \ref{hist_spp} we histogram the output of these simulations overlayed with the position of the coefficient obtained from the real data.

Fig. \ref{hist_ind} and Fig. \ref{hist_spp} clearly demonstrate the ability of the fitting procedure to account for intrinsic scatter and, on average, return unbiased coefficients. The standard deviations of the distributions presented in both figures are slightly larger than the errors given in Tables \ref{indradtab} and \ref{sppradtab} because here the intrinsic scatter is resampled as well as the measurement uncertainties. Nevertheless, the signifiance of the C4668, H$\beta$ and Mgb index trends and log(age), [Mg/Fe], [C/Fe] and [N/Fe] stellar population parameter trends is apparent.

\section{Data Tables}
\label{appendixd}

The index and velocity dispersion measurements used in this work are tabulated in Table \ref{data}.

\begin{table*}
\centering
\caption{Absorption-line index and velocity dispersion measurements used in this work. Galaxies are identified by their position on the sky (J2000) and, where available, their GMP number. The quoted S/N is per \AA\ and the velocity dispersion measurements are not aperture corrected. The indices and their errors are given in conventional units, i.e. magnitudes for CN1, CN2, Mg1 and Mg2, angstroms for all others. The flag column is binary with 0 indicating no detected emission with A/N = 4 and 1 denoted galaxies with detected emission. The full version of this table will be provided in the electronic version of the journal.}
\label{data}
\begin{tabular}{ccccccccccccc}
\hline
RA&DEC&GMP&S/N&$\sigma$&$\sigma$$_{err}$&H$\delta$A&H$\delta$F&CN1&CN2&Ca4227 \\ 
G4300&H$\gamma$A&H$\gamma$F&Fe4383&Ca4455&Fe4531&C4668&H$\beta$&Fe5015&Mg1&Mg2 \\ 
Mgb&Fe5270&Fe5335&Fe5406&Fe5709&Fe5782&H$\delta$A$_{err}$&H$\delta$F$_{err}$&CN1$_{err}$&CN2$_{err}$&Ca4227$_{err}$ \\ 
G4300$_{err}$&H$\gamma$A$_{err}$&H$\gamma$F$_{err}$&Fe4383$_{err}$&Ca4455$_{err}$&Fe4531$_{err}$&C4668$_{err}$&H$\beta$$_{err}$&Fe5015$_{err}$&Mg1$_{err}$&Mg2$_{err}$ \\ 
Mgb$_{err}$&Fe5270$_{err}$&Fe5335$_{err}$&Fe5406$_{err}$&Fe5709$_{err}$&Fe5782$_{err}$&Flag \\ 
\hline
13:00:08.13&+27:58:37.0&2921&46.7&386.9&6.9&-2.63&-0.04&0.136&0.179&1.14 \\ 
5.40&-6.83&-2.04&5.01&1.68&3.84&7.91&1.44&5.30&0.169&0.325 \\ 
5.19&3.06&2.96&1.97&0.96&0.95&0.21&0.14&0.006&0.007&0.11 \\ 
0.18&0.22&0.14&0.28&0.16&0.24&0.36&0.16&0.34&0.004&0.004 \\ 
0.16&0.19&0.23&0.18&0.13&0.11&0 \\ 
13:04:58.38&+29:07:20.1&282&43.1&299.7&6.4&-1.34&0.64&0.087&0.112&1.03 \\ 
5.08&-5.85&-1.64&5.00&1.30&3.68&7.90&1.71&4.77&0.144&0.288 \\ 
4.68&2.76&2.74&2.00&1.06&0.76&0.25&0.17&0.007&0.008&0.13 \\ 
0.21&0.24&0.15&0.31&0.17&0.26&0.39&0.17&0.37&0.004&0.005 \\ 
0.18&0.21&0.25&0.19&0.15&0.12&0 \\ 
12:57:24.36&+27:29:52.1&4928&48.3&278.4&4.5&-1.93&0.47&0.103&0.141&1.13 \\ 
5.53&-6.27&-1.78&5.16&1.38&3.58&7.86&1.28&5.48&0.147&0.299 \\ 
4.82&2.88&2.70&1.93&0.97&0.76&0.22&0.15&0.006&0.007&0.12 \\ 
0.19&0.23&0.14&0.29&0.16&0.23&0.35&0.20&0.34&0.003&0.004 \\ 
0.16&0.19&0.22&0.16&0.12&0.10&0 \\ 
13:01:53.73&+27:37:28.0&1750&48.6&277.1&4.4&-2.20&0.19&0.124&0.161&1.19 \\ 
5.48&-6.80&-2.00&4.93&1.47&3.43&8.01&1.50&5.55&0.158&0.316 \\ 
4.89&3.19&3.07&2.05&0.91&0.94&0.19&0.13&0.005&0.006&0.10 \\ 
0.17&0.21&0.13&0.26&0.15&0.22&0.34&0.15&0.33&0.003&0.004 \\ 
0.15&0.19&0.22&0.17&0.12&0.10&0 \\ 
13:01:57.57&+28:00:21.0&1715&40.9&277.0&6.1&-2.71&-0.23&0.134&0.171&1.03 \\ 
5.05&-6.09&-1.92&4.47&1.04&3.44&8.44&1.83&4.80&0.165&0.316 \\ 
4.96&3.13&2.95&1.71&0.77&0.99&0.27&0.18&0.007&0.009&0.14 \\ 
0.22&0.26&0.16&0.34&0.18&0.27&0.40&0.18&0.39&0.004&0.005 \\ 
0.19&0.22&0.26&0.20&0.17&0.14&1 \\ 
12:59:35.71&+27:57:33.3&3329&43.7&273.5&5.1&-2.72&0.14&0.125&0.166&1.23 \\ 
5.49&-6.24&-1.86&5.00&1.36&3.65&6.87&1.54&5.44&0.147&0.292 \\ 
4.80&2.99&2.81&1.95&0.84&0.65&0.26&0.18&0.007&0.009&0.13 \\ 
0.22&0.26&0.16&0.32&0.17&0.26&0.39&0.17&0.37&0.004&0.005 \\ 
0.17&0.20&0.24&0.18&0.15&0.13&0 \\ 
12:59:03.90&+28:07:25.3&3792&46.0&273.2&4.8&-2.32&0.14&0.137&0.175&1.13 \\ 
5.58&-6.83&-2.19&5.23&1.35&3.39&8.33&1.23&5.83&0.163&0.327 \\ 
5.20&3.06&2.77&1.86&0.86&0.84&0.22&0.15&0.006&0.007&0.11 \\ 
0.19&0.23&0.14&0.28&0.16&0.24&0.36&0.16&0.35&0.004&0.004 \\ 
0.16&0.19&0.23&0.18&0.13&0.11&0 \\ 
13:01:33.60&+29:07:50.1&1990&40.3&269.6&5.9&-2.25&0.19&0.113&0.145&1.06 \\ 
5.80&-6.36&-1.84&4.56&1.52&3.57&6.01&1.66&4.56&0.132&0.279 \\ 
4.76&2.75&2.72&1.87&0.82&0.72&0.29&0.19&0.008&0.009&0.15 \\ 
0.24&0.28&0.18&0.36&0.19&0.28&0.43&0.18&0.40&0.004&0.005 \\ 
0.19&0.22&0.26&0.20&0.15&0.13&1 \\ 
12:56:18.60&+26:21:32.0& &48.2&263.2&4.2&-2.15&0.16&0.111&0.151&1.15 \\ 
5.16&-6.12&-1.70&4.83&1.32&3.58&8.70&1.37&5.33&0.151&0.298 \\ 
4.98&3.09&2.64&2.01&0.89&0.90&0.24&0.16&0.006&0.008&0.12 \\ 
0.20&0.23&0.14&0.29&0.16&0.24&0.34&0.15&0.33&0.004&0.004 \\ 
0.16&0.18&0.22&0.16&0.13&0.11&1 \\ 
12:56:43.52&+27:10:43.7&5279&52.4&259.9&2.7&-2.22&0.24&0.118&0.152&1.00 \\ 
5.20&-6.26&-2.01&4.84&1.45&3.73&8.64&1.48&5.53&0.155&0.306 \\ 
4.96&3.13&2.81&1.84&0.85&0.80&0.19&0.13&0.005&0.006&0.10 \\ 
0.17&0.20&0.13&0.26&0.14&0.21&0.32&0.14&0.30&0.003&0.004 \\ 
0.14&0.17&0.20&0.15&0.11&0.09&0 \\ 
12:55:41.30&+27:15:02.7&5886&51.5&258.3&2.7&-2.14&-0.07&0.110&0.150&1.02 \\ 
5.30&-6.21&-1.86&5.00&1.27&3.46&8.95&1.49&5.32&0.160&0.309 \\ 
4.91&2.92&2.78&1.93&0.80&0.93&0.20&0.13&0.005&0.007&0.10 \\ 
0.17&0.20&0.13&0.26&0.14&0.22&0.32&0.14&0.31&0.003&0.004 \\ 
0.14&0.17&0.21&0.16&0.12&0.10&0 \\  
\hline 
\end{tabular}
\end{table*}



\end{document}